%% file: main.tex
\documentclass[%
 reprint,
 amsmath,amssymb,
 aps,
pra,
]{revtex4-2}
\usepackage{upgreek}
\usepackage{amsmath}
\usepackage{graphicx}
\usepackage{dcolumn}
\usepackage{bm}
\usepackage{xcolor}
\usepackage[normalem]{ulem}
\usepackage{cancel}
\usepackage{multibib}
\usepackage{hyperref}

\begin{document}

\preprint{APS/123-QED}

\title{Defect-Assisted Recombination in Semiconductors and Photovoltaic Device Parameters from First Principles}

\author{Jiban Kangsabanik${}^{1,2}$}
\email{jibka@dtu.dk}
\author{Kristian S. Thygesen${}^{1}$}
\email{thygesen@fysik.dtu.dk}

\affiliation{${}^{1}$CAMD, Computational Atomic-Scale Materials Design, Department of Physics, Technical University of Denmark, 2800 Kgs. Lyngby, Denmark\\
${}^{2}$Department of Condensed Matter and Materials Physics, S. N. Bose National Centre for Basic Sciences, Kolkata 700106, India}

\begin{abstract}
We introduce a method to calculate defect-assisted Shockley-Read-Hall (SRH) recombination rates in imperfect semiconductors from first principles. The method accounts for the steady state recombination dynamics under given non-equilibrium conditions (split quasi Fermi levels), by invoking a full solution to the rate equations describing transitions across the band gap via all possible charge states of the defect. Transition rates due to radiative and non-radiative multi-phonon emission processes are calculated from first principles. The method is used to evaluate the effect of selected defects on the photovoltaic device parameters of seven emergent photovoltaic semiconductors. These examples clearly highlight the limitations of commonly employed approximations to the recombination dynamics. Our work advances the description and understanding of defect-induced losses in photovoltaics and provides a basis for developing the important concept of defect tolerant semiconductors and to discover high-performance photovoltaic materials computationally.
\end{abstract}

\maketitle

\section{Introduction}
Deep-level defects play a crucial role in modern semiconductor device technologies including transistors, photovoltaics, and quantum optics.\cite{Freysoldt2014, Alkauskas2016tut, Park2018, Dreyer2018} By virtue of their localized electronic states positioned deep inside the semiconductor band gap, deep-level point defects can capture free charge carriers and facilitate recombination of electrons and holes across the band gap. The latter type of processes are particularly relevant for photonic devices such as light emitting diodes and solar cells, where they lead to efficiency losses beyond the thermodynamic limit of the perfect (i.e. defect free) semiconductor. 

Carrier recombination processes involve two subsequent transitions, each involving a delocalised Bloch state and a localised defect state.\cite{shockley1952statistics} 
The physical interaction driving the transitions can be either coupling to the light field (radiative recombination) or coupling to phonons (non-radiative recombination)\cite{Dreyer2020, alkauskas2014first,Park2018} 
The carrier recombination dynamics of point defects was first investigated by Shockley, Read and Hall (SRH) during 1950’s and their empirical theory has been very successful in explaining various phenomena in semiconductor physics.\cite{shockley1952statistics, hall1952electron, henry1977nonradiative, rosier1971thermal} Recently, fully \emph{ab initio} approaches to calculate the carrier capture rates of point defects via radiative and non-radiative multiphonon emission have been developed \cite{Shi2012, alkauskas2014first, Wang2019, Dreyer2020} and successfully used to obtain microscopic interpretations of experimental observations made for photovoltaic materials such as kesterites and perovskites.\cite{Kim2020, Zhang2021inorganicpv, Dou2023, Wang2024}

Leaving aside the problem of calculating the rate of a specific defect transition, previous computational work on SRH recombination in photovoltaic materials have relied on some seemingly innocent assumptions about the recombination dynamics. It is commonly assumed that the net capture rate is governed by a single transition (the rate limiting step). For defects with only a single charge transition level (CTL) in the band gap this is motivated by the expectation that the capture of a minority carrier is much less likely than the capture of a majority carrier. For an $n$-type semiconductor this is true if $C_n n \gg C_p p$ ($C_{n/p}$ and $n/p$ are the capture coefficient and density of electrons/holes). However, since the ratio of the capture coefficients can be many orders of magnitude and may exceed the ratio of the carrier densities (under light illumination), this \emph{minority carrier approximation} becomes questionable.   

For defects with multiple CTLs in the gap the situation is more complicated. In such cases, two different approximations have been used. One invokes the minority carrier approximation and assigns a single transition as the rate limiting step based on the position of the defect level to the minority band. This is based on the fact that the transition rate due to phonon emission is exponentially suppressed with the transition energy. The other common approximation, assumes that the carrier densities are equal ($n=p$). This assumption leads to significant simplification of the the rate equations, but cannot be justified in general. In fact, it only holds for intrinsic (undoped) semiconductors, which are rare in practice and not well suited for photovoltaics, or in the limit of large quasi Fermi level splittings where the excited carrier density dominates the intrinsic carrier densities. The latter situation, however, corresponds to low losses and thus tacitly assumes that the net recombination rate (which is the objective of the calculation) is small. As such, a general formalism to calculate SRH recombination rates and photovoltaic parameters without implicit assumptions about the charge dynamics, is still lacking.\\  

In this work, we develop a methodology to compute the photovoltaic parameters of semiconductors without making any assumptions regarding the carrier concentrations, number of defect levels in the band gap, or the relative importance of different transitions. The total recombination current in the semiconductor is calculated taking as input the absorption coefficient of the pristine semiconductor, the equilibrium electron and hole carrier concentrations, the defect concentrations, and the capture rate of electrons and holes for all charge states of the defects. For defects with multiple charge transition levels, we show the importance of going beyond the common approximations and solve the full set of coupled charge transition rate equations. We introduce an effective capture coefficient that allows us to analyze the importance of the individual capture processes and to identify rate limiting steps at different doping concentrations and operating conditions.  

To illustrate our formalism, we provide several examples of relevance for emerging photovoltaic materials, and present an in-depth analysis of two specific systems. First, we consider the defect Sn$_{\text{Zn}}$ in Cu$_2$ZnSnS$_4$, which is known to be crucial for kesterite photovoltaics\cite{Yee2015, Kim2018, Li2019}. This defect has two charge transition levels inside the band gap and a harmonic defect potential energy surface. The second defect is Sn$_{\text{Se}}$ in trigonal selenium, which has four charge transition levels inside the band gap and a highly anharmonic defect potential energy surface.\cite{todorov2017ultrathin, Nielsen2024} These examples illustrate how our methodology advances previous schemes with important consequences for the predicted photovoltaic parameters. In addition, we discuss the importance of accounting for both radiative and non-radiative (multi-phonon) SRH carrier capture. In particular, we show how the interplay between the two recombination channels can influence the total recombination rate in non-trivial ways. \\

\begin{figure*}[!t]
    \centering
    \includegraphics[width=0.95\linewidth]{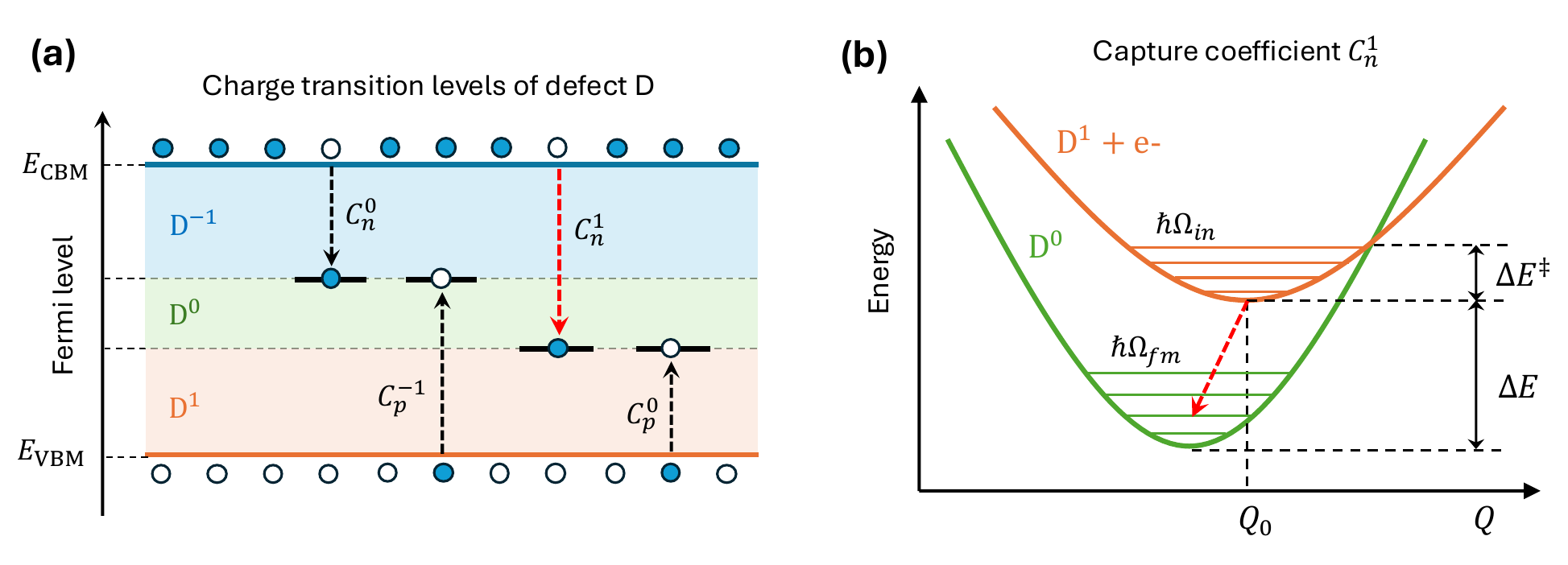}
    \caption{(a) Illustration of SRH carrier capture processes for a defect (D) with three possible charge states ($q=1$, $q=0$ and $q=-1$) and thus two charge transition levels. (b) Illustration of a 1D configuration coordinate (CC) diagram. For simplicity, the potential energy surfaces of the initial and final states have been assumed to have parabolic form, although the formalism can treat general shapes. The transition energy, $\Delta E$, the semiclassical transition barrier height, $\Delta E^{\ddagger}$, and the equilibrium configuration of the initial state, $Q_0$, are indicated. The horisontal lines indicate the discrete vibrational energy levels in the initial and final states, respectively.}
    \label{fig:Schematic}
\end{figure*}

\section{Methods}
\subsection{Rate equations}
The $R_\text{SRH}$ parameter for a deep level defect with a single charge transition level, $q/q-1$, can be expressed as (see SI for derivation), 
\begin{equation}
R_\text{SRH}=\frac{N_DC_n^qC_p^{q-1}(np-n_0p_0)}{[C_n(n+n^q)+C_p(p+p^q)]}
\label{eqSR7}
\end{equation}
Here $N_D$ is the defect concentration and $C_n^q$ ($C_p^{q-1}$) is the electron (hole) capture coefficient of the defect in initial charge state $q$ ($q-1$), see the leftmost transition in Fig. \ref{fig:Schematic}. In the above equation, $n=n_0+\Delta n$ and $p=p_0+\Delta p$ denote the electron and hole densities under steady state operating conditions while $n_0$ and $p_0$ denote the same quantities at equilibrium and $\Delta n$ and $\Delta p$ are the concentrations of photo-excited carriers. We note that $\Delta n=\Delta p$. The quantities $n^q$ and $p^q$ represent the electron and hole carrier densities with the Fermi level set to the charge transition energy $q/q-1$ of the defect ($q$ being the larger charge). 

The photo-excited carrier density, $\Delta n$, is obtained by equating the operating voltage $V$ with the difference between the electron and hole quasi Fermi levels.\cite{kim2020upper}
A simplified expression is\cite{kim2020upper},
\begin{equation}
	\Delta n(V)=\frac{1}{2}[-n_0-p_0+\sqrt{(n_0+p_0)^2-4n_0p_0(1-e^{\frac{eV}{k_\text{B}T}})]}
	\label{eqSR8}
\end{equation}

Before deriving the expression for $R_\text{SRH}$ for defects with multiple CTLs, we will first show an example for a deep level defect D with a single charge transition level, $q=1 \rightarrow q=0$. Under steady state operating conditions, during electron capture at charge $q$=1, the net electron capture rate can be expressed as,
\begin{equation}
R_n=N_1C_n^1n-N_{0}C_n^1n^1,
\label{eqmulti1}
\end{equation}
where $N_0$ and $N_1$ are the concentrations of the defect D in the two charge states, respectively. Obviously we have  
\begin{equation}
N_D=N_0+N_{1}
\label{eqmulti3}
\end{equation}
Similarly, the net hole capture rate with the defect is in charge state $q=0$, can be written as, 
\begin{equation}
R_p=N_{0}C_p^{0}p-N_1C_p^{0}p^1
\label{eqmulti2}
\end{equation}
In Eqs. (\ref{eqmulti1}) and (\ref{eqmulti2}) the rightmost terms represent the emission of carriers from the defect level to the bands. As we will show, these terms are often small, but can be crucial to include. They were considered in the original SRH theory, but have been neglected in previous first principles work.\cite{kim2020upper, Zhang2021inorganicpv}

In steady state it holds that $R_n=R_p$, which leads to
\begin{equation}
N_1C_n^1n-N_{0}C_n^1n^1=N_{0}C_p^{0}p-N_1C_p^{0}p^1
\label{eqmulti4}
\end{equation}
Solving Eqs. (\ref{eqmulti3}-\ref{eqmulti4}) we obtain the net recombination rate
\begin{widetext}
\begin{equation}
R_\text{SRH}=N_1C_n^1n-N_{0}C_n^1n^1=\frac{N_DC_n^1C_p^0(np-n^1p^1)}{[C_n^1(n+n^1)+C_p^0(p+p^1)]}=\frac{N_DC_n^1C_p^0(np-n_i^2)}{[C_n^1(n+n^1)+C_p^0(p+p^1)]}
\label{eqmulti5}
\end{equation}
\end{widetext}
which is same as Eq. (\ref{eqSR7}).

We follow a similar approach to derive the expression for the SRH recombination rate for a defect with two CTLs. For example, a defect with $q+1/q$ and $q/q-1$ charge transitions in the band gap, we obtain 
\begin{equation}
N_{q+1}C_n^{q+1}n-N_{q}C_n^{q+1}n^{q+1}=N_{q}C_p^{q}p-N_{q+1}C_p^{q}p^{q+1}
\label{eqmulti6}
\end{equation}
\begin{equation}
N_qC_n^qn-N_{q-1}C_n^qn^q=N_{q-1}C_p^{q-1}p-N_qC_p^{q-1}p^q
\label{eqmulti7}
\end{equation}
\begin{equation}
N_D=N_{q+1}+N_q+N_{q-1}
\label{eqmulti8}
\end{equation}
Here $n^{q+1}$, $p^{q+1}$ are the free electron and hole concentration if the Fermi level lies in the trap energy at CTL$(q+1/q)$. Similarly $n^q$, $p^q$ are the same if the Fermi level lies in the trap energy at CTL$(q/q-1)$. It follows from the mass-action law that
\begin{equation*}
n^{q+1}p^{q+1}=n^qp^q=n_0 p_0.
\end{equation*}
The net rate of recombination via the CTLs $(q+1/q \; \vee \; q/q-1)$ is
\begin{widetext}
\begin{equation}
R_\text{SRH}=N_{q+1}C_n^{q+1}n-N_{q}C_n^{q+1}n^{q+1}+N_qC_n^qn-N_{q-1}C_n^qn^q,
\end{equation}
which can be solved to yield
\begin{equation}
R_\text{SRH}=\frac{N_D}{G}[\frac{C_n^qC_p^{q-1}(np-n^qp^q)}{C_p^{q-1}p+C_n^qn^q} + \frac{C_p^qC_n^{q+1}(np-n^{q+1}p^{q+1})}{C_n^{q+1}n+C_p^qp^{q+1}}],
\label{eqSRH22}
\end{equation}
with
\begin{equation*}
G=[1+\frac{C_p^qp+C_n^{q+1}n^{q+1}}{C_n^{q+1}n+C_p^qp^{q+1}}+\frac{C_n^qn+C_p^{q-1}p^q}{C_p^{q-1}p+C_n^qn^q}].
\end{equation*}
\end{widetext}
In the same manner, closed expressions for $R_\text{SRH}$ for defects with multiple CTLs inside the band gap can be derived as shown in the SI for the case of three and four CTLs.


\subsection{Capture coefficients}
As mentioned earlier, the carrier capture process can be either radiative (involving photons) or non-radiative (involving phonons).
For photovoltaic operation, both radiative and non-radiative carrier capture will contribute to carrier loss, and the total capture coefficient, $C^q$, can be expressed as,
\begin{equation}
    C^q=C_{\mathrm{rad}}^q+C_{\mathrm{nonrad}}^q
    \label{eqcboth}
\end{equation}

The radiative capture coefficient can be calculated using Fermi's golden rule as,
\begin{equation}
	C_{\mathrm{rad}}^q=\frac{VE_{if}^{3}n_{\mathrm{ref}}\lvert \textbf{D}_{if}\rvert^{2}}{3\pi\epsilon_0\hbar^4c^3}
	\label{eqrad}
\end{equation}
Here, $V$ is the supercell volume, $n_{\mathrm{ref}}$ is the refractive index of the material, and $E_{if}$ is the emission energy and $n_{\mathrm{ref}}$ is the refractive index of the material. $\textbf{D}_{if}$ is the transition dipole moment between initial and final electronic states. The latter can be expressed as,\\
\begin{equation}
\textbf{D}_{if}=\langle\psi_f\lvert\hat{\textbf{r}}\rvert\psi_i\rangle=\frac{i\hbar}  {m_0}\frac{\langle\psi_f\lvert\hat{\textbf{p}}\rvert\psi_i\rangle}{\varepsilon_f-\varepsilon_i},
\end{equation}
where $\hat{\textbf{r}}$ and $\hat{\textbf{p}}$ are the dipole and momentum operators, respectively, and $m_0$ is the electron mass.

Using Fermi's golden rule from second order perturbation theory, the non-radiative capture coefficient can be calculated as,\cite{alkauskas2014first} 

\begin{widetext}
\begin{equation}
	C_{\mathrm{nonrad}}^q=\eta_s\frac{2\pi}{\hbar}gV|W_{if}|^2\sum_{m}w_m\sum_{n}\lvert\langle\chi_{im}\lvert\hat{Q}-Q_0\rvert\chi_{fn}\rangle\rvert^2 \delta(E_{if}+m\hbar\Omega_{im}-n\hbar\Omega_{fn})
	\label{eqCC2}
	\end{equation}
\end{widetext}

 Here, $\eta_s$ is the Sommerfeld factor, $g$ and $V$ are the degeneracy factor of the defect and volume of the supercell, respectively, $W_{if}$ is the electron-phonon coupling matrix elements between the initial and final electronic state while $\sum_{m}w_m\sum_{n}\lvert\langle\chi_{im}\lvert\hat{Q}-Q_0\rvert\chi_{fn}\rangle\rvert^2$ accounts for the overlap between the initial and final vibronic states defined by the 1D potential energy diagrams. The vibronic states are found by numerical solution of the 1D-Schrödinger equation in order to
 account for the anharmonicity of the defect potential energy surfaces, which can have significant effect on the calculated carrier capture coefficients.\cite{Kim2019} The occupations factors, $w_m$, of the initial state vibronic levels are taken to follow a Boltzmann distribution. The $\delta$-functions (replaced by Gaussians of width 0.8$\hbar\Omega_f$) expresses the energy conservation: The energy difference between the initial and the final states in their equilibrium geometry, $E_{if}$, must be matched by the difference in vibrational energy.

\section{Results and Discussion}
When the Shockley-Read-Hall (SRH) recombination due to deep-level defects is included in the photovoltaic efficiency equation, the defect limited power conversion efficiency\cite{shockley1961detailed, tiedje1984limiting, kim2020upper} can be expressed as,
\begin{equation}
	\eta=\frac{\max(V[J_\text{sc}-J_{r}^\text{rad}(e^{\beta eV}-1)-eR_\text{SRH}(V)d)])}{\int_{0}^{\infty}\hbar^2\omega I_\text{sun}(\omega) d\omega}
	\label{eqPV5}
	\end{equation}
In this expression, the denominator is the input power density, $P_\text{in}$, where the typical AM1.5G solar spectrum at 298 K is taken as incident solar spectrum, $I_\text{sun}(\omega)$. The numerator is the highest possible output power density corresponding to the maximum of $P=JV$. The output current density, $J$, at a given output voltage, $V$, consists of the short-circuit current density, $J_\text{sc}$, the radiative band-to-band recombination current density in the dark, $J_{r}^{\text{rad}}$, and the non-radiative recombination current density, $J_r^{\mathrm{nonrad}}=e R_\text{SRH}d$, where $R_\text{SRH}$ is the SRH recombination rate and $d$ the thickness of the absorber material. A detailed derivation of Eq. (\ref{eqPV5}) is given in the supplementary information (SI).

The initial condition (before light illumination) of the absorber is defined by the equilibrium Fermi level, which in turn is set by the doping level of the semiconductor. Upon light illumination, the population of electrons in the conduction band and holes in the valence band are described by the quasi Fermi levels. The difference between the quasi Fermi levels equals the output voltage, $eV$, and defines the operating condition. 

The calculation of PV parameters based on Eq. (\ref{eqPV5}) proceeds as follows: For a given set of initial and operating conditions, Eq. (\ref{eqSR8}) yields the photo-excited carrier concentrations, $\Delta n$. Together with the equilibrium carrier concentrations ($n_0$,$p_0$) this determines the total carrier densities ($n,p$) under operating conditions. These densities, together with the capture coefficients (defined below), are used to solve the rate equations for the steady state occupations of the different defect charge states, $N_q$. With these at hand, the SRH recombination rate, $R_{\mathrm{SRH}}$, can be obtained. The SRH recombination current together with the radiative band-to-band recombination current yields the total recombination current. By substracting the latter from the short-circuit current (equal to the number of photons absorbed per second) one obtains the total output current, $J$. The output current will decrease with increasing voltage, because the recombination current will increase as a result of the larger population of photo-excited carriers. The voltage at which $J$ becomes zero, defines the open circuit voltage, $V_{\mathrm{oc}}$. The voltage at which $JV$ is maximal, defines the defect limited power conversion efficiency, $\eta$.

\begin{table*}[t!]
\caption{\label{tab:table_1}
Key parameters describing the host semiconductors and defects explored in the work. These include: The experimental and calculated band gap ($E_g$) of the host materials, The exchange mixing parameter ($\alpha$) used for the HSE06 calculations, The charge transition levels (CTLs), The carrier capture coefficients for electron and hole capture in different charge states ($C_{n/p}^q$), The semiclassical capture energy barrier ($\Delta E^\ddagger$) related to defect assisted carrier capture and carrier capture coefficients for different host semiconductors and associated defects. The Total capture coefficient were calculated by adding the radiative and non-radiative contributions.}
\begin{ruledtabular}
\begin{tabular}{ccccccccccc}
Host & $E_g$ & $E_g$ & $\alpha$ & Defect (D) & CTL &   & & & Capture coefficients &    \\[5.6pt] \cline{7-11}
 & (expt.) & (HSE06) & (HSE06) & & & $C_{n/p}^q$ & $\Delta E^\ddagger$ & Non-rad. & Radiative & Total \\[5.6pt]
& (eV) & (eV) & & & (eV) & & (eV) & (cm$^3$s$^{-1}$) & (cm$^3$s$^{-1}$) & (cm$^3$s$^{-1}$) \\[5.6pt]
\hline\\[1pt]
CuInSe$_2$ & 1.01 & 1.03 & 0.28 & In$_\mathrm{Cu}$ & (2/1) = 0.50 & $C_n^2$ & 1.84 & 1.1x10$^{-18}$ & 3.0x10$^{-18}$ & 4.1x10$^{-18}$  \\[5.6pt]
& & & & & & $C_p^1$ & $>$5 & 2.1x10$^{-43}$ & 2.5x10$^{-16}$ & 2.5x10$^{-16}$ \\[5.6pt]
& & & & & (1/0) = 0.70 & $C_n^1$ & $>$5 & 1.3x10$^{-27}$ & 3.6x10$^{-19}$ & 3.6x10$^{-19}$ \\[5.6pt]
& & & & & & $C_p^0$ & $>$5 & 2.8x10$^{-41}$ & 3.3x10$^{-13}$ & 3.3x10$^{-13}$ \\[5.6pt]
CuGaSe$_2$ & 1.70 & 1.69 & 0.295 & Ga$_\mathrm{Cu}$ & (2/1) = 0.96 & $C_n^2$ & $>$5 & 2.3x10$^{-33}$ & 6.5x10$^{-16}$ & 6.5x10$^{-16}$  \\[5.6pt]
& & & & & & $C_p^1$ & 1.06 & 1.4x10$^{-28}$ & 1.0x10$^{-14}$ & 1.0x10$^{-14}$ \\[5.6pt]
& & & & & (1/0) = 1.20  & $C_n^1$ & $>$5 & 4.4x10$^{-32}$ & 9.8x10$^{-17}$ & 9.8x10$^{-17}$ \\[5.6pt]
& & & & & & $C_p^0$ & 1.34 & 2.5x10$^{-31}$ & 4.5x10$^{-13}$ & 4.5x10$^{-13}$ \\[5.6pt]
Cu$_2$ZnSnS$_4$ & 1.50 & 1.48 & 0.255 & Sn$_\mathrm{Zn}$ & (2/1) = 0.36 & $C_n^2$ & $>$5 & 6.6x10$^{-30}$ & 2.4x10$^{-15}$ & 2.4x10$^{-15}$  \\[5.6pt]
& & & & & & $C_p^1$ & 0.0007 & 2.9x10$^{-13}$ & 6.1x10$^{-17}$ & 2.9x10$^{-13}$ \\[5.6pt]
& & & & & (1/0) = 0.80 & $C_n^1$ & 0.34 & 2.3x10$^{-08}$ & 2.6x10$^{-16}$ & 2.3x10$^{-08}$ \\[5.6pt]
& & & & & & $C_p^0$ & 0.14 & 8.6x10$^{-11}$ & 8.3x10$^{-14}$ & 8.6x10$^{-11}$ \\[5.6pt]
Cu$_2$ZnGeS$_4$ & 2.15 & 2.19 & 0.27 & Ge$_\mathrm{Zn}$ & (2/1) = 0.38 & $C_n^2$ & $>$5 & 8.3x10$^{-30}$ & 1.4x10$^{-13}$ & 1.4x10$^{-13}$ \\[5.6pt]
& & & & & & $C_p^1$ & 0.03 & 2.6x10$^{-12}$ & 8.4x10$^{-19}$ & 2.6x10$^{-12}$ \\[5.6pt]
& & & & & (1/0) = 0.76 & $C_n^1$ & 0.83 & 9.4x10$^{-15}$ & 3.5x10$^{-14}$ & 4.4x10$^{-14}$ \\[5.6pt]
& & & & & & $C_p^0$ & 0.16 & 4.1x10$^{-10}$ & 1.7x10$^{-14}$ & 4.1x10$^{-10}$ \\[5.6pt]
Cu$_2$ZnSnSe$_4$ & 1.00 & 1.02 & 0.27 & Sn$_\mathrm{Zn}$ & (2/1) = 0.41 & $C_n^2$ & $>$5 & 6.0x10$^{-29}$ & 3.0x10$^{-17}$ & 3.0x10$^{-17}$  \\[5.6pt]
& & & & & & $C_p^1$ & 0.10 & 1.6x10$^{-13}$ & 2.4x10$^{-15}$ & 1.6x10$^{-13}$ \\[5.6pt]
& & & & & (1/0) = 0.67 & $C_n^1$ & $>$5 & 4.1x10$^{-21}$ & 2.8x10$^{-18}$ & 2.9x10$^{-18}$ \\[5.6pt]
& & & & & & $C_p^0$ & 0.47 & 3.1x10$^{-16}$ & 2.4x10$^{-13}$ & 2.4x10$^{-13}$ \\[5.6pt]
Cu$_2$ZnGeSe$_4$ & 1.36 & 1.35 & 0.275 & Ge$_\mathrm{Zn}$ & (2/1) = 0.24 & $C_n^2$ & $>$5 & 6.8x10$^{-32}$ & 2.8x10$^{-14}$ & 2.8x10$^{-14}$ \\[5.6pt]
& & & & & & $C_p^1$ & 0.0060 & 2.6x10$^{-11}$ & 9.8x10$^{-17}$ & 2.6x10$^{-11}$ \\[5.6pt]
& & & & & (1/0) = 0.62 & $C_n^1$ & 0.65 & 1.5x10$^{-12}$ & 4.0x10$^{-15}$ & 1.5x10$^{-12}$ \\[5.6pt]
& & & & & & $C_p^0$ & 0.28 & 4.7x10$^{-12}$ & 7.7x10$^{-14}$ & 4.8x10$^{-12}$ \\[5.6pt]
Se & 1.85 & 1.84 & 0.26 & Sn$_\mathrm{Se}$ & (2/1) = 0.17 & $C_n^2$ & $>$5 & 1.6x10$^{-126}$ & 1.8x10$^{-12}$ & 1.8x10$^{-12}$ \\[5.6pt]
& & & & & & $C_p^1$ & 0.03 & 2.9x10$^{-9}$ & 8.2x10$^{-19}$ & 2.9x10$^{-9}$ \\[5.6pt]
& & & & & (1/0) = 0.48 & $C_n^1$ & $>$5 & 8.1x10$^{-127}$ & 4.9x10$^{-13}$ & 4.9x10$^{-13}$ \\[5.6pt]
& & & & & & $C_p^0$ & 2.84 & 8.6x10$^{-17}$ & 1.9x10$^{-15}$ & 2.0x10$^{-15}$ \\[5.6pt]
& & & & & (0/-1) = 1.28 & $C_n^0$ & 0.0004 & 2.4x10$^{-7}$ & 5.0x10$^{-15}$ & 2.4x10$^{-7}$ \\[5.6pt]
& & & & & & $C_p^{-1}$ & $>$5 & 1.6x10$^{-16}$ & 5.1x10$^{-13}$ & 5.1x10$^{-13}$ \\[5.6pt]
& & & & & (-1/-2) = 1.57 & $C_n^{-1}$ & 0.01 & 1.6x10$^{-8}$ & 1.3x10$^{-17}$ & 1.6x10$^{-8}$ \\[5.6pt]
& & & & & & $C_p^{-2}$ & $>$5 & 2.0x10$^{-126}$ & 1.9x10$^{-12}$ & 1.9x10$^{-12}$ \\[5.6pt]
\end{tabular}
\end{ruledtabular}
\end{table*}

\clearpage

\subsection{SRH carrier capture}
Figure \ref{fig:Schematic}(a) shows a schematic diagram of carrier capture by a defect D with two charge transition levels (CTLs), 1/0 and 0/-1, inside the band gap. Electron capture can happen along two possible pathways: One is where the defect is initially in change state $q=1$, captures an electron from the conduction band and goes to charge state $q=0$. Alternatively, the defect can start in charge state $q=0$, capture an electron and go to charge state $q=-1$. The capture coefficients corresponding to these pathways are denoted $C_n^{1}$ and $C_n^{0}$, respectively. Similarly, the hole capture from the valence band can also proceed along two different pathways, namely $0/1$ ($C_p^{0}$) and $-1/0$ ($C_p^{-1}$). For defects with three CTLs there will be three pathways for both electron and hole capture, and so on. 

In this work we consider two different physical interactions driving the carrier capture process, namely electron-light (radiative) and electron-phonon (non-radiative) interactions, respectively. While the radiative transition rate is essentially independent of temperature and increases with the transition energy as $E^3$, the non-radiative transition rate is strongly temperature dependent and decreases exponentially with the transition energy. As such the two types of interactions impact the carrier dynamics in distinctly different ways and can exhibit a non-trivial interplay.  

Calculation of the radiative capture coefficient is a straightforward application of Fermi's golden rule involving a dipole transition matrix element and the photonic density of states, see Eq. (\ref{eqrad}). The non-radiative capture via multi-phonon emission is more involved. The full expression, see Eq. (\ref{eqCC2}), involves an electron-phonon matrix element weighted by a term containing vibronic matrix elements. The situation is illustrated by the one-dimensional (1D) adiabatic configuration coordinate (CC) diagram in Figure \ref{fig:Schematic}(b). The two curves represent the total energy in the initial and final state along the 1D path connecting the initial and final state configurations. Although the curves are shown as parabolas, the formalism applies to CC diagrams of any shape. This is ensured by obtaining the vibrational states and energies as the numerically exact solutions to the 1D Schrödinger equation. More details on the methodology can be found under Methods and in the SI.

\subsection{Effective capture coefficient}
The individual capture coefficients are essentially transition rates between initial and final states. As such they are independent of both the initial and operating conditions.  In contrast, the SRH recombination parameter, $R_{\mathrm{SRH}}$, depends on the recombination dynamics via the occupation numbers of the individual charge states ($N_q$) and the free carrier concentrations ($n$ and $p$). It is thus a function of both the initial and operating conditions. To calculate $R_{\mathrm{SRH}}$, we solve the rate equations under steady state conditions for a given equilibrium Fermi level and output voltage, see Methods.  

As mentioned in the introduction, previous work has obtained $R_{\mathrm{SRH}}$ using various approximations of the rate equations, e.g. minority carrier approximation, neglect of emission processes, and an $n=p$ assumption. In the SI, we show how these arise from the general formalism and in Sec. \ref{methodSRH} we give examples to illustrate their quantitative limitations for practical PV calculations.

It is useful to define an effective, defect-specific capture coefficient, $C_{\mathrm{eff}}$, as
\begin{equation}
    R_{\mathrm{SRH}}=N_DC_{\mathrm{eff}}\Delta n,
    \label{eqc_eff}
\end{equation}
where $N_D$ is the concentration of the considered defect and $\Delta n$ is the density of photo-excited carriers. By analyzing how the individual capture rates via the different charge states contributes to $C_{\mathrm{eff}}$, we can identify and rationalize the rate limiting step under different operating conditions.

\subsection{The defect systems}
Copper indium gallium selenide (CIGS) and kesterites are popular semiconductor materials for thin film photovoltaics. They possess band gaps in the range of 1-1.5 eV, making them ideal for single junction PV applications.\cite{Dou2023} While CIGS solar cells show high photovoltaic device efficiency close to 23\%, they suffer from material scarcity due to requirement of Indium (In). Kesterites, on the other hand, can be seen as derivatives of CIGS where In$^{+3}$ are replaced by Zn$^{+2}$ and Sn$^{+4}$ (to balance the charge neutrality) both of which are relatively inexpensive and abundant in nature. Now, the kesterite Cu$_{2}$ZnSnS$_{4}$ exhibit relatively low device efficiency in the range of 10-12\% despite having a band gap close to the ideal Shockley-Queisser region (1.5 eV).\cite{Crovetto2020} Poor defect tolerance has been proposed to be the reason behind, and Sn$_{\mathrm{Zn}}$ was proposed to be an important detrimental defect.\cite{Kim2018, kim2020upper} To increase the efficiency, it has been proposed to replace Sn with Ge and/or S with Se, in order to tailor the the band gap and improve the defect tolerance.\cite{scaffidi2023ge} To illustrate our formalism, we investigate six compounds in this series, namely CuInSe$_{2}$, CuGaSe$_{2}$, Cu$_{2}$ZnSnS$_{4}$, Cu$_{2}$ZnGeS$_{4}$, Cu$_{2}$ZnSnSe$_{4}$ and Cu$_{2}$ZnGeSe$_{4}$. For each compound we study the most important (detrimental) defect identified in earlier studies, namely In$_{\mathrm{Cu}}$, Ga$_{\mathrm{Cu}}$, Sn$_{\mathrm{Zn}}$, Ge$_{\mathrm{Zn}}$, Sn$_{\mathrm{Zn}}$, and Ge$_{\mathrm{Zn}}$, respectively. For each of these defect systems we calculate all carrier capture rates, the total SRH recombination rates, and the photovoltaic device parameters, in order to determine the importance of the considered defects on the PV performance. Our results will thus also shed light on the question whether elemental chemical substitution in the Cu$_{2}$ZnSnS$_{4}$ can be used to improve its defect tolerance.

\begin{figure*}[t!]
    \centering
    \includegraphics[width=0.95\linewidth]{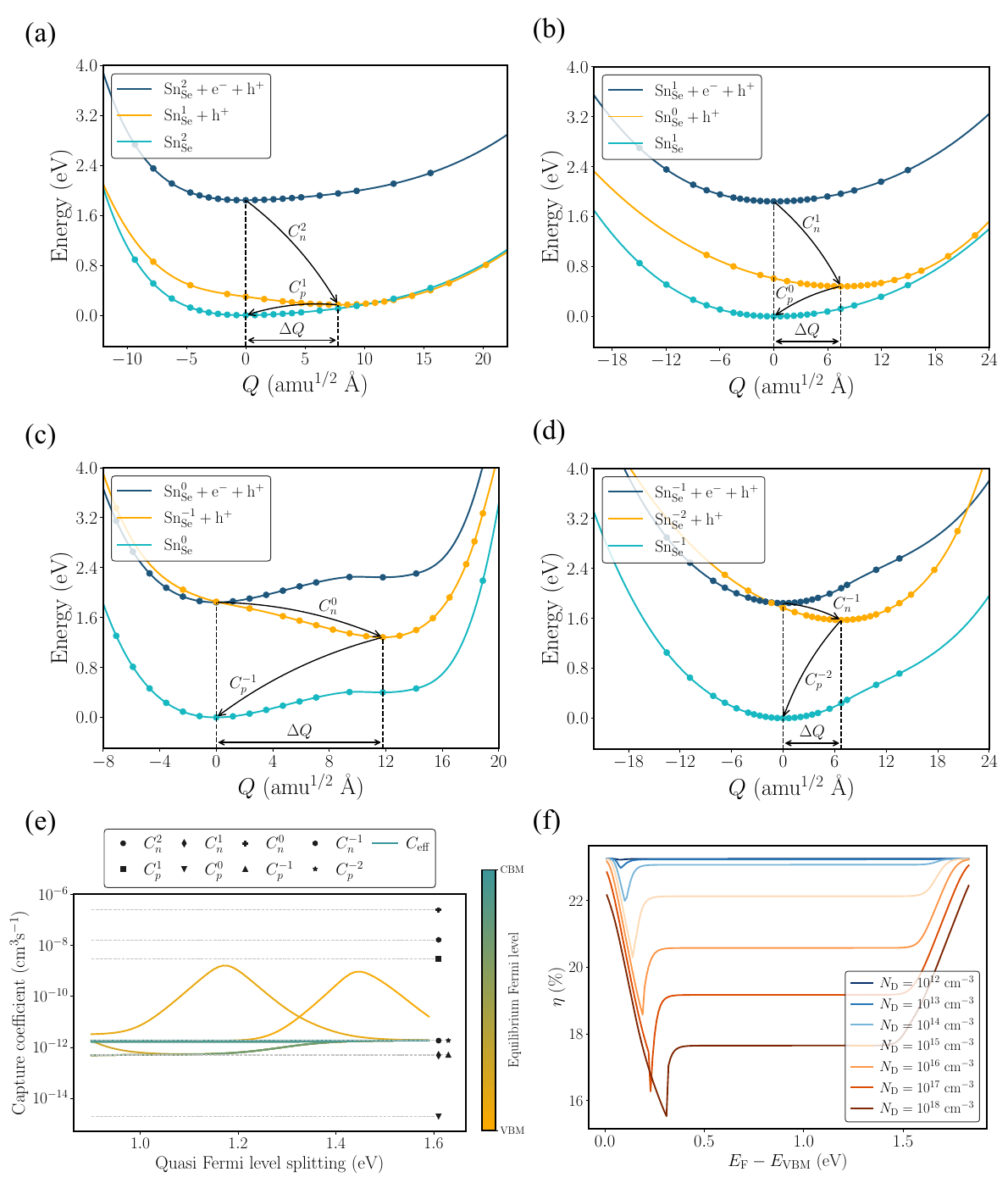}
    \caption{(a)-(d) One-dimensional configuration coordinate diagrams for the $2/1$, $1/0$, $0/-1$, $-1/-2$ charge state transitions of the Sn$_{\mathrm{Se}}$ defect in t-Se, respectively. Solid circles denote the data points calculated using HSE06 functional and solid lines are obtained by fitting them with quadratic spline functions accounting for anharmonicity. (e) Carrier capture coefficients  (including both radiative and nonradiative capture processes) at different charge states and associated effective capture coefficient ($C_{\mathrm{eff}}$) with respect to quasi-Fermi level splitting at different majority carrier types (equilibrium Fermi levels denoted by the color scale) at 300 K.(f) Photovoltaic device efficiency ($\mathrm{\eta}$) with respect to the equilibrium Fermi level position at different defect concentrations at 300 K and 500 nm device thickness.}
    \label{fig:Sn_Se}
\end{figure*}

The fact that the CIGS and kesterite semiconductors have relatively narrow band gaps, restricts the number of possible CTLs, which in practice rarely exceeds two. Large band gap semiconductors, which are relevant for tandem cell applications, photocatalysis, and for indoor photovoltaics, may contain defects with more CTLs inside the band gap. One such example is trigonal Selenium (t-Se), which has a band gap of 1.85 eV making it ideal for applications in tandem and indoor photovoltaics. The extrinsic defect Sn$_{\mathrm{Se}}$ is believed to form during synthesis of t-Se (due to the employed substrate). \cite{Liu2023} This defect gives rise to four different CTLs inside the band gap some of which have highly anharmonic PESs (see Fig. \ref{fig:Sn_Se}). We study this defect system as an example of a particularly complex and theoretically challenging case.

All the defect systems, together with their calculated CTLs levels and capture rates, are shown in Table \ref{tab:table_1}. A graphical overview of the CTLs in the seven defect systems can be found in the SI, see in Figure \ref{fig:ctls}.

\begin{figure*}[t!]
    \centering
    \includegraphics[width=0.95\linewidth]{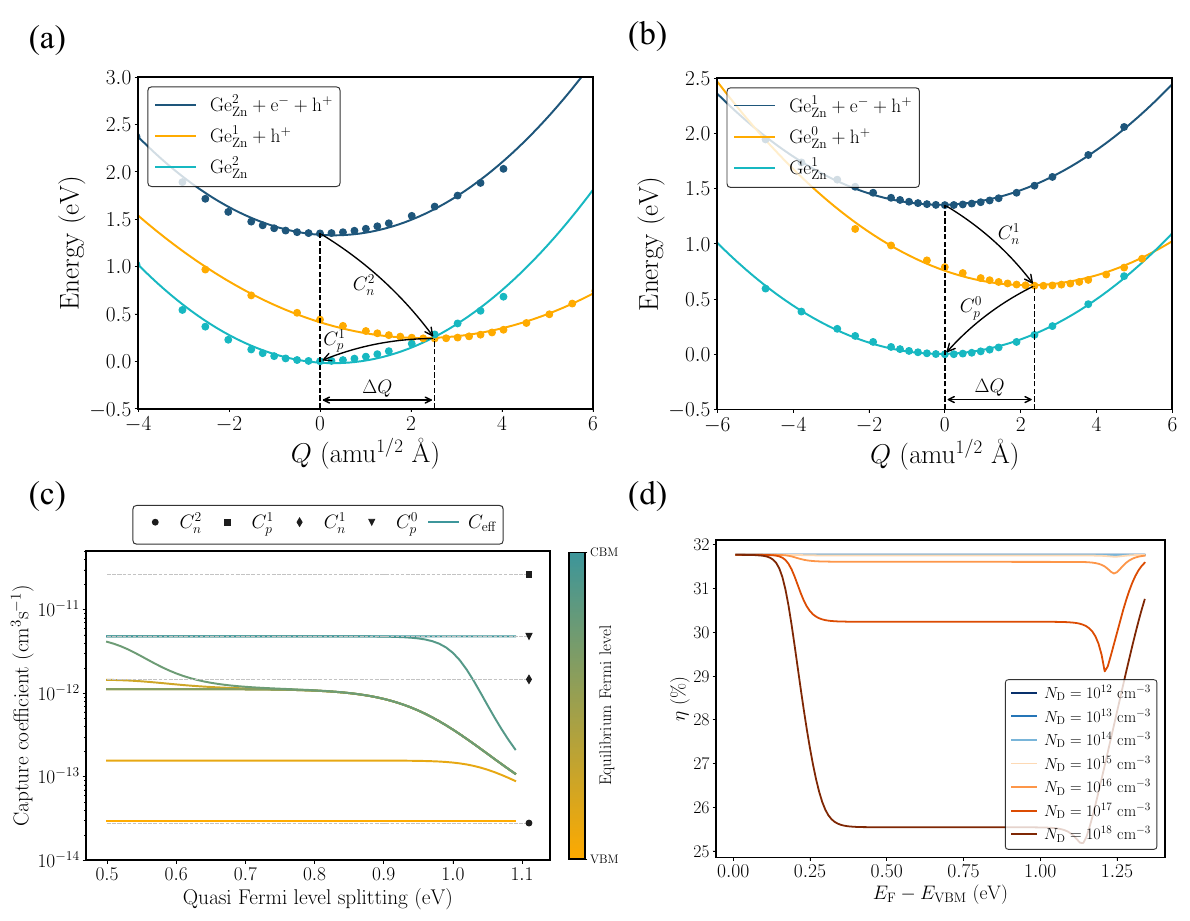}
    \caption{(a), (b) One-dimensional configuration coordinate diagrams for the $2/1$ and $1/0$ charge state transitions in the Ge$_{\mathrm{Zn}}$ defect in Cu$_{2}$ZnGeSe$_{4}$. Solid circles denote the data points calculated using the HSE06 functional and solid lines represent parabolic fits. (c) Carrier capture coefficients (including both radiative and non-radiative capture processes) for different charge states are shown as black symbols. The colored lines show the effective capture coefficient as a function of the quasi-Fermi level splitting at different majority carrier types (equilibrium Fermi levels denoted by the color scale) at 300 K.(d) Photovoltaic device efficiency with respect to equlibrium Fermi level position at different defect concentrations at 300 K and 500 nm film thickness.}
    \label{fig:CZGSe}
\end{figure*}

\subsection{Ge$_{\mathrm{Zn}}$ in Cu$_{2}$ZnGeSe$_{4}$}
From the kesterite defect series we consider the case of Ge$_{\mathrm{Zn}}$ in Cu$_{2}$ZnGeSe$_{4}$ as a representative example. This defect, like the other kesterite defect systems considered, have two CTLs in the band gap.  The $2/1$ CTL is positioned at 0.24 eV above the valence band maximum (VBM) while the $1/0$ CTL occurs at 0.62 eV above the VBM. These values are in reasonable agreement with existing literature.\cite{kim2020upper} We emphasize that an accurate calculation of the CTL energies is crucial as the non-radiative capture coefficients are very sensitive to the transition energies. Here, we have used the HSE06 functional with the exchange mixing parameter optimized to reproduce the experimental band gap, see Table \ref{tab:table_1}.

Figure \ref{fig:CZGSe}(a,b) shows the CC diagrams for the $2/1$ and $1/0$ charge state transitions, respectively. 
It is noted that for this defect all the CC curves are harmonic to a good approximation within the relevant energy range. In the next section we consider an example where this is not the case. We further note that the shape of the CC curves only depend on the charge state of the defect and not on the presence of an electron in the CBM or VBM. In particular, the CC curve for $\mathrm{D}^q+\mathrm{e}^-+\mathrm{h}^+$ (before recombination) and $\mathrm{D}^q$ (after recombination), are identical. This because the presence of the delocalized electron/hole in the conduction/valence bands, to a good approximation, do not exert any force on the atoms.  

To understand the recombination dynamics in this system, we first consider some basic features of the two CC diagrams separately. 
From Figure \ref{fig:CZGSe}(a) it can be seen that because the $2/1$ CTL sits much closer to the VBM than the CBM, the (semi-classical) energy barrier for hole capture is much smaller than the barrier for electron capture, see Table \ref{tab:table_1}. In fact, the CC curves describing the electron capture do not even intersect, resulting in an infinite barrier and a negligible non-radiative capture coefficient ($C_n^{2}\approx10^{-32}$). In comparison, the hole capture coefficient is much larger ($C_p^{1}\approx10^{-11}$). From Figure \ref{fig:CZGSe}(b) we can see that in the case of the $1/0$ charge transition, the defect level is in the mid gap region. Consequently, the capture barriers for electron capture (0.65 eV) and hole capture (0.28 ev), are similar. The small difference can be attributed to the facts that the CTL is not exactly in the middle of the band gap and that the CC curves are not perfectly harmonic. As expected from the similar transition energies and barriers, the non-radiative capture coefficients, $C_n^{1}$ and $C_p^{0}$, are very close ($\sim 10^{-12}$). We note in passing that the Sommerfeld factor \cite{passler1976relationships}, which accounts for the increase/decrease in the local amplitude of the free carrier wave function around a charged defect, also influence the capture coefficients, although only to a moderate degree (see the SI).

The above considerations mainly concern the non-radiative capture via multi-phonon emission.\cite{alkauskas2014first} The total capture coefficient also includes the radiative mechanism, see Eq. (\ref{eqcboth}). As can be seen from Table \ref{tab:table_1}, the radiative capture coefficients generally vary much less from defect to defect and charge state to charge state, than does the non-radiative ones. As previously noted, this is due to the weaker $E^3$-dependence of the radiative transition rate on the transition energy as compared to the exponential dependence of the non-radiative transition rate.

Figure \ref{fig:CZGSe}(c) shows the total (radiative plus non-radiative) capture coefficients of the individual processes (black symbols) together with the effective capture coefficient, $C_{\mathrm{eff}}$, defined in Eq. (\ref{eqc_eff}) (colored lines). The latter is plotted as a function of the quasi Fermi level splitting and the equilibrium Fermi level (color coded).  
It can be seen from Figure \ref{fig:CZGSe}(c) that for heavy p-type doping (yellow), $C_{\mathrm{eff}}$ equals $C_n^2$. Thus the transition $2\to 1$ becomes the rate limiting step. For heavy p-type doping, the number of electrons in the conduction band is much smaller than the number of holes in the valence band. Thus it is expected that the rate limiting step will be of the electron capture type. This is the basis of the minority carrier approximation for a single CTL. Now, in the present case of two CTLs, there are two electron capture processes, $2\to 1$ and $1\to 0$. The former CTL is further away form the CBM than the latter. Based on the general decreasing behavior of the capture rate with transition energy, one would thus expect the $2\to 1$ transition to be rate limiting. From Table \ref{tab:table_1} it follows that indeed $C_n^2 < C_n^1$. Similar arguments can be used to rationalize that the transition $0\to 1$ becomes rate limiting under heavy n-type doping conditions ($C_p^0 < C_p^1$). We stress that the above arguments do not always hold. For example, for the In$_\mathrm{Cu}$ defect in CuInSe$_2$, we have $C_n^2>C_n^1$ (see Table \ref{tab:table_1}), even though the position of the CTLs relative to the CBM would suggest the opposite. This happens because the structural rearrangements during the $1/0$ transition are such that the transition becomes highly unlikely with a large energy barrier of $>5$eV. The example shows that simple arguments based on carrier concentrations and defect energies must be used with care when predicting 
recombination dynamics. A more detailed discussion of the commonly used approximations is given Sec.\ref{methodSRH}).

Above we discussed examples where the carrier density and the relative size of the capture coefficients, determine the rate limiting transition. However, another important role is played by the densities of the defect charge states. In steady state, these are determined by the capture coefficients and the initial and operating conditions. Figure \ref{fig:ND6} in the SI shows the densities of the charge states of all the defect systems as function of equilibrium Fermi level. It can be seen that the densities of the different charge states vary with the doping level and the operating conditions. This further illustrates the difficulty of predicting the rate limiting transition without solving the full set of rate equations.   

Figure \ref{fig:CZGSe}(d) shows the device efficiency at room temperature for a $0.5\mu\mathrm{m}$ thick  Cu$_{2}$ZnGeSe$_{4}$ film. Specifically, we show the maximum defect limited conversion efficiency, $\eta$, as function of the equilibrium Fermi level and for different concentrations of the  Ge$_{\mathrm{Zn}}$ defect. The defect concentration ranges from very low (10$^{12}$ cm$^{-3}$) to very high (10$^{18}$ cm$^{-3}$). While a defect concentration corresponding to a thermal equilibrium distribution is often assumed in theoretical works, we find it more instructive to consider a broad range of defect concentrations. This is because the actual defect concentrations in synthesized samples depend strongly on the growth conditions and can deviate significantly from the equilibrium concentration. For all, but the highest defect concentrations, the efficiency stays close to the theoretical maximum given by the Shockley-Queisser limit. This shows that the Ge$_{\mathrm{Zn}}$ defect does not have a particular detrimental effect on the PV performance. 

From Figure \ref{fig:CZGSe}(d) we also note that the efficiency is higher for n- or p-type doping compared to the intrinsic semiconductor. This is a general trend found for all the defects studied in this work and can be ascribed to the following effect: When the equilibrium Fermi level sits close to the VBM or CBM, a given quasi Fermi level splitting, i.e. voltage, will produce a lower SRH recombination current as compared to the case where the equilibrium Fermi level lies in the mid-gap region. This is because of the way $R_{\mathrm{SRH}}$ depends on the carrier densities. For example, in the simplest case of a single CTL, the dependence is $(np-n_0p_0)/(n+p)$ (see e.g. Eq. (\ref{eqSR7})). While the numerator is independent of the equilibrium Fermi level, the denominator ($n+p$) will be smaller (by orders of magnitude) when the equilibrium Fermi level is at mid gap compared to when it is close to VBM or CBM, and consequently $R_{\mathrm{SRH}}$ will be larger. Referring to Eq. (1), the short-circuit current, $J_{\mathrm{sc}}$, is independent of the equilibrium Fermi level (apart from Pauli blocking effects, which are minimal under PV conditions, see Sec. \ref{methodology}(E)). Also the radiative recombination current is independent of the equilibrium Fermi level (it depends on carrier densities as $np$). However, as argued above, the magnitude of the third term, $R_{\mathrm{SRH}}$, will decrease for larger doping concentrations, resulting in a higher efficiency.

\subsection{Sn$_{\mathrm{Se}}$ in t-Se}
The defect Sn$_{\mathrm{Se}}$ in t-Se shows four CTLs deep inside the band gap. The lowest positioned is the $2/1$ CTL (0.17 eV above VBM) and the highest is the $-1/-2$ CTL (0.27 eV below CBM). The other two levels $1/0$ and $0/-1$ are positioned at 0.48 eV and 1.28 eV above the VBM respectively. Figure \ref{fig:Sn_Se} (a-d) shows the 1-D CC diagrams for the the four CTLs. We notice that unlike the kesterites, the defect potential energy curves show significant anharmonicity, most prominently for the $0/-1$ transition (panel (c)).

Next, we discuss how the size of the carrier capture coefficients for each CTL is related to its position in the band gap, the shape of its potential energy curve, and the initial charge state of the defect. A numerical overview can be found at the bottom of Table \ref{tab:table_1}. Because the $2/1$ and $1/0$ CTLs are positioned far away from the CBM (more than 1 eV), the energy barrier,$\Delta E^\ddagger$, for electron capture is huge ($>5$ eV). Consequently, the their non-radiative electron capture coefficients are essentially zero. Oppositely, the $-1/0$ and $-2/-1$ levels are far from the VBM (more than 1 eV) and consequently their barriers for hole capture are huge and capture coefficients vanishingly small.

The radiative capture coefficient also remains low. Further, Sn$_{\mathrm{Se}}^{+2}$, being an attractive center for electrons, the Sommerfeld factor is high and as such $C_n^{2}$ shows a low-moderate value of $\approx$ 10$^{-12}$ cm$^{3}$ s$^{-1}$ (See Table \ref{tab:table_1}). Similar trend follows for $C_n^{-1}$.Whereas, for $C_n^{0}$ and $C_n^{-1}$, the positions of the corresponding CTLs remain close to CBM, resulting in lower capture barrier. This contributes to a higher non-radiative capture rate ($\approx$ 10$^{-7}$-10$^{-8}$ cm$^{3}$ s$^{-1}$) (See Table \ref{tab:table_1}) despite them being neutral and electron repulsive centres respectively. Similarly, for hole capture, the position of the CTL makes the largest impact here and $C_p^{1}$ ($\approx$ 10$^{-9}$ cm$^{3}$ s$^{-1}$) remains the highest. For the other charge transition levels, as we go further away from the VBM, the hole capture becomes low ($\approx$ 10$^{-11}$-10$^{-13}$ cm$^{3}$ s$^{-1}$) as the $\Delta E^\ddagger$ increases (See Table \ref{tab:table_1}), with radiative capture being the main contributor. 

With a four level defect such as this one, it becomes very interesting and complex to intuitively understand which defect level consists the rate limiting transition. It can be seen from figure \ref{fig:Sn_Se} (e), that at p-type condition (t-Se is known to be synthesized as p-type material),\cite{Nielsen2022} the effective capture rate varies very much depending on the initial and operating conditions which cannot be captured by the simpler approximations, like, minority carrier approximation or $n=p$ approximation. We plot the defect charge state occupation during operation (See Figure \ref{fig:ND7}), where one can see that at p-type conditions ($p_0$ $\approx$ 10$^{16}$ cm$^{-3}$) at low quasi-Fermi level splitting the Sn$_{\mathrm{Se}}$, defect remains similarly distributed between +1 and +2 charge state, and as a result, $C_n^{1}$ and $C_n^{2}$ becomes rate limiting steps and both being low the effect of the defect is small. The picture changes during operation and at close to 1.2 eV quasi-Fermi level splitting the defect levels become almost equally occupied (See Figure \ref{fig:ND7}) and as a result the effective capture rate suddenly jumps towards the higher values as $C_n^{0}$, $C_n^{-1}$ take prominent role in recombination. This explains that at p-type conditions at moderate (10$^{16}$ cm$^{-3}$) to high (10$^{18}$ cm$^{-3}$) defect concentrations, Sn$_{\mathrm{Se}}$ can become detrimental in the case of t-Se photovoltaics (See Figure \ref{fig:Sn_Se} (f)). On the other hand for, n-type and intrinsic conditions, at all quasi-Fermi level splitting, the defect remains at -2 charge state and as such the rate limiting step remains $C_p^{-2}$ which is low.

Here, also the effect of including the radiative capture can be effectively seen from comparing the Figure \ref{fig:radiativec}(c) and \ref{fig:radiativec}(d). Inclusion of radiative capture channel has important effect on the total defect capture and hence defect charge state distribution (as clear from Eq \ref{eq_N2N1}-\ref{eq_N1N0}) during operation. As a result, the net capture rate can vary significantly (four orders of magnitude here), affecting the conclusions regarding defect tolerance.

\begin{figure*}[t!]
    \centering
    \includegraphics[width=0.95\linewidth]{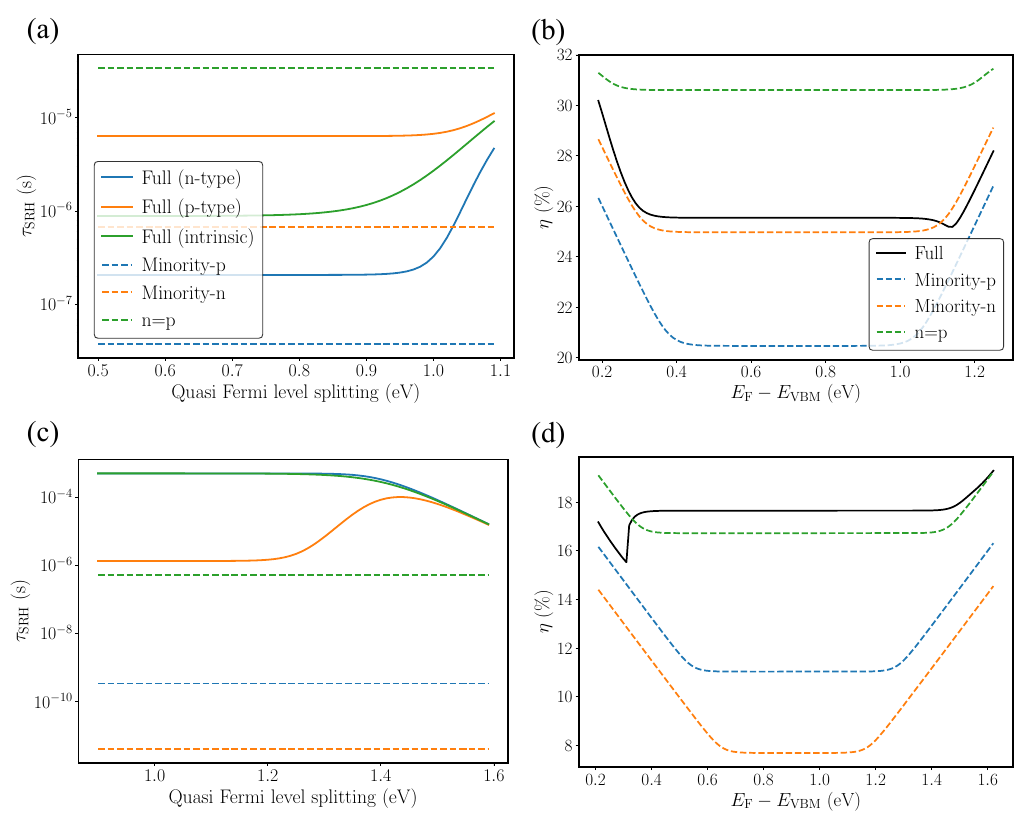}
    \caption{(a) Comparison of effective carrier lifetime ($\tau _{\mathrm{SRH}}$) at different majority carrier types (p-type, n-type, intrinsic) between our full formalism (Full), minority carrier approximation (Minority), and $n=p$ approximation (n=p) for Ge$_{\mathrm{Zn}}$ defect in Cu$_{2}$ZnGeSe$_{4}$. (b) Photovoltaic device efficiency ($\mathrm{\eta}$) with respect to equilibrium Fermi level with different approximations at $N_{\mathrm{D}}$=10$^{18}$ cm$^{-3}$, T=300 K, and 500 nm device thickness for Ge$_{\mathrm{Zn}}$ defect in Cu$_{2}$ZnGeSe$_{4}$.  (c)-(d) same as (a), (b), respectively for Sn$_{\mathrm{Se}}$ defect in t-Se.}
    \label{fig:comparison}
\end{figure*}

\subsection{CIGS and kesterites}
In this section we use our formalism to evaluate the effect of SRH recombination for some of the most common point defects in the emerging CIGS and kesterite PV materials. Since the central concepts and relation between microscopic quantities and device parameters have already been introduced in the previous sections, the discussion will focus on the results with fewer details on the underlying microscopic mechanisms.

\subsubsection{CIGS}
For CIGS solar cells, the antisite defects In$_{\mathrm{Cu}}$ and Ga$_{\mathrm{Cu}}$ (known as III$_\mathrm{I}$ antisite defects) are considered to be responsible for non-radiative recombination in CuInSe$_{2}$ and CuGaSe$_{2}$, respectively. However, first-principles calculations by Dou \emph{et al.}\cite{Dou2023} show that these defects themselves cannot explain the observed recombination rates. It can be seen in Table \ref{tab:table_1} that the In$_{\mathrm{Cu}}$ defect in CuInSe$_{2}$ and the Ga$_{\mathrm{Cu}}$ defect in CuGaSe$_{2}$ introduce two CTLs inside the band gap. Inspection of the CC diagrams show that the defect potential energy surfaces are almost parallel to each other (see Figures \ref{fig:CISPV}(a, b) and \ref{fig:CGSPV}(a, b) in SI), and as a consequence the capture energy barrier, $\Delta E^\ddagger$, becomes very high. This results in low total SRH carrier capture coefficients for all the transitions. From Figures \ref{fig:CISPV}(c) and S4(c) in the SI, we can see that the effective capture rates remain low at all doping conditions. As a result, the PV properties of CuInSe$_{2}$ and CuGaSe$_{2}$ are insensitive to the considered defects, even for very high defect concentrations. Our conclusion is in agreement with results of Dou \emph{et al.}\cite{Dou2023}. To explain the experimental data, Dou \emph{et al.} proposed a mechanism involving a thermally activated distortion of the defect structure in the neutral charge state for CuGaSe$_{2}$, which would enable hole capture, i.e. increase $C^0_p$. We did not explore this mechanism in the current work.

\subsubsection{Kesterites}
For kesterites, there has been experimental reports of deep level defects (around 0.8 eV above the VBM) for Cu$_{2}$ZnSnS$_{x}$Se$_{4-x}$.\cite{miller2012electronically} Islam \emph{et al.} reported the presence of a deep level defect 1.1 eV above the VBM for Cu$_{2}$ZnSnS$_{4}$ (CZTS).\cite{islam2015determination} This particular defect was not found in Cu$_{2}$ZnSnSe$_{4}$, resulting in better device performance than both CZTS and Cu$_{2}$ZnSnS$_{x}$Se$_{4-x}$. There are relatively few high-level calculations of SRH recombination for the kesterites (I$_{2}$-II-IV-VI$_{4}$). Most calculations employed a small 64 atom supercell and did not tune the $\alpha$ parameter of the HSE functional in order to match the experimental band gap.\cite{ratz2022relevance, Li2019, Xu2021, Han2013}  For CZTS the only high quality studies are reported by Kim \emph{et al.}\cite{kim2020upper, Kim2018, kim2019lone}  From their work, it is however, unclear whether the investigated Sn$_{\mathrm{Zn}}$ defect has a role in non-radiative recombination. In particular, the CTL energies reported in their papers are inconsistent. As a result, their conclusions regarding the size and importance of the non-radiative capture coefficients vary significantly. Moreover, only one carrier capture process at one CTL (namely, the electron capture $C_n^{2}$) was considered whereas hole capture as well as the other CTL were ignored. As we have shown in the current work, it can be crucial to consider all the carrier capture processes and their interplay during operation. 

Using our full SRH formalism, we have revisited the role of the kesterite IV$_\mathrm{II}$ antisite defects (an intrinsic element with +4 oxidation state substitutes one with +2 oxidation state, e.g. Sn$_{\mathrm{Zn}}$). Based on our HSE06 calculations with optimized $\alpha$-parameter and full solution of the SRH rate equations, we conclude that for CZTS, Sn$_{\mathrm{Zn}}$ has a moderate effect on the efficiency, while for other stoichiometric kesterites, this defect only has a limited effect (see Figures \ref{fig:CZGSe}(d), \ref{fig:CZTS}(d), \ref{fig:CZGS}(d), \ref{fig:CZTSe}(d)). This agrees well with recent experimental reports where defect clusters are reported to be responsible for non-radiative recombination losses in Cu$_{2}$ZnSn(S,Se)$_{4}$.\cite{Crovetto2020}
Below we elaborate on our results for the IV$_\mathrm{II}$ antisite defects in kesterites.

For the Sn$_{\mathrm{Zn}}$ in Cu$_{2}$ZnSnS$_{4}$ (CZTS), we can see from Figures \ref{fig:CZTS}(a) and \ref{fig:CZTS}(b) that $C_n^{2}$ has a very large energy barrier, whereas $C_p^{1}$, $C_p^{0}$ and $C_n^{1}$ all have relatively smaller barriers. In a typical CZTS solar cell the majority carriers are of p-type and the carrier concentration varies from 10$^{15}$-10$^{17}$ cm$^{-3}$.\cite{kim2020upper, Crovetto2020} From Fig. \ref{fig:ND3}, we can see that when we go to a moderate p-type region ($E_F$$\approx$0.25 eV above the VBM corresponding to around 10$^{16}$ cm$^{-3}$ p-type carrier concentration), all the defect levels are similarly occupied. This happens, due to two factors.  If we use the full formalism as in our methodology, we can derive the defect charge state occupation ratios as given in Eq. \ref{eq_N2N1}-\ref{eq_N1N0}. At the above operating condition, p$>>$n and as the $2/1$ level is close to VBM (0.24 eV above VBM) so (p$^2$$>>$n) and (p$>>$n$^2$). As such, N$_2$/N$_1$ is dominated by p/p$^2$ (p and p$^2$ are similar at the above operating condition).  For N$_1$/N$_0$, as $C_n^{1}$ $>>$ $C_p^{0}$ that makes numerator and denominator similar at the given condition even if p$>>$n$^1$ and p$^1$$>>$n. In this scenario, 1/0 CTL pathway becomes dominant and the $C_{\mathrm{eff}}$ varies from $C_n^{1}$ to $C_p^{0}$. As a result we see a dip in device efficiency (see Figure S5(d)).  At n-type conditions $C_p^{0}$ becomes the rate limiting step and the defect will be stuck in neutral change state (See Figure S17). With $C_p^{0}$ being $\approx$ 10$^{-11}$ cm$^{3}$ s$^{-1}$, a high-quality sample of this material will likely be defect tolerant under n-type conditions. 

When Sn in the CZTS is replaced by Ge to form Cu$_{2}$ZnGeS$_{4}$ (CZGS), the IV$_\mathrm{II}$ antisite defect Ge$_{\mathrm{Zn}}$ is also a two level deep defect. However, since the $1/0$ level is now much further down from the conduction band maximum (CBM) (1.43 eV compared to 0.68 eV for Sn$_{\mathrm{Zn}}$ in CZTS), the $1\to 0$ transition has a higher semiclassical energy barrier (0.83 eV compared to 0.34 eV) resulting in the electron capture coefficient $C_n^{1}$ being reduced by five orders of magnitude. Although, $C_p^{0}$ remains slightly higher for CZGS compared to CZTS, the very low  $C_n^{1}$ ensures that CZGS remains defect tolerant with respect to Ge$_{\mathrm{Zn}}$ at all p-type conditions. Under  n-type doping, $C_p^{0}$ becomes the rate limiting transition and the defect will have a larger, yet still small, effect on the device parameters (see figure S6(d)). 

With Se substitution of S in Cu$_{2}$ZnSnSe$_{4}$ (CZTSe) the band gap reduces and as such, the defect levels in general come closer to the band edges. In particular, the $1/0$ CTL level of Sn$_{\mathrm{Zn}}$ moves closer to the CBM. But in this case the reorganization of the defect upon electron capture is relatively small resulting in a smaller $\Delta Q$ (1.76 $\sqrt{\mathrm{amu}}$\r{A} compared 2.23 $\sqrt{\mathrm{amu}}$\r{A} in CZTS). In CZTSe, the CC curves are more parallel to each other resulting in a very high energy barrier and low electron capture rate compared to CZTS. This holds true also for hole capture. These result in smaller carrier capture coefficients and as such CZTSe seems to be defect tolerant with respect to Sn$_{\mathrm{Zn}}$ defect. 

As discussed earlier in our results, CZGSe, where Ge and Se replace Sn and S, respectively, can be defect tolerant if the defect concentration is kept below 10$^{17}$ cm$^{-3}$. This shows that element substitution can be an effective strategy for new designing materials with similar electronic structures as the parent compound but improved defect properties, e.g. lower recombination rates. In our discussion of the CIGS and kesterites, we have focused on specific defects that are believed to be the most important. For a more complete analysis of the degree of defect tolerance of these materials one should consider all the possible deep level defects.

\subsection{Full theory vs. approximate models}
In this section, we compare the full SRH formalism with the approximations commonly used in the recent literature. As we have already mentioned (and discuss further in the SI) previous work on non-radiative SRH recombination apply either the minority carrier approximation or assumes that $n=p$. Both of these approximations ignore defect emission processes. Our results show that while the minority carrier approximation can provide results in agreement with the full formalism (under certain initial/operating conditions) the second approximation is too simple and should be avoided. A clear deficiency of both approximations is that they neither depend on the equilibrium Fermi energy (equivalent to the doping level) nor on the quasi-Fermi level splitting (equivalent to the voltage). As we have demonstrated throughout this work, the recombination dynamics can be strongly affected by these conditions and must be accounted for when solving the SRH rate equations for a quantitatively accurate determination of $R_\text{SRH}$.

In Figure \ref{fig:comparison} (a) and \ref{fig:comparison} (c) we compare the effective carrier lifetime, $\tau _{\mathrm{SRH}}=1/(N_{\mathrm{D}}C_{\mathrm{eff}})$, evaluated within the full formalism and the two approximations, for the case of Ge$_{\mathrm{Zn}}$ in CZGSe and Sn$_{\mathrm{Se}}$ in t-Se, respectively. Results are shown for an equilibrium Fermi level corresponding to n-type, p-type, and intrinsic. The first thing to note is that both the minority carrier and $n=p$ approximations imply a constant carrier lifetime independent of the initial and operating conditions; thus the horizontal dashed lines in panels (a) and (c). To set a scale, we mention that a condition of $\tau_{\mathrm{SRH}}\lesssim$ 10$^{-7}$s is sometimes used as a rule-of-thumb for a detrimental defect. Using this condition, we can see that the different approximations lead to different conclusions regarding the detrimental nature of the two defects. 

For both both p-type and n-type doping, the minority carrier approximation greatly underestimates the effective carrier lifetime. Recall that the 'Minority-n' approximation assumes the rate limiting step to be an electron capture process. On this basis, one could naively expect that the lifetime obtained with the Minority-n approximation would become accurate/exact for large p-doping and thus would agree with the 'Full (p-type)' result. Similarly, the Minority-p approximation would agree with the Full (n-type) result. This is clearly not the case and can be attributed to the fact that the minority approximation involves only a single capture rate assumed to be the rate limiting step. Moreover, the transition selected as the rate limiting is often based on the defect level positions. As we have seen in the previous sections one cannot always deduce the relative size of the capture rates based on the CTL positions. Also the size of the energy barrier, $\Delta E^\ddagger$, the degree of anharmonicity, and the initial and operating conditions play a role. Even when the transition considered within the minority approximation is the dominant/rate-limiting one, the neglect of other transitions will generally lead to an overestimation of the capture rate and thus underestimation of the carrier lifetime.

The $n=p$ approximation overestimates the carrier lifetime for CZGSe but underestimates it for t-Se. In general, the $n=p$ approximation may work well only for weak nonradiative defect centers. As in only those cases, the quasi-Fermi level splitting at the operating condition will be large and the photo-excited carrier concentration will dominate over the equilibrium carrier concentration (at least for moderate doping levels), and the $n=p$ condition is approximately fulfilled.

In Figure \ref{fig:comparison} (b),(d), we plot the device efficiency for the two defects, comparing the approximations and the full formalism. 
 For CZGSe (Figure \ref{fig:comparison} (b)), the minority carrier approximation underestimates $\tau_{\mathrm{SRH}}$ by an order of magnitude and thus underestimates $\mathrm{\eta}$. Wheres, $n=p$ approximation overestimates the carrier lifetime by almost two orders of magnitude and largely overestimates $\mathrm{\eta}$. For Sn$_{\mathrm{Se}}$, (Figure \ref{fig:comparison} (d)) the minority carrier approximation largely underestimates defect tolerance. Here the $n=p$ approximation shows close values to full formalism mainly because the carrier lifetime remains higher that 1 $\mu$s in both the cases. In short, qualitatively, $n=p$ approximation may find the defect tolerant material but also may predict some defect intolerant materials to be defect tolerant, whereas minority carrier approximation will do the opposite.

\section{Conclusions}
We have presented a methodology to calculate Shockley-Read-Hall (SRH) recombination rates in semiconductors from first-principles. Our approach goes beyond previous works by taking the effects of defect emission, equilibrium carrier concentrations and non-equilibrium photovoltaic (PV) operating conditions into account via a full solution of the coupled rate equations for all relevant charge states of the defect. Our results not only provide high-accuracy benchmark results, but also allow us to analyze how different capture processes (radiative and non-radiative processes via different charge states) contribute to the total recombination rate and how anharmonicity of the defect potential energy surface affects the carrier recombination dynamics. On this basis we conclude that it is always recommended to use the full formalism as it may be difficult to predict the validity of different approximations on before hand. 

We have used the SRH methodology to evaluate how different point defects influence the performance of a number of emerging (CIGS and kesterites) and prospective (selenium) PV materials. These results show that the widely used minority carrier and $n=p$ approximations can produce carrier lifetimes deviating by orders of magnitude from the full solution of the rate equations for realistic defect concentrations. The approximate schemes become particularly problematic for complex defects with multiple charge transition levels in the band gap with large consequences for the predicted PV efficiencies and the degree of defect tolerance. 

Our work sets a new standard for quantitative calculations of SRH recombination rates and PV device parameters from first principles. As such, our methodology will be useful for advancing the concept of defect tolerant semiconductors, including the determination of optimal doping strategies and growth conditions. In addition, its significance extends to other application areas where SRH recombination plays a dominant role, such as advanced optical (meta)materials, optoelectronic devices, and quantum light sources.  

\section{Data availability}
The data supporting this article have been included as part of the Supplementary Information. The codes used in this work are openly available as referred in the manuscript. 

\section{Conflict of interest}
The authors declare no conflict of interest.

\section{Acknowledgement}
The authors acknowledge funding from the Novo Nordisk Foundation Data Science Research Infrastructure 2022 Grant:  A high-performance computing infrastructure for data-driven research on sustainable energy materials (Grant no. NNF22OC0078009) and the Novo Nordisk Foundation Challenge Programme 2021: Smart nanomaterials for applications in life-science, BIOMAG (Grant No. NNF21OC0066526). KST is a Villum Investigator supported by VILLUM FONDEN (Grant No. 37789). JK acknowledges the Department of Science and Technology, India, for financial support through the INSPIRE faculty fellowship (IFA-23-PH-302).

\section{Computational details}
The atomic structures of CuInSe$_{2}$ and CuGaSe$_{2}$ were represented in a 16-atom tetragonal unit cell with I$\Bar{4}$2d (\#122) space group. The atomic structures of Cu$_{2}$ZnSnS$_{4}$, Cu$_{2}$ZnGeS$_{4}$, Cu$_{2}$ZnSnSe$_{4}$, and Cu$_{2}$ZnGeSe$_{4}$ were modeled using a 16-atom unit cell with space group I$\Bar{4}$ (\#82). Trigonal selenium (t-Se) was represented in a 3-atom unit cell with space group P$3_1$21 (\#152).

All first principles calculations were performed using density functional theory (DFT) within the projector augmented wave formalism (PAW) as implemented in the GPAW code\cite{mortensen2024gpaw}, in combination with Atomic Simulation Environment (ASE)\cite{larsen2017atomic}. The Heyd-Scuseria-Ernzerhoff (HSE06) exchange-correlation (xc)functional [REF] with the exchange mixing parameter ($\alpha$) optimized to reproduce the experimental band gaps for each compounds (see Table \ref{tab:table_1}). Optical absorption spectra were calculated within the random phase approximation (RPA) with single-particle wave functions and energies from a Perdew-Burke-Ernzerhoff (PBE) calculation. A scissors operator was applied to match the HSE06 band gap. Phonon assisted absorption at the indirect band gap (for t-Se) was computed using the methodology proposed in our previous work \cite{kangsabanik2022indirect}, which involves calculation of the phonons and electron-phonon coupling. The latter was calculated using the PBE xc-functional and a 16x16x16 k-point mesh. All DFT calculations were performed with a Fermi-Dirac smearing of 0.05 eV.

For point defect calculations we used a 2x3x1 (96 atom) supercell (CuInSe$_{2}$, CuGaSe$_{2}$, Cu$_{2}$ZnSnS$_{4}$, Cu$_{2}$ZnGeS$_{4}$, Cu$_{2}$ZnSnSe$_{4}$ and Cu$_{2}$ZnGeSe$_{4}$) and a 3x3x3 (81 atom) supercell (t-Se), respectively. We considered the following defects:  In$_{\mathrm{Cu}}$ (CuInSe$_{2}$), Ga$_{\mathrm{Cu}}$ (CuGaSe$_{2}$), Sn$_{\mathrm{Zn}}$ (Cu$_{2}$ZnSnS$_{4}$, Cu$_{2}$ZnSnSe$_{4}$), Ge$_{\mathrm{Zn}}$ (Cu$_{2}$ZnGeS$_{4}$, Cu$_{2}$ZnGeSe$_{4}$), Sn$_{\mathrm{Se}}$ (t-Se).  For all the defects we considered $q$= 2, 1, 0, -1, -2 charge states (in units of the elementary charge). All the defect supercells were relaxed with HSE06 functional\cite{heyd2003hybrid} (with the optimized $\alpha$ parameter as shown in Table \ref{tab:table_1}). A plane wave cutoff of 800 eV and Gamma-point sampling was used for the all the defect calculations. All atoms in the supercells were relaxed until the maximum force was below 0.01 eV/Å.  The electrostatic Freysoldt-Neugebauer-van de Walle (FNV) correction scheme was used to correct the defect formation energies.\cite{freysoldt2009fully} The computational defect and PV workflows were created using Atomic Simulation Recipes\cite{gjerding2021atomic} and executed using the Myqueue task scheduler frontend \cite{mortensen2020myqueue}.

The 1D configuration coordinate (CC) diagrams were calculated by evaluating the total energy with the HSE06 functional along a linear path connecting the initial and final configurations. The discrete data points were fitted using spline interpolation. The 1D-Schrödinger equation was then solved numerically for the potential energy of the initial and final states, repspectively, and the overlap between the vibronic wave functions was calculated. Anharmonicity of the 1D potentials were thus fully accounted for.  The dipole transition matrix element (for radiative transitions) and the electron-phonon coupling (for non-radiative transitions) were calculated with GPAW\cite{mortensen2024gpaw} using the methods described in Refs.\cite{alkauskas2014first, zhang2017first,turiansky2021nonrad, kim2020carriercapture}

\bibliography{main}

\clearpage
\onecolumngrid
\input{supp.tex}

\end{document}

%% file: supp.tex
\setcounter{figure}{0}
\setcounter{table}{0}
\setcounter{section}{0}
\setcounter{equation}{0}
\renewcommand{\thesection}{S\arabic{section}}
\renewcommand{\thetable}{S\arabic{table}}
\renewcommand{\thefigure}{S\arabic{figure}}
\renewcommand{\theequation}{S\arabic{equation}}
\renewcommand{\arraystretch}{1.2}

\begingroup
\centering
\textbf{Supplemental Material:\\
Defect-Assisted Recombination in Semiconductors and Photovoltaic Device Parameters from First Principles}\\[1em]

Jiban Kangsabanik${}^{1,2}$\thanks{jibka@dtu.dk},
Kristian S. Thygesen${}^{1}$\thanks{thygesen@fysik.dtu.dk} \\[0.5em]

\textit{
${}^{1}$CAMD, Computational Atomic-Scale Materials Design, Department of Physics, Technical University of Denmark, 2800 Kgs. Lyngby, Denmark\\
${}^{2}$Department of Condensed Matter and Materials Physics, S. N. Bose National Centre for Basic Sciences, Kolkata 700106, India
}
\par
\vspace{1em}
\endgroup
\vspace{-1.0em}
\onecolumngrid
\section{Notation}
Table \ref{tab:notation} provides an overview of the most important symbols used in the theory of photovoltaics and Shockley-Read-Hall recombination. 
\begin{table}[h!]
    \caption{\footnotesize Overview of the most important symbols used in the theory of photovoltaics and Shockley-Read-Hall recombination.}
    \label{tab:notation}
    \vspace{0.2em}
    \centering
    \renewcommand{\arraystretch}{0.65}
    \begin{tabular}{|c|l|c|}
      \hline
        \bf{Symbol} & \bf{Description} & \bf{Unit}  \\[1.9pt]
        \hline
      $k_B$ & Boltzmann's constant & eVK$^{-1}$ \\ [1.9pt]
      $T$ & Temperature & K\\[1.9pt]
      $\beta$ & $1/k_B T$ & eV$^{-1}$\\[1.9pt]
      $E_v$ & Valence band maximum (VBM) & eV \\[1.9pt]
       $E_c$ & Conduction band maximum (CBM) & eV \\[1.9pt]
       $\mathrm{D}^q$ & Defect in charge state $q$ &  \\[1.9pt]
       $E_D$ & Defect energy $\sim$ charge transition level  & eV \\[1.9pt]
       $E^f[\mathrm{D}^q]$ & Formation energy of defect D$^q$ & eV \\[1.9pt]
       $\mu_0$ & Fermi level at equilibrium & eV\\[1.9pt]
       $\mu_n$ & Quasi-Fermi level of electrons & eV \\[1.9pt]
        $\mu_p$ & Quasi-Fermi level of holes & eV \\[1.9pt]
      $\mu_D$ & Fermi level of defect & eV\\[1.9pt]
      $n_0$ & Equilibrium electron concentration & cm$^{-3}$\\[1.9pt]
           $p_0$ & Equilibrium hole concentration & cm$^{-3}$\\[1.9pt]
             $n$ & Electron conc. under operating cond. & cm$^{-3}$\\[1.9pt]
           $p$ & Hole conc. under operating cond. & cm$^{-3}$\\[1.9pt]
            $\Delta n$ & Conc. of photo-excited electrons & cm$^{-3}$\\[1.9pt]
           $\Delta p$ & Conc. of photo-excited holes& cm$^{-3}$\\[1.9pt]
           
           $n^q$ & Electron density for Fermi level at $E_D$ & eV\\[1.9pt]
                     $p^q$ & Hole density for Fermi level at $E_D$ & eV\\[1.9pt]
     $N_D$ & Defect concentration & cm$^{-3}$\\[1.9pt]
     $N_q$ & Conc. of defects in charge state $q$& cm$^{-3}$ \\[1.9pt]
     $N_c$ & Effective DOS of conduction band & cm$^{-3}$eV$^{-1}$\\[1.9pt]
      $N_v$ & Effective DOS of valence band & cm$^{-3}$eV$^{-1}$\\[1.9pt]
      $\eta_s$ & Sommerfeld factor &  \\[1.9pt]
      $f(E)$ & Fermi-Dirac distribution function &  \\[1.9pt]
      $f_D$ & Defect state occupation, $0\leq f_D\leq 1$ &  \\[1.9pt]
      $C_n$ & Capture coefficient: CBM$\to D$ & cm$^{3}$s$^{-1}$ \\[1.9pt]
     $\overline C_n$ & Capture coefficient: $D\to \mathrm{CBM}$  & cm$^{3}$s$^{-1}$ \\[1.9pt]
      $C_p$ & Capture coefficient: $D\to \mathrm{VBM}$& cm$^{3}$s$^{-1}$ \\[1.9pt]
     $\overline C_p$ & Capture coefficient: $\mathrm{VBM}\to D$  & cm$^{3}$s$^{-1}$ \\[1.9pt]
      $R_{n,C}$ & Electron capture rate (no emission) & cm$^{-3}$s$^{-1}$ \\[1.9pt]
    $R_{n,E}$ & Electron emission rate  & cm$^{-3}$s$^{-1}$ \\[1.9pt]
 $R_{p,C}$ & Hole capture rate (no emission) & cm$^{-3}$s$^{-1}$ \\[1.9pt]
    $R_{p,E}$ & Hole emission rate  & cm$^{-3}$s$^{-1}$ \\[1.9pt]
    $R_n$ & Electron capture rate & cm$^{-3}$s$^{-1}$ \\[1.9pt]
    $R_p$ & Hole capture rate & cm$^{-3}$s$^{-1}$ \\[1.9pt]

        \hline
    \end{tabular}
\end{table}

\clearpage
\section{Methodology}\label{methodology}

\subsection*{A. Defect formation energy and equilibrium concentration}\label{methodDF}

The formation energy of a defect $\mathrm{D}^q$ is written as
\begin{equation}\label{eq:eform}
\begin{split}
E^{f}\left[\mathrm{D}^q\right] = E_{\mathrm{tot}}\left[\mathrm{D}^q\right] - E_{\mathrm{tot}}^\mathrm{bulk} - \sum_a n_{a}\mu_a + \\
q(E_v + \Delta \mu)+ E_\mathrm{corr},
\end{split}
\end{equation}
where $E_{\mathrm{tot}}\left[\mathrm{D}^q\right]$ is the total energy of the supercell with defect D in charge state $q$. $E_{\mathrm{tot}}^\mathrm{bulk}$ is the total energy of the pristine supercell with same size. $\mu_a$ is the chemical potential of the defect species, where $n_a$ represents the number of such atoms added ($n_a > 0$) or removed ($n_a <0$) from the pristine supercell to create the defect supercell. The next term represents the chemical potential of the electrons added or removed to create the charge state $q$, where $E_v$ is the energy of the valence band maximum (VBM) and $\Delta \mu$ is the Fermi level with respect to the VBM (which can vary from 0 to $E_{\mathrm{gap}}$). $E_\mathrm{corr}$ is the correction term removing the spurious electrostatic interaction between image charges of the periodic repetitions of the supercell as well as the compensating homogeneous background charge and the potential alignment corrections for charged defects. 

For a defect D, the charge transition level (CTL) between charge state $q$ and $q'$ can be calculated as,
\begin{equation}\label{eq:ectl}
\begin{split}
E^{q/q'}= \frac{E^{f}\left[\mathrm{D}^q\right]-E^{f}\left[\mathrm{D}^{q'}\right]}{q'-q}.
\end{split}
\end{equation}
For a given Fermi level position, the equilibrium concentration of a defect at temperature $T$ can be calculated as,
\begin{equation}\label{eq:defc}
N\left[\mathrm{D}^q\right]= N_{\mathrm{site}}g_q \exp(-\beta E^{f}\left[\mathrm{D}^q\right]),
\end{equation}
where $\beta=1/k_BT$ and $N_{\mathrm{site}}$ and $g_q$ are the number of equivalent crystal defect sites and the orbital and spin degeneracy of the charge state $q$, respectively. Because the defect formation energy depends on the Fermi level, the concentration of charged defects will also depend on it. To ensure charge neutrality, the equilibrium Fermi level ($\mu_0$) and defect concentrations must be determined self-consistently. 

The charge neutrality equation can be expressed as, 
\begin{equation}\label{eq:neut}
\sum_{D,q}qN\left[\mathrm{D}^q\right]=n_0-p_0,
\end{equation}
where $n_0$ and $p_0$ are the equilibrium concentrations of free electrons and holes, respectively. These can be calculated using the density of states, $N(E)$, of the pristine system (assuming a non-degenerate electron gas),
\begin{equation}\label{eq:freen}
n_0=e^{-\beta (E_c-\mu_0)}N_c
\end{equation}
and
\begin{equation}\label{eq:freep}
n_0=e^{\beta (E_v-\mu_0)}N_v,
\end{equation}
where we have introduced the effective density of states for the conduction and valence bands, 
\begin{equation}\label{eq:freenc}
N_c=\int_{E_c}^{\infty}e^{-\beta (E-E_c)}N(E)dE
\end{equation}
\begin{equation}\label{eq:freepc}
N_v=\int_{-\infty}^{E_v} e^{-\beta (E_v-E)} N(E)dE
\end{equation}
It is important to stress that the equilibrium Fermi level calculated in this way can be misleading due to the incomplete knowledge of possible intrinsic and extrinsic defects during synthesis. As such, it would often be more useful to set the Fermi level manually based on experimental information about majority carrier type and concentration, and then determine the equilibrium defect concentrations using Eq. (\ref{eq:defc}). For example, Selenium and CZTS are well known p-type semiconductors with majority carrier concentration of around 10$^{16}$. For a general defect in a an experimentally unknown material, we would recommend to calculate the defect concentrations and associated recombination rates by varying the equilibrium Fermi level, in order to consider different majority carrier types and concentrations.

\begin{figure*}[t]
    \centering
    \includegraphics[width=0.95\linewidth]{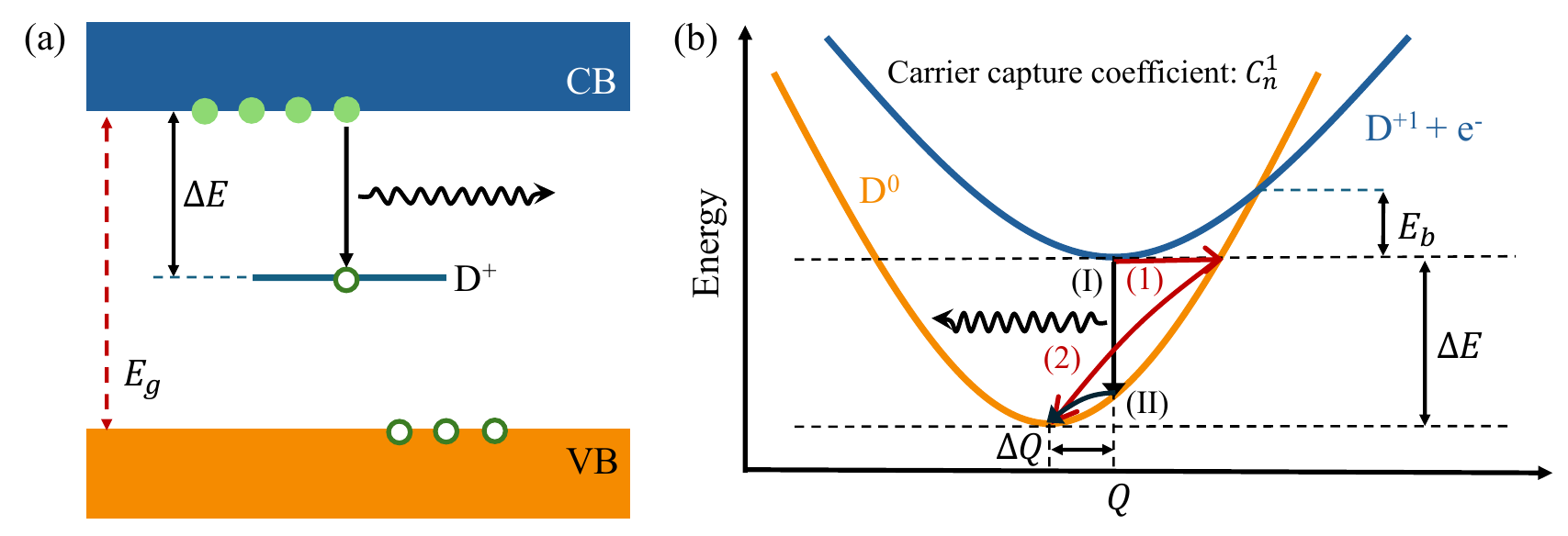}
    \caption{\footnotesize (a) Schematic illustration of a deep level defect inside the semiconductor band gap and associated carrier capture. (b) Schematic illustration of a configuration coordinate diagram for a electron capture process by a positively charged defect $\text{D}^{+1}$. The related capture coefficient is $C_n^{1}$. Here the initial state is $\text{D}^{+1}$ with an electron ($e^-$) inside the conduction band and the final state corresponds to a neutral defect $\text{D}^{0}$. $\Delta E$ corresponds to the transition energy and $\Delta Q$ is the difference between the two geometries at their respective equilibrium. The electron capture is a two step process-(1) an energy conserving electronic transition and (2) vibrational relaxation. $E_b$ is the semiclassical energy barrier for the capture process.}
    \label{fig:cc_diagram}
\end{figure*}

\subsection*{B. Recombination via carrier capture: Radiative and  
 nonradiative multi-phonon emissions}\label{methodCC}

In this section, we describe the formalism used to calculate the defect carrier capture rates. Depending on the mechanism, carrier capture by a defect can be of two types: i) Radiative capture via emission of a photon, ii) Nonradiative capture via (multi)-phonon emission. In general, both radiative and non-radiative carrier capture will result in loss of photogenerated carriers. Consequently, the total capture rate should be used when   
calculating the SRH recombination rate. In equilibrium the principle of detailed balance dictates that a capture process (electron looses energy) must be accompanied by a corresponding emission process (electron gains energy). We will consider both capture and emission processes in the next section.

The radiative capture coefficient giving the transition rate from an initial state, $\psi_i$, to a final state, $\psi_f$, can be calculated using Fermi's golden rule as,
\begin{equation}
C^{\mathrm{rad}}_{if}=\frac{V E_{if}^{3}n_{\mathrm{ref}}\lvert \mathbf D_{if}\rvert^{2}}{3\pi\epsilon_0\hbar^4c^3}
\label{eqradSI}
\end{equation}
Here, $V$ is the supercell volume, $E_{if}=E_f-E_i$ is the emission energy equal to the electronic transition energy, $c$ is the speed of light, and $n_{\mathrm{ref}}$ is the refractive index of the material. $\mathbf D_{if}$ is the transition dipole moment between the initial and final electronic states, which  can be expressed as
\begin{equation}
    \mathbf D_{if}=\langle\psi_f\lvert\hat{\mathbf r}\rvert\psi_i\rangle=\frac{i\hbar}{m_0}\frac{\langle\psi_f\lvert\hat{\mathbf p}\rvert\psi_i\rangle}{\varepsilon_f-\varepsilon_i},
\end{equation}
where $\hat{\mathbf r}$ and $\hat{\mathbf p}$ are the dipole and momentum operators, respectively, and $m_0$ is the electron mass. $\varepsilon_i$ and $\varepsilon_f$ are the single-particle energies of the Kohn-Sham orbitals $\psi_i$ and $\psi_f$. Note that these differ from $E_i$ and $E_f$, as the latter are total energies. 
 
 To calculate the non-radiative carrier capture we use the approach proposed by Alkauskas \emph{et al.}\cite{S1} which models the non-radiative SRH carrier capture as a multi-phonon emission process. In Fig. \ref{fig:cc_diagram} a schematic diagram of an electron capture by a positively charged defect $\text{D}^+$ is shown. This results in transition to a neutral charge state $\text{D}^0$. Fig. \ref{fig:cc_diagram} represents the potential energy surfaces of a defect in different charge states, in terms of a generalized 1-dimensional configuration coordinate $Q$,

 \begin{equation}
	Q^2=\sum_{a}m_i\Delta \mathbf R_a^2
	\label{eqCC1}.
\end{equation}
 Here the sum runs over all atoms in the supercell with masses $m_a$ and $\Delta \mathbf R_a$ is the displacement vector of atom $a$ between the initial and final configurations. 
  
Within this formalism, the non-radiative capture can be seen as a two-step process: The first step is an energy conserving transition in which energy is transferred from the electron system to the phonon system via emission of phonons. The next step is a lattice relaxation via multi-phonon emission. It is assumed that the electron transfer step (step 1) is the rate determining transition. Using Fermi's golden rule this multi-phonon (m-ph) capture rate can be calculated as,\cite{S1}

\begin{equation*}
	C^{\mathrm{m-ph}}_{if}=\eta_s \frac{2\pi}{\hbar}gV|W_{if}|^2\sum_{m}w_m\sum_{n}\lvert\langle\chi_{im}\lvert\hat{Q}-Q_0\rvert\chi_{fn}\rangle\rvert^2 \delta( E_{if} +m\hbar\Omega_{im}-n\hbar\Omega_{fn}).
	\label{eqCC2SI}
\end{equation*}

Here, $\eta_s$ is the Sommerfeld factor (see below), $g_q$ and $V$ are the orbital degeneracy of the defect (in the considered charge state) and the volume of the supercell, respectively. $W_{if}$ is the electron-phonon coupling matrix elements between the initial and final electronic states while $\sum_{m}w_m\sum_{n}\lvert\langle\chi_{im}\lvert\hat{Q}-Q_0\rvert\chi_{fn}\rangle\rvert^2$ accounts for the overlap between the initial and final vibronic states defined by the 1D potential energy diagrams. The latter are assumed to be parabolas (the harmonic approximation), but we allow for different curvatures for the initial and final states.  The occupations factors, $w_m$, of the initial state vibronic levels are taken as the Bose-Einstein distribution. The $\delta$-functions (in practice represented by Gaussians of width 0.8$\hbar\Omega_f$ where $\Omega_f$ are the vibration frequencies in the final state) expresses the energy conservation: The difference in electronic energy between the initial and the final states, $E_{if}$, must be matched by the difference in vibrational energy, see Fig. \ref{fig:cc_diagram}b.
 
 The Sommerfeld factor, $\eta_s$, accounts for the Coulomb interaction between a charged defect and a free carrier\cite{S2}. For attractive centers $\eta_s$ can be expressed as,
\begin{equation}
	\eta_s=4\lvert Z \rvert(\pi E_R/k_B T)^{1/2}
	\label{eqCC3}
	\end{equation}
 For repulsive centers,
\begin{equation}
	\eta_s=8/\sqrt{3}(\pi^2Z^2E_R/k_B T)^{2/3}e^{-3(\pi^2Z^2E_R/k_B T)^{1/3}}
	\label{eqCC4}
	\end{equation}
Here, $Z$ is the ratio of defect charge and the carrier charge and as such will be negative for an attractive center and positive for a repulsive one. Further, 
$E_R=(m^*e^4/(2\hbar^2\epsilon^2))$ is effective Rydberg energy with $m^*$ and $\epsilon$ being the carrier effective mass and the static dielectric constant, respectively.

\onecolumngrid

\subsection*{C. SRH statistics and approximations: Single-level defects}\label{methodSRH}

Let D represent a deep level trap acting as a recombination center in a semiconductor. In this case, there will be four possible processes contributing to the recombination cycle, namely (i) electron capture from the conduction band, (ii) electron emission to the conduction band, (iii) hole capture from the valence band, iv) hole emission to the valence band. In this section we will focus on the electron capture/emission processes. Similar considerations apply to the hole processes.

The rate of electron capture from the conduction band to the trap D in a unit volume of material can be written as,
\begin{equation}
R_{n,C}=[1-f_D] N_D f(E_c)N_cC_n
\end{equation}
Here $f_D$ is the probability of the trap being occupied, $N_{D}$ is the density of traps, $N_c$ is the effective density of states of the conduction band, $f(E)$ is the Fermi-Dirac distribution function, and $C_n$ is the electron capture coefficient. The latter is a transition rate of the form $C_{if}$ where the initial state is the CBM and the final state is the trap state. 

Similarly the rate of emission from the trap to the conduction band can be expressed as, 
\begin{equation}
R_{n,E}=f_D N_D [1-f(E_c)]N_c\overline C_n,
\end{equation}
where $\overline C_n$ is the transition rate for an electron from the defect state to the CBM. 

In thermal equilibrium, the principle of detailed balance necessitates that the rate of capture should equal the rate of emission. Equating $R_{n,C}$ and $R_{n,E}$ yields
\begin{equation}
 [1-f_D]f(E_c)C_n=f_D[1-f(E_c)]\bar C_n,
\label{eqSR1}
\end{equation}
which can be solved to obtain
\begin{equation*}
\overline C_n/C_n=e^{-\beta (E_c-E_D)}.
\end{equation*}
To derive the above equation, it is useful to write the defect occupation as
\begin{equation}
f_D = \frac{1}{1+e^{\beta (E_D-\mu_D)}} 
\end{equation}
from which it follows that 
\begin{equation}
1-f_D = f_D e^{\beta (E_D-\mu_D)} 
\end{equation}
Moreover, the distribution of electrons is described by the distribution function
\begin{equation}
f(E)= \frac{1}{1+e^{\beta (E-\mu_n)}} 
\end{equation}
from which it follows that
\begin{equation}
1-f(E)=f(E) e^{\beta (E-\mu_n)} 
\end{equation}
The hole distribution is assumed to take a similar form with the chemical potential $\mu_h$.
In thermal equilibrium it holds that $\mu_D=\mu_n=\mu_h=\mu_0$.

In general, we consider a steady state situation where the chemical potentials take constant but different values. Under such conditions, the net electron capture rate can be written,
\begin{eqnarray}
R_n&=&R_{n,C}-R_{n,E}\\&=&[1-e^{\beta (\mu_D-\mu_n)}] [1-f_D]N_{D}f(E_c)N_c C_n
\label{eqSR2}
\end{eqnarray}
Similar expression for hole capture can be derived.

Elaborating on Eq. (\ref{eqSR2}) assuming the electrons in the conduction band is a non-degenerate gas, we can write
\begin{equation*}
e^{\beta (\mu_D-\mu_n)}f_{D}e^{\beta (E_{D}-\mu_D)}N_{c}e^{\beta (\mu_n-E_{c})}N_{D}C_n
\end{equation*}
\begin{equation*}
=N_{D}f_{D}N_{c}e^{\beta(E_{D}-E_{c})}C_n
\end{equation*}
\begin{equation*}
=N_{D}f_{D}n^qC_n
\end{equation*}
In the last equation we have introduced $n^q$ to denote the electron density for a Fermi level at the trap energy, $E_D$, which is equivalent to charge transition level $q/q-1$.

Now we can write the net electron capture rate as,
\begin{equation}
R_n=C_n([1-f_D]n-f_{D}n^q)N_D
\label{eqSR4}
\end{equation}
Similarly the net hole capture rate can be written as, 
\begin{equation}
R_p=C_p(f_{D}p-[1-f_D]p^q)N_D
\label{eqSR5}
\end{equation}
where $C_p$ is the hole capture coefficient, i.e. a transition rate of the form $C_{if}$ where the initial state is the trap state and the final state is the top of the valence band. $p^q$ is the number of free holes if the Fermi level equals the trap energy, $E_D$.

In steady state, the net rate of capture of electrons ($R_n$) should be equal to the net rate capture of holes ($R_p$). Thus 
\begin{equation}
C_n([1-f_D]n-f_{D}n^q)=C_p(f_{D}p-[1-f_D]p^q)
\label{eqSR6}
\end{equation}
After solving for $f_D$ we can express
the net capture rate as,
\begin{equation}
R_{n}=R_{p}=\frac{N_DC_nC_p(np-n_0p_0)}{[C_n(n+n^q)+C_p(p+p^q)]}
\label{eqSR7SI}
\end{equation}
Here $n_0$ and $p_0$ are the free electron and hole density at equilibrium while $n$ and $p$ are the same at operating steady state conditions. For the latter we write $n=n_0+\Delta n$ and $p=p_0+\Delta n$.
$\Delta n$ is the photo-excited carrier density, which is obtained by equating the operating voltage $V$ with the difference between the electron and hole quasi Fermi levels, $eV=\mu_n-\mu_p$. A simplified expression\cite{S3} can be written as, 
\begin{equation}
	\Delta n(V)=\frac{1}{2}[-n_0-p_0+\sqrt{(n_0+p_0)^2-4n_0p_0(1-e^{\beta eV})}]
	\label{eqSR8SI}
\end{equation}

\subsubsection*{Approximation I: Minority carrier approximation}
In case of a p-type material where $p_0 \gg n_0$, it has been proposed \cite{S3} that minority carrier capture will be the rate limiting step and as such, 
\begin{equation}
R_n\approx N_D\Delta nC_n=\frac{\Delta n}{\tau_n},
\label{eqSR9}
\end{equation}
where the minority carrier lifetime, $\tau_n=1/(N_DC_n)$, has been introduced. 
It is, however, clear from Eq. (\ref{eqSR7SI}) that the condition $p_0 \gg n_0$ is not sufficient. For example, the approximation can break down if the defect energy level is close to the CBM (in which case $n^q$ can become large) and/or the photo-generated carrier concentration is large compared to $p_0$ (in which case $n\sim p$). The minority carrier approximation should thus be used with care. 

The above considerations were made for a p-type material, but completely analogous arguments apply to an n-type semiconductor where the minority carrier approximation reads $R_n\approx \Delta n/\tau_p$.

\subsubsection*{Approximation II: Negligible emission}
We consider a deep level defect D with a charge transition $q=0\rightarrow q=-1$. The defect concentration is split into concentrations of the neutral and charged species,
\begin{equation*}
N_D=N_0+N_{-1} = [1-f_D]N_D + f_D N_D
\end{equation*}
If the CTL lies sufficiently far from both band edges we can assume $n \gg n_q$ and $p \gg p_q$ under PV operating conditions. From Eq. (\ref{eqSR4}) the net electron capture rate can then be approximated as
\begin{equation*}\label{eq:R_n_approx}
R_n\approx N_0C_nn
\end{equation*}
Similarly the net hole capture rate (where defect is in $q=-1$) can be written as, 
\begin{equation*}
R_p\approx N_{-1}C_pp
\end{equation*}
In these expressions the emission term for electrons (holes) has been neglected.

Under steady state steady state condition, $N_0C_nn=N_{-1}C_pp$, which yields
\begin{equation*}
N_{-1}=\frac{N_0C_nn}{C_pp}
\end{equation*}
and
\begin{equation*}
N_{D}=\frac{N_0(C_pp+C_nn)}{C_pp}.
\end{equation*}
The net capture rate can then be expressed as
\begin{equation}
R_n\approx N_0C_nn=\frac{N_{D}C_nnC_pp}{C_pp+C_nn}
\label{eqSR10}
\end{equation}
This expression of course holds for any charge transitions $q\rightarrow q-1$ as long as the stated conditions are satisfied.

\subsubsection*{Approximation III: Negligible emission and $n=p$}
We now discuss a further approximation that has been used in the recent literature to calculate total capture coefficients of defect with multiple charge states in the band gap. Here for simplicity we consider it for the case of a defect with a single level in the band gap. 

The $n=p$ approximation assumes that the photo-generated carrier densities are much higher than the equilibrium densities, i.e. $\Delta n \gg \max\{n_0, p_0\}$. Now Eq. (\ref{eqSR10}) can be written  
\begin{equation}
R_n\approx N_0C_nn\approx \frac{N_{D}nC_nC_p}{C_p+C_n}\approx\frac{N_{D}\Delta nC_nC_p}{C_p+C_n}
\label{eqSR11}
\end{equation}

\vspace{10pt}

\subsection*{D. Multiple charge transition levels}
In this section we derive expressions for the SRH recombination rates for a deep level defect D with multiple charge transition levels (CTLs). To introduce some notation we begin by re-deriving Eq. (\ref{eqSR7SI}) for the capture rate of a single CTL. Subsequently, the expression is generalised to the case of defects with two, three and four CTLs. 

\subsubsection{Single CTL}
We first consider a defect with the single CTL $q=1\rightarrow q=0$ inside the band gap. Under PV steady state operating conditions Eqs. (\ref{eqSR4}) and (\ref{eqSR5}) gives the electron and hole capture rates,  
\begin{eqnarray}
R_n&=&N_1C_n^1n-N_{0}C_n^1n^1\\
R_p&=&N_{0}C_p^{0}p-N_1C_p^{0}p^1
\end{eqnarray}
Here $N_0$ and $N_{1}$ are the concentration of defect D in charge states $q=0$ and $q=1$, thus
\begin{equation}
N_D=N_0+N_{1}
\label{eqmulti3SI}
\end{equation}
Under steady state conditions, $R_n=R_p$, which gives
\begin{equation}
N_1C_n^1n-N_{0}C_n^1n^1=N_{0}C_p^{0}p-N_1C_p^{0}p^1
\label{eqmulti4SI}
\end{equation}
Solving Eqs. (\ref{eqmulti3SI}-\ref{eqmulti4SI}) we obtain the net recombination rate
for CTL (1/0),
\begin{equation*}
R_\text{SRH}=R_n=N_1C_n^1n-N_{0}C_n^1n^1=\frac{N_DC_n^1C_p^0(np-n^1p^1)}{[C_n^1(n+n^1)+C_p^0(p+p^1)]}=\frac{N_DC_n^1C_p^0(np-n_i^2)}{[C_n^1(n+n^1)+C_p^0(p+p^1)]},
\label{eqmulti5SI}
\end{equation*}
which is same as Eq. (\ref{eqSR7SI}).

\subsubsection{Two CTLs}
Next, we use a similar approach for an amphoteric defect with 1/0 and 0/-1 CTLs inside the band gap. For such a defect we can write 
\begin{equation}
N_1C_n^1n-N_{0}C_n^1n^1=N_{0}C_p^{0}p-N_1C_p^{0}p^1
\label{eqmulti6SI}
\end{equation}
\begin{equation}
N_0C_n^0n-N_{-1}C_n^0n^0=N_{-1}C_p^{-1}p-N_0C_p^{-1}p^0
\label{eqmulti7SI}
\end{equation}
\begin{equation}
N_D=N_1+N_0+N_{-1}
\label{eqmulti8SI}
\end{equation}
Here $n^1$, $p^1$ are the free electron and hole concentrations if the Fermi level lies in the trap energy at CTL(1/0). Similarly $n^0$, $p^0$ are the electron and hole concentrations if the Fermi level lies in the trap energy at CTL(0/-1). As such
\begin{equation*}
n^1p^1=n^0p^0=n_i^2
\end{equation*}
The recombination rate for CTL (1/0 | 0/-1) is given by
\begin{equation}
R_\text{SRH}=N_1C_n^1n-N_{0}C_n^1n^1+N_0C_n^0n-N_{-1}C_n^0n^0.
\label{eqtwoctl}
\end{equation}
Eliminating the charge dependent defect concentrations, $N_i$, using the steady state Eqs. (\ref{eqmulti6SI}-\ref{eqmulti8SI}) it can be written:

\begin{equation*}
R_\text{SRH}=\frac{N_D}{G}[\frac{C_n^0C_p^{-1}(np-n^0p^0)}{C_p^{-1}p+C_n^0n^0} + \frac{C_p^0C_n^1(np-n^1p^1)}{C_n^1n+C_p^0p^1}]
\end{equation*}
\begin{equation*}
G=[1+\frac{C_p^0p+C_n^1n^1}{C_n^1n+C_p^0p^1}+\frac{C_n^0n+C_p^{-1}p^0}{C_p^{-1}p+C_n^0n^0}]
\end{equation*}

Similarly, the expression for the SRH recombination rate for a defect with the CTLs (2/1 | 1/0) becomes
\begin{equation*}
R_\text{SRH}=\frac{N_D}{G}[\frac{C_n^1C_p^0(np-n^1p^1)}{C_p^0p+C_n^1n^1} + \frac{C_p^1C_n^2(np-n^2p^2)}{C_n^2n+C_p^1p^2}]
\end{equation*}
\begin{equation*}
G=[1+\frac{C_p^1p+C_n^2n^2}{C_n^2n+C_p^1p^2}+\frac{C_n^1n+C_p^0p^1}{C_p^0p+C_n^1n^1}]
\end{equation*}
\subsubsection{Three CTLs}
For a defect with the CTLs (1/0 | 0/-1 | -1/-2) inside the band gap, the SRH recombination rate reads
\begin{equation}
\begin{split}
R_\text{SRH}=\frac{N_D}{G}[\frac{C_p^{-1}C_p^{0}C_n^1(npp-n^1p^1p)+C_n^0C_p^{0}C_n^1(npn^0-n^1p^1n^0)}{(C_n^{0}n+C_p^{-1}p^0)(C_n^{1}n+C_p^0p^1)} + \\\frac{C_n^0C_p^{-1}(np-n^0p^0)}{C_n^{0}n+C_p^{-1}p^0} + \frac{C_n^{-1}C_p^{-2}(np-n^{-1}p^{-1})}{C_p^{-2}p+C_n^{-1}n^{-1}}]
\end{split}
\end{equation}
\begin{equation*}
G=[1+\frac{C_p^{-1}p+C_n^0n^0}{C_n^0n+C_p^{-1}p^0}+\frac{C_n^{-1}n+C_p^{-2}p^{-1}}{C_p^{-2}p+C_n^{-1}n^{-1}}+\frac{(C_p^{-1}p+C_n^0n^0)(C_p^0p+C_n^1n^1)}{(C_n^0n+C_p^{-1}p^0)(C_n^1n+C_p^0p^1)}]
\end{equation*}

Similarly, for a defect with the CTLs (2/1 | 1/0 | 0/-1), the SRH recombination rate becomes
\begin{equation}
\begin{split}
R_\text{SRH}=\frac{N_D}{G}[\frac{C_p^0C_p^1C_n^2(npp-n^2p^2p)+C_n^1C_p^1C_n^2(npn^1-n^2p^2n^1)}{(C_n^1n+C_p^0p^1)(C_n^2n+C_p^1p^2)} + \\ \frac{C_n^1C_p^0(np-n^1p^1)}{C_n^1n+C_p^0p^1} + \frac{C_n^0C_p^{-1}(np-n^0p^0)}{C_p^{-1}p+C_n^0n^0}]
\end{split}
\end{equation}
\begin{equation*}
G=[1+\frac{C_n^0n+C_p^{-1}p^0}{C_p^{-1}p+C_n^0n^0}+\frac{C_p^0p+C_n^1n^1}{C_n^{1}n+C_p^0p^1}+\frac{(C_p^0p+C_n^1n^1)(C_p^1p+C_n^2n^2)}{(C_n^1n+C_p^0p^1)(C_n^2n+C_p^1p^2)}]
\end{equation*}
\subsubsection{Four CTLs}
For a defect with the four CTLs (2/1 | 1/0 | 0/-1 | -1/-2) inside the band gap, the expression for the SRH recombination rate is

\begin{equation*}
\begin{split}
R_\text{SRH}=\frac{N_D}{G}[\frac{C_p^0C_p^1C_n^2(npp-n^2p^2p)+C_n^1C_p^1C_n^2(npn^1-n^2p^2n^1)}{(C_n^1n+C_p^0p^1)(C_n^2n+C_p^1p^2)} + \frac{C_n^1C_p^0(np-n^1p^1)}{C_n^1n+C_p^0p^1} + \\
\frac{C_n^0C_p^{-1}(np-n^0p^0)}{C_p^{-1}p+C_n^0n^0} +
\frac{C_n^0C_n^{-1}C_p^{-2}(npn-n^{-1}p^{-1}n)+C_p^{-1}C_n^{-1}C_p^{-2}(npp^0-n^{-1}p^{-1}p^0)}{(C_p^{-1}p+C_n^0n^0)(C_p^{-2}p+C_n^{-1}n^{-1})}]
\end{split}
\end{equation*}
\begin{equation*}
\begin{split}
G=[1+\frac{C_p^0p+C_n^1n^1}{C_n^1n+C_p^0p^1}+\frac{C_n^0n+C_p^{-1}p^0}{C_p^{-1}p+C_n^0n^0}+\frac{(C_p^0p+C_n^1n^1)(C_p^1p+C_n^2n^2)}{(C_n^1n+C_p^0p^1)(C_n^2n+C_p^1p^2)}+\\
\frac{(C_n^0n+C_p^{-1}p^0)(C_n^{-1}n+C_p^{-2}p^{-1})}{(C_p^{-1}p+C_n^0n^0)(C_p^{-2}p+C_n^{-1}n^{-1})}]
\end{split}
\end{equation*}

\onecolumngrid
\subsection*{E. Photovoltaic device parameters}\label{methodPV}
 The widely used Shockley-Queisser (SQ) efficiency formalism \cite{S4} is modified in our model to calculate photovoltaic efficiency by including a thickness dependent absorptivity that includes the material's absorption coefficient ($\alpha$) and film thickness ($d$), rather than a step-like function that only depends on the band gap. In the radiative limit, the photovoltaic power conversion efficiency (PCE) ($\eta_\text{rad}$) is expressed as,
\begin{equation}
	\eta_\text{rad}=\frac{\max(V[J_\text{sc}-J_{r}^\text{rad}(e^{\beta eV}-1)])}{\int_{0}^{\infty}\hbar^2\omega I_\text{sun}(\omega) d\omega}
	\label{eqPV1}
	\end{equation}

 In this expression, the denominator is the input power density ($P_\text{in}$), where the typical AM1.5G solar spectrum at 298 K is taken as incident solar spectrum, $I_\text{sun}(\omega)$. The numerator has the power density of a material with the highest feasible output, which corresponds to the maximal power density $P=JV$. The current is written as the short-circuit current $J_\text{sc}$ subtracted the radiative recombination current density. The latter equals $J_{r}^{\text{rad}}$ in the dark and under finite quasi Fermi level, $\mu_n-\mu_p=eV$, splitting is increased by a factor $np=\exp(\beta eV)$.
 
 The short circuit current density $J_\text{sc}$ can be expressed as,
	\begin{equation}
	J_\text{sc}=e\hbar\int_{0}^{\infty}a(\omega, d) I_\text{sun}(\omega) d\omega
	\label{eqPV2}
	\end{equation}
	Here, absorptivity $a(\omega, d)$ is defined as $a(\omega, d)=1-e^{-2\alpha(\omega)d}$, for a thin film with a front surface with zero reflectivity and a back surface with unity reflectivity. $\alpha(\omega)$ is the absorption coefficient of the semiconductor absorber and $d$ is the film thickness. The total absorption coefficient is determined as the sum of the direct and indirect (phonon-assisted) contributions.\cite{S5} For device parameter calculations, we average $\alpha(\omega)$ over the three polarization directions.

 In equilibrium (in the dark), absorption and emission through the cell surface must be equal. Thus $J_{r}^\text{rad}$ can be derived from the black-body photons absorbed from surrounding thermal bath as \cite{S6},
	\begin{equation}
	J_{r}^\text{rad}=e\pi\hbar\int_{0}^{\infty}a(\omega, d) I_\text{bb}(\omega,T) d\omega
    \label{eqPV3}
	\end{equation} 
	Here $I_\text{bb}(\omega,T)$ is the photon flux from the black-body spectrum (the refractive index of the medium is set to 1) at temperature $T$ (in this work 298 K),
	\begin{equation}
	I_\text{bb}(\omega,T)=\frac{1}{2\pi^2}\omega^2h^{-1}c^{-2}[e^{\hbar\omega/{k_\text{B}T}}-1]^{-1}
	\label{eqPV4}
	\end{equation}

 When the non-radiative recombination due to deep-level traps is taken into account, an additional current loss of size $eR_\text{SRH}d$ appears in the efficiency expression. The defect limited conversion efficiency is then given by\cite{S3},
 \begin{equation}
	\eta=\frac{\max(V[J_\text{sc}-J_{r}^\text{rad}(e^{\frac{eV}{k_\text{B}T}}-1)-eR_\text{SRH}(V)d)])}{\int_{0}^{\infty}\hbar^2\omega I_\text{sun}(\omega) d\omega}
	\label{eqPV5SI}
	\end{equation}
where $R_\text{SRH}$ is calculated as discussed in the previous sections. Note that the $V$-dependence is not explicitly shown in the expressions for $R_\text{SRH}$. It enters implicitly via the photo-generated carrier density $\Delta n$, which is determined from $V$ by Eq. (\ref{eqSR8SI}). 

At high initial doping conditions (equlibrium Fermi level) Pauli blocking might have an effect on the optical absorption. In order to observe the Pauli blocking effect in the PV device equations we tested it by calculating the modified absorption $\alpha(E)$ at different initial equlibrium Fermi levels using the following expressions,
 \begin{equation}
	    \alpha(E)=P(E){\rm JDOS}(E)
	\label{eqPV6}
	\end{equation}
Here $P(E)$ can be seen as the transition strength. 
\begin{equation}
	    {\rm JDOS}^{\mathrm {Pauli}}(E)=\sum_{v, c} \delta(E-(E_c-E_v))f_v (1-f_c)
	\label{eqPV7}
	\end{equation}
$f_v$ and $f_c$ are Fermi occupation factors.
\begin{equation}
	    \alpha^{\rm mod}(E)=P(E){\rm JDOS}^{\mathrm {Pauli}}(E)
	\label{eqPV8}
	\end{equation}
We see a very minor change in the $J_{\rm sc}$ and $\eta$ when including the Pauli blocking effect and as such this was neglected for further PV device operation calculations.

\clearpage

\section{Results}\label{results}

\subsection{Defect Charge transition levels}\label{ctls}

In this section we show the defect charge transition levels for the considered systems. 

\begin{figure*}[!htbp]
    \centering
    \includegraphics[width=0.7\linewidth]{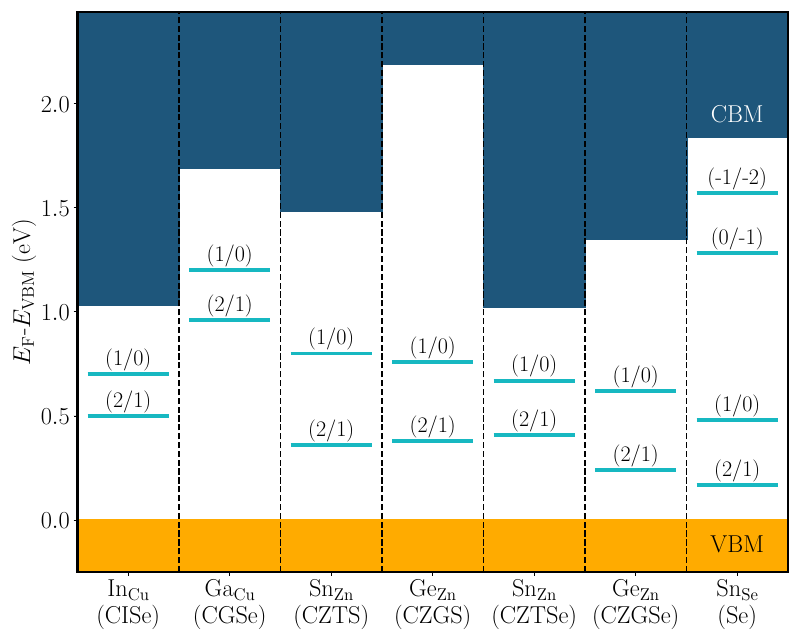}
    \caption{\footnotesize Defect charge transition level diagram for the considered defects in the corresponding host materials, CuInSe$_{2}$ (CISe), CuGaSe$_{2}$ (CGSe), Cu$_{2}$ZnSnS$_{4}$ (CZTS), Cu$_{2}$ZnGeS$_{4}$ (CZGS), Cu$_{2}$ZnSnSe$_{4}$ (CZTSe), Cu$_{2}$ZnGeSe$_{4}$ (CZGSe), and t-Se (Se).}
    \label{fig:ctls}
\end{figure*}

\clearpage

\subsection{CC diagrams, carrier capture coefficients, photovoltaic device parameters}\label{ccpv}

In this section,  we show the 1D configuration co-ordinate diagrams, carrier capture coefficients and photovoltaic device parameters for the remaining systems.

\begin{figure*}[!htbp]
    \centering
    \includegraphics[width=1\linewidth]{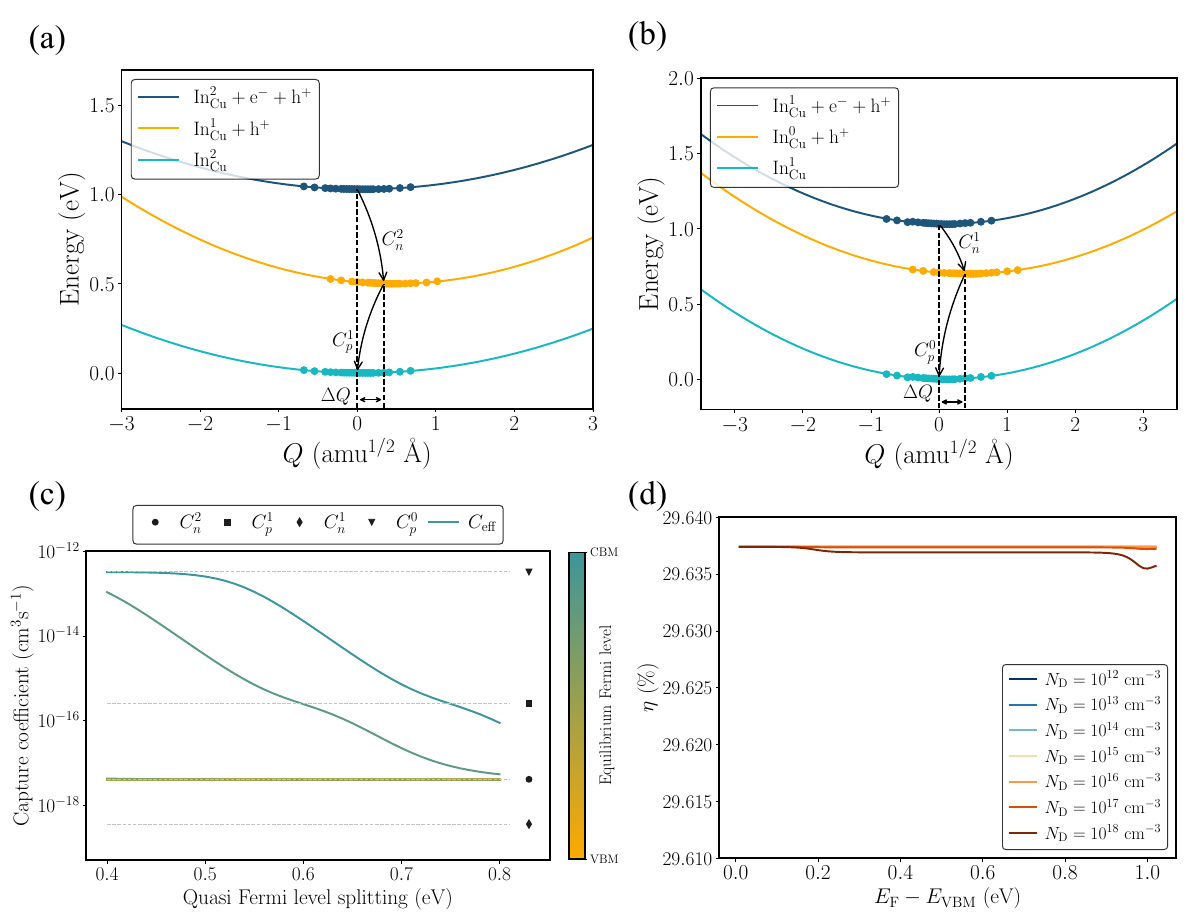}
    \caption{\footnotesize (a), (b) One-dimensional configuration coordinate diagrams for the $2/1$ and $1/0$ charge state transitions in the In$_{\mathrm{Cu}}$ defect in CuInSe$_{2}$. Solid circles denote the data points calculated using the HSE06 functional and solid lines represent parabolic fits. (c) Carrier capture coefficients (including both radiative and non-radiative capture processes) for different charge states are shown as black symbols. The colored lines show the effective capture coefficient as a function of the quasi-Fermi level splitting and doping concentrations (color bar) at 300 K.(d) Photovoltaic device efficiency with respect to equlibrium Fermi level position at different defect concentrations at 300 K and 500 nm film thickness.}
    \label{fig:CISPV}
\end{figure*}

\begin{figure*}
    \centering
    \includegraphics[width=1\linewidth]{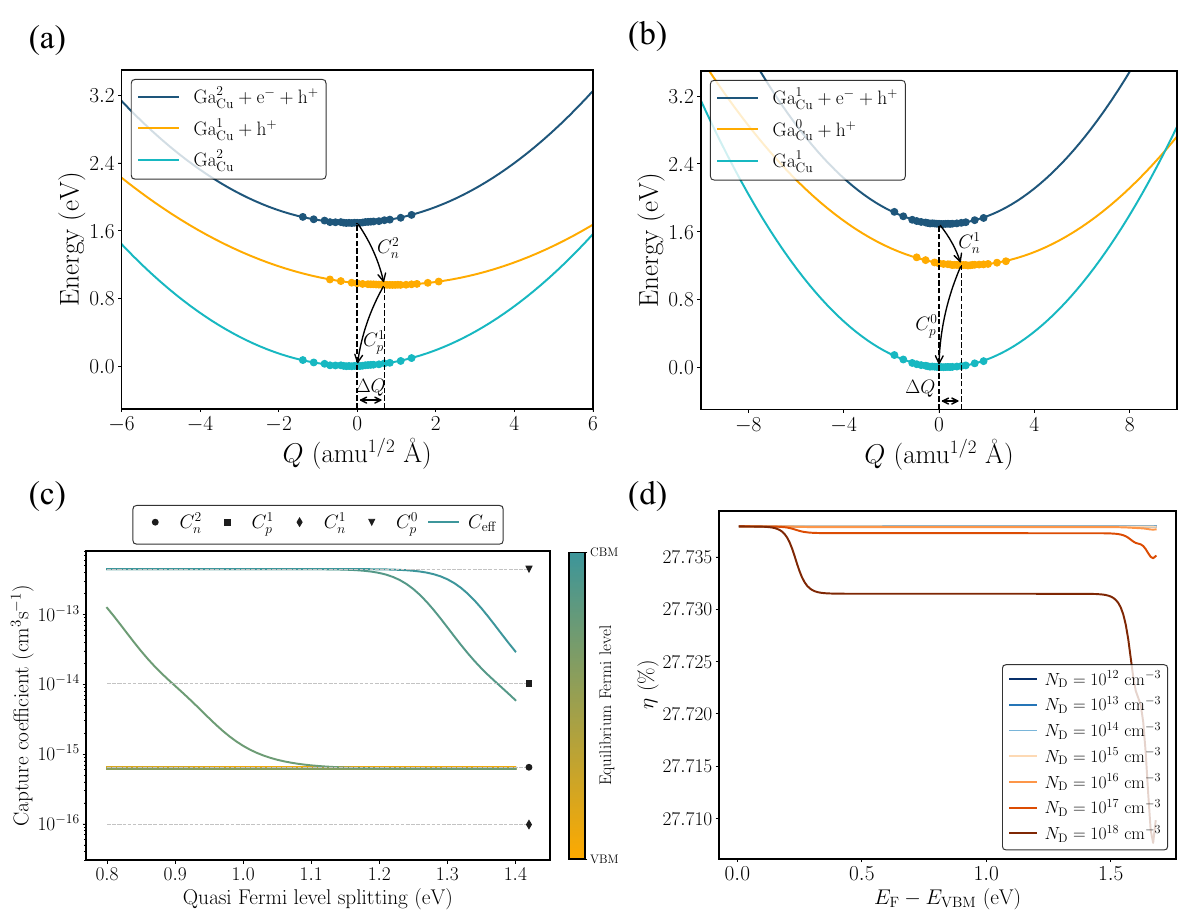}
    \caption{\footnotesize (a), (b) One-dimensional configuration coordinate diagrams for the $2/1$ and $1/0$ charge state transitions in the Ga$_{\mathrm{Cu}}$ defect in CuGaSe$_{2}$. Solid circles denote the data points calculated using the HSE06 functional and solid lines represent parabolic fits. (c) Carrier capture coefficients (including both radiative and non-radiative capture processes) for different charge states are shown as black symbols. The colored lines show the effective capture coefficient as a function of the quasi-Fermi level splitting and doping concentrations (color bar) at 300 K.(d) Photovoltaic device efficiency with respect to equlibrium Fermi level position at different defect concentrations at 300 K and 500 nm film thickness.}
    \label{fig:CGSPV}
\end{figure*}

\begin{figure*}
    \centering
    \includegraphics[width=1\linewidth]{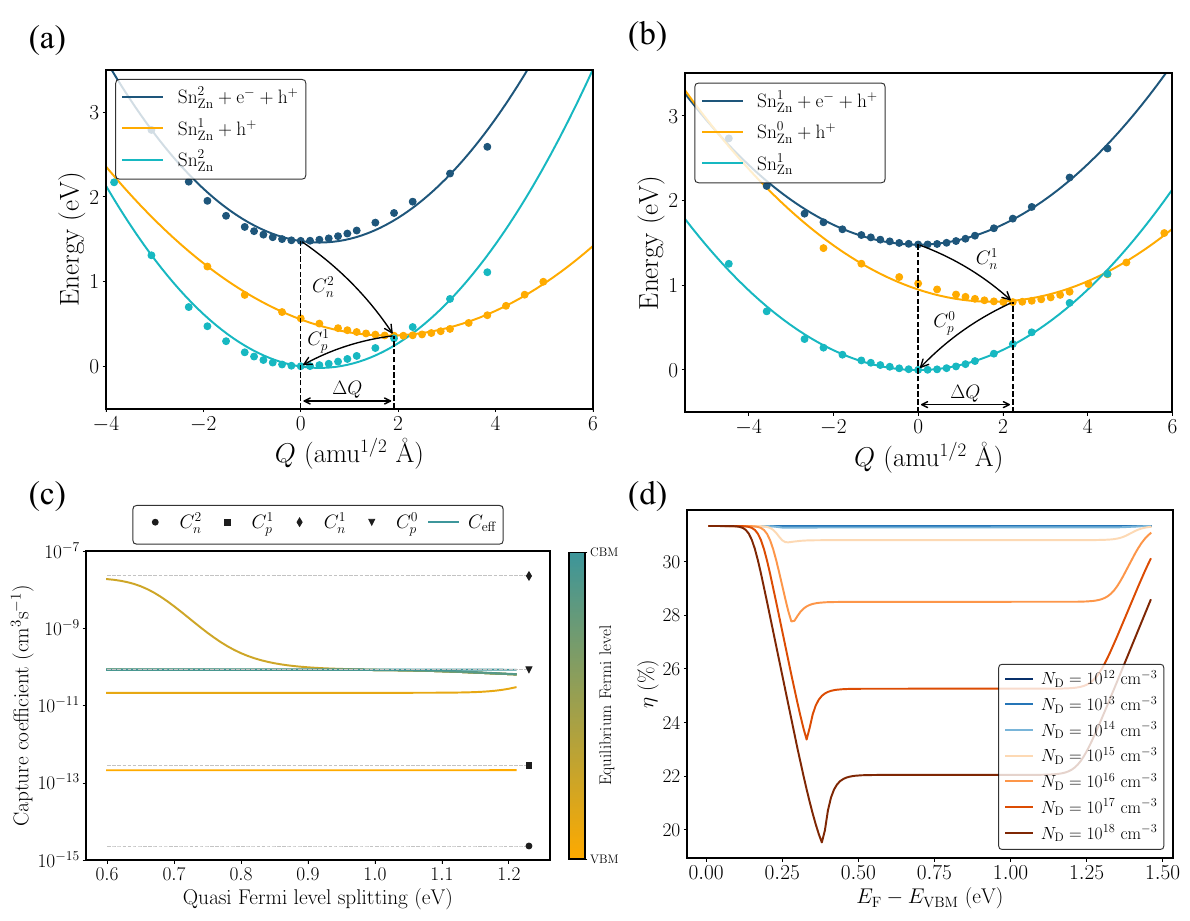}
    \caption{\footnotesize (a), (b) One-dimensional configuration coordinate diagrams for the $2/1$ and $1/0$ charge state transitions in the Sn$_{\mathrm{Zn}}$ defect in Cu$_{2}$ZnSnS$_{4}$. Solid circles denote the data points calculated using the HSE06 functional and solid lines represent parabolic fits. (c) Carrier capture coefficients (including both radiative and non-radiative capture processes) for different charge states are shown as black symbols. The colored lines show the effective capture coefficient as a function of the quasi-Fermi level splitting and doping concentrations (color bar) at 300 K.(d) Photovoltaic device efficiency with respect to equlibrium Fermi level position at different defect concentrations at 300 K and 500 nm film thickness.}
    \label{fig:CZTS}
\end{figure*}

\begin{figure*}
    \centering
    \includegraphics[width=1\linewidth]{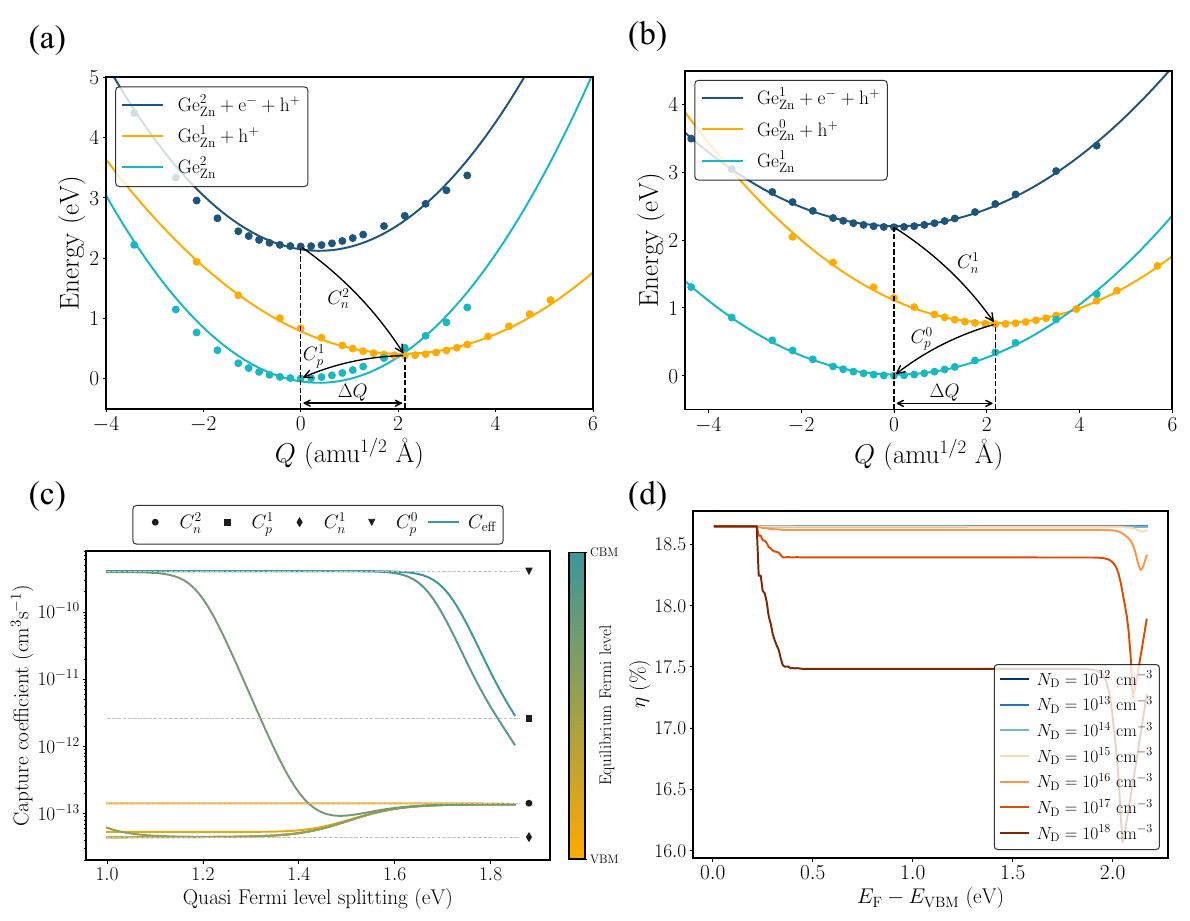}
    \caption{\footnotesize (a), (b) One-dimensional configuration coordinate diagrams for the $2/1$ and $1/0$ charge state transitions in the Ge$_{\mathrm{Zn}}$ defect in Cu$_{2}$ZnGeS$_{4}$. Solid circles denote the data points calculated using the HSE06 functional and solid lines represent parabolic fits. (c) Carrier capture coefficients (including both radiative and non-radiative capture processes) for different charge states are shown as black symbols. The colored lines show the effective capture coefficient as a function of the quasi-Fermi level splitting and doping concentrations (color bar) at 300 K.(d) Photovoltaic device efficiency with respect to equlibrium Fermi level position at different defect concentrations at 300 K and 500 nm film thickness.}
    \label{fig:CZGS}
\end{figure*}

\clearpage

\begin{figure*}
    \centering
    \includegraphics[width=1\linewidth]{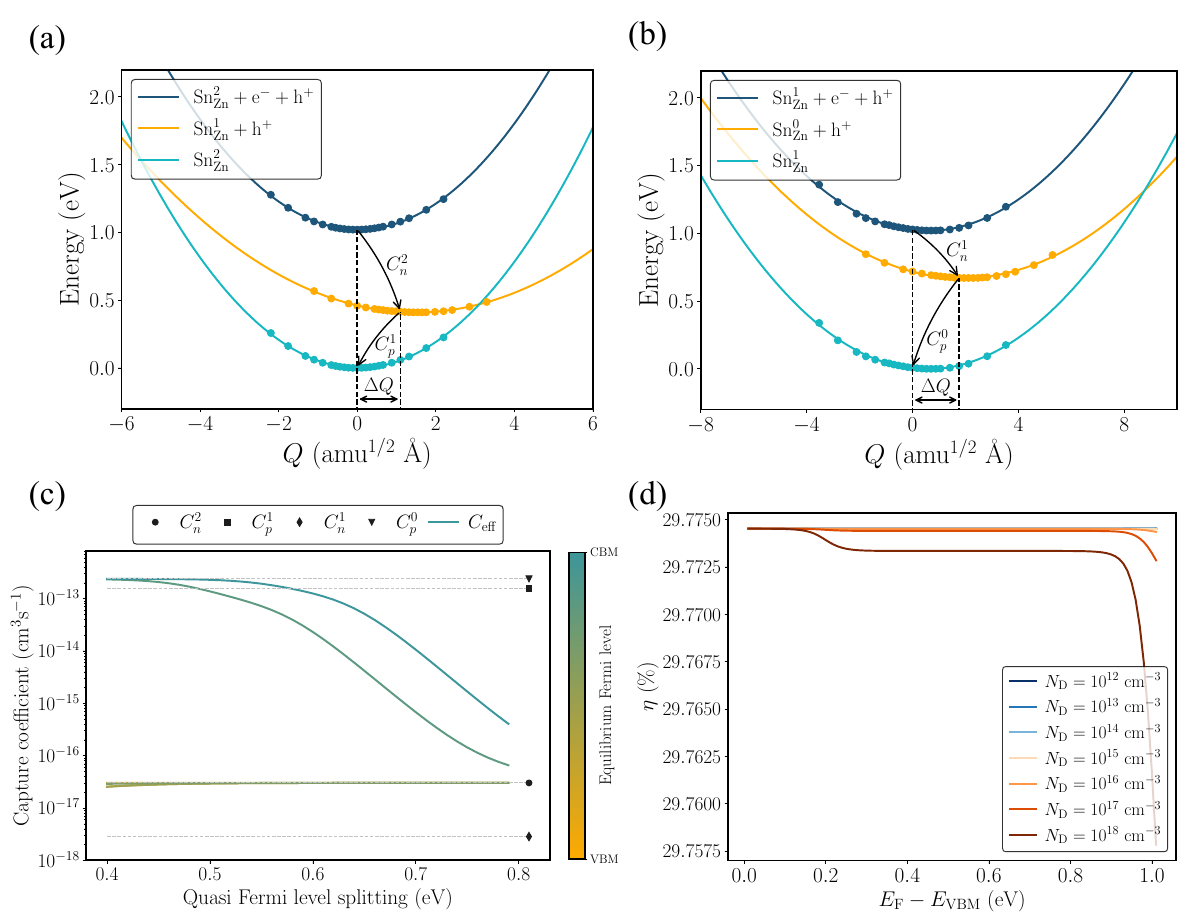}
    \caption{\footnotesize (a), (b) One-dimensional configuration coordinate diagrams for the $2/1$ and $1/0$ charge state transitions in the Sn$_{\mathrm{Zn}}$ defect in Cu$_{2}$ZnSnSe$_{4}$. Solid circles denote the data points calculated using the HSE06 functional and solid lines represent parabolic fits. (c) Carrier capture coefficients (including both radiative and non-radiative capture processes) for different charge states are shown as black symbols. The colored lines show the effective capture coefficient as a function of the quasi-Fermi level splitting and doping concentrations (color bar) at 300 K.(d) Photovoltaic device efficiency with respect to equlibrium Fermi level position at different defect concentrations at 300 K and 500 nm film thickness.}
    \label{fig:CZTSe}
\end{figure*}

\clearpage

\subsection{PV device parameters with a tolerance in CTLs}\label{pvtol}
In this section, we plot the PV device parameters ($\eta$ and $V_{\mathrm{oc}}$) for the systems at different defect concentrations and at three different initial conditions (p-type, intrinsic and n-type). Here we have used a tolerance of 0.2 eV for the defect energy levels. We rigidly shifted the defect levels (both above and below the calculated levels) and calculated the the carrier capture coefficients and associated device parameters which is shown via the error bars. Typical HSE calculations in literature has an accuracy of 0.2 eV for defect level determination with respect to experimental data. From the plots we can see for some compounds (CuInSe$_{2}$, CuGaSe$_{2}$, mostly defect tolerant ones) it is not very sensitive to defect level position, but for Cu$_{2}$ZnSnS$_{4}$ it is highly sensitive in ptype and intrinsic initial conditions. This plot shows why it is important to accurately evaluate the defect energy levels in order to conclude about the defect tolerance. 

\begin{figure*}[!htbp]
    \centering
    \includegraphics[width=0.83\linewidth]{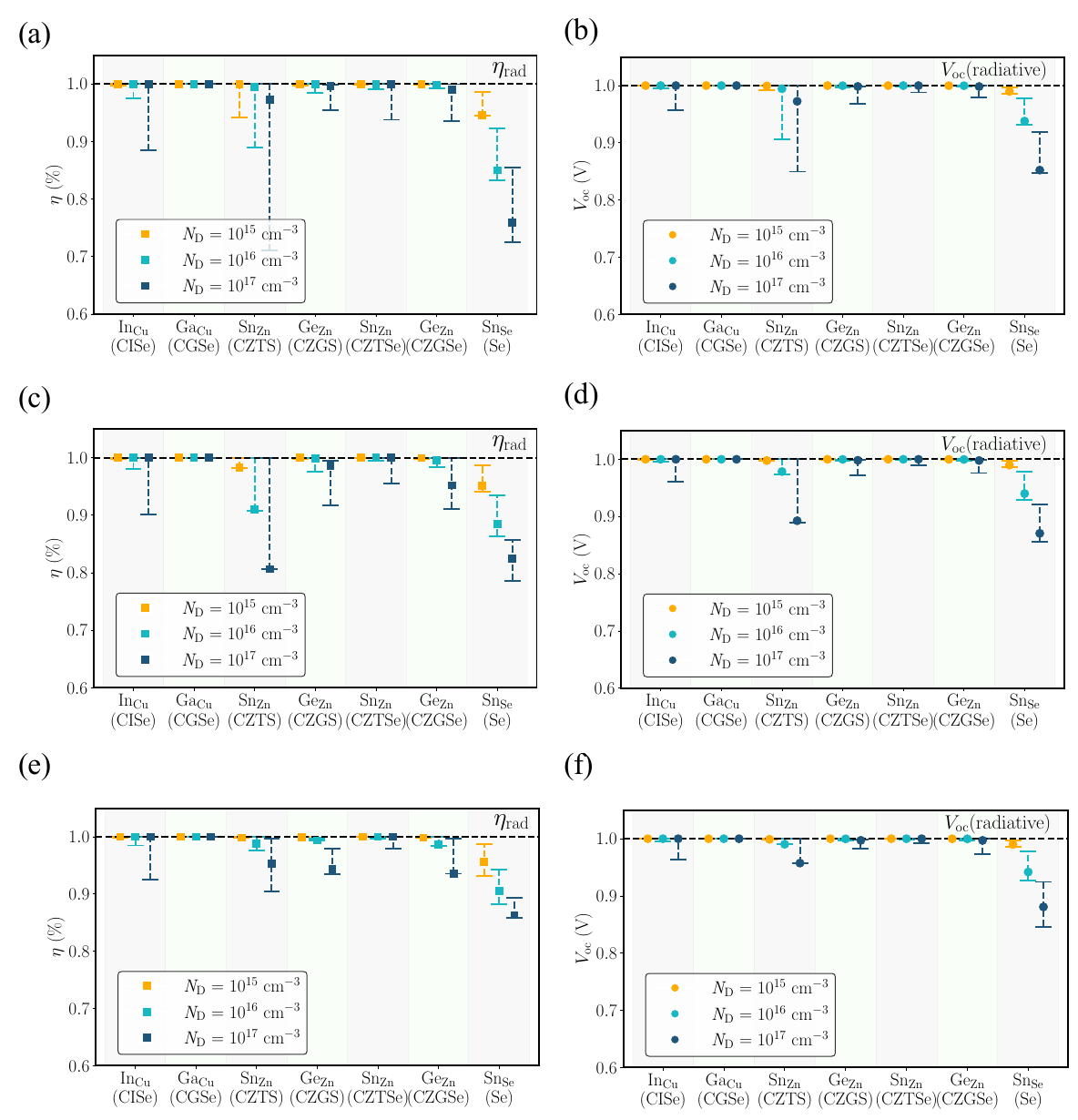}
    \caption{\footnotesize (a), (c), (e) Photovoltaic device efficiency $\eta$ for the host materials, CuInSe$_{2}$ (CISe), CuGaSe$_{2}$ (CGSe), Cu$_{2}$ZnSnS$_{4}$ (CZTS), Cu$_{2}$ZnGeS$_{4}$ (CZGS), Cu$_{2}$ZnSnSe$_{4}$ (CZTSe), Cu$_{2}$ZnGeSe$_{4}$ (CZGSe), and t-Se (Se), considering the effect of considered defects, at different defect concentrations (10$^{15}$, 10$^{16}$, and 10$^{17}$ cm$^{-3}$) and at three different initial equilibrium Fermi levels (p-type, intrinsic and n-type, respectively). (b), (d), (f) Photovoltaic open-circuit voltage ($V_{\mathrm{oc}}$) for the same as above. PV device parameters are plotted at 300 K and 500 nm film thickness. Here we have considered a tolerance of 0.2 eV in the defect charge transition levels. The corresponding carrier capture coefficients are calculated by rigid shifting the CC curves at the considered levels. Associated device parameters are shown as error bars.}
    \label{fig:allPV}
\end{figure*}

\clearpage

\subsection{Effect of including radiative carrier capture}\label{cradl}
In this section, we show the effect of including radiative carrier capture in the net capture rate for Ge$_{\mathrm{Zn}}$ in Cu$_{2}$ZnGeSe$_{4}$ and Sn$_{\mathrm{Se}}$ in t-Se. For photovoltaics both radiative and nonradiative defect-assisted capture contributes to loss and as such should be added to account for total capture. Radiative carrier capture coefficient independently is very small and cannot independently account for efficiency loss. But combined with nonradiative loss, here we can see that depending on whether we include or exclude radiative capture, the effective capture rate can change majorly (upto four orders of magnitude here for Sn$_{\mathrm{Se}}$), which will contribute to the conclusion of defect tolerance both quantitatively and qualitatively. It is therefor essential to include both the capture pathways when accounting for defect assisted carrier recombination.    

\begin{figure*}[!htbp]
    \centering
    \includegraphics[width=1\linewidth]{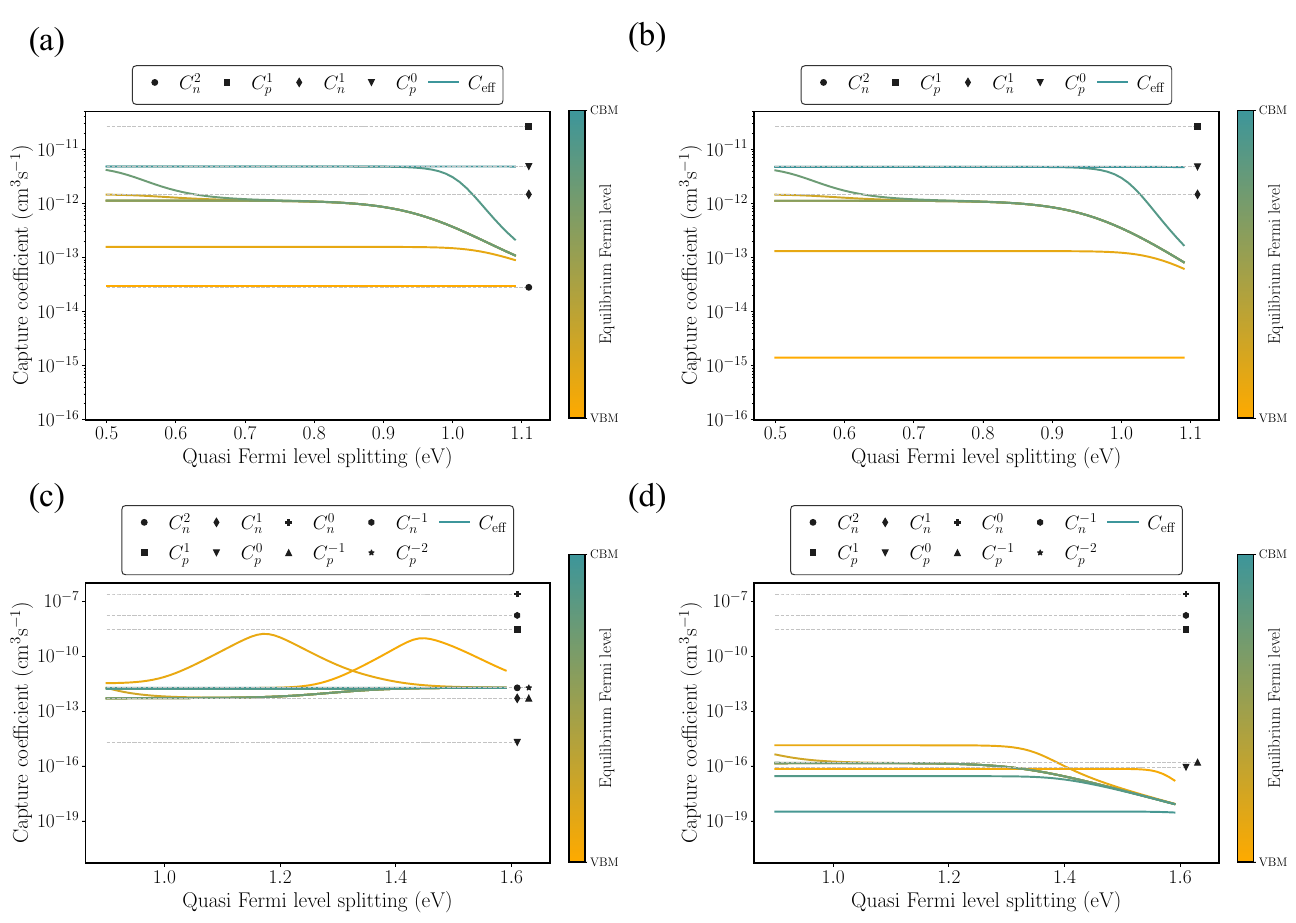}
    \caption{\footnotesize (a), (c) Carrier capture coefficients (including both radiative and non-radiative capture processes) for different charge states are shown as black symbols. The colored lines show the effective capture coefficient as a function of the quasi-Fermi level splitting and doping concentrations (color bar) at 300 K. for Ge$_{\mathrm{Zn}}$ in Cu$_{2}$ZnGeSe$_{4}$ and Sn$_{\mathrm{Se}}$ in t-Se, respectively. (b), (d) same without including the radiative capture process.}
    \label{fig:radiativec}
\end{figure*}

\clearpage

\subsection{Carrier lifetime and PV device parameter comparison between full formalism and different approximations}\label{compfullapproxSI}
In this section, we show the comparison of effective carrier lifetimes, PV device parameters, between Full formalism and commonly used approximations, at different intial and operating conditions for the remaining systems.

\onecolumngrid
\begin{figure*}[!htbp]
    \centering
    \includegraphics[width=1.00\linewidth]{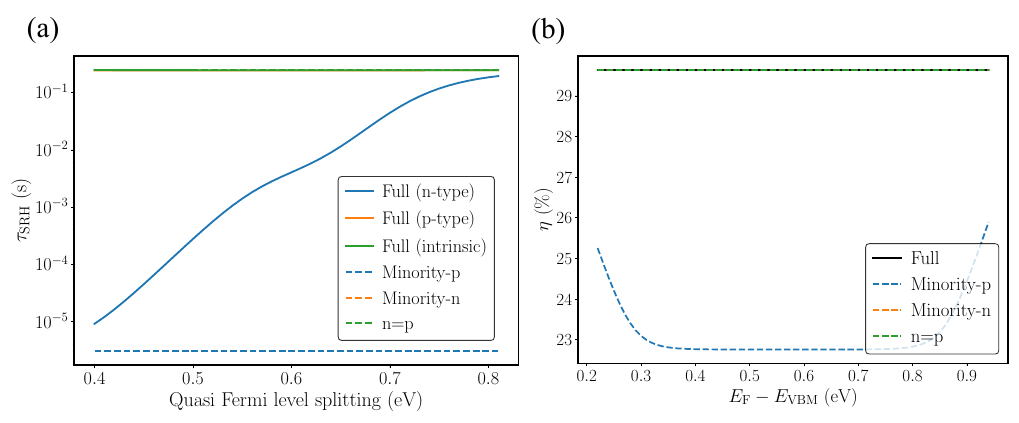}
    \caption{\footnotesize (a) Comparison of effective carrier lifetime ($\tau _{\mathrm{SRH}}$) at different majority carrier types (p-type, n-type, intrinsic) between our full formalism (Full), minority carrier approximation (minority), and $n=p$ approximation (np) for In$_{\mathrm{Cu}}$ defect in CuInSe$_{2}$. (b) Photovoltaic device efficiency ($\mathrm{\eta}$) with respect to equilibrium Fermi level with different approximations at N$_{\mathrm{D}}$=10$^{18}$ cm$^{-3}$, T=300 K, and 500 nm device thickness.}
    \label{fig:CISecomp}
\end{figure*}

\begin{figure*}[!htbp]
    \centering
    \includegraphics[width=1.00\linewidth]{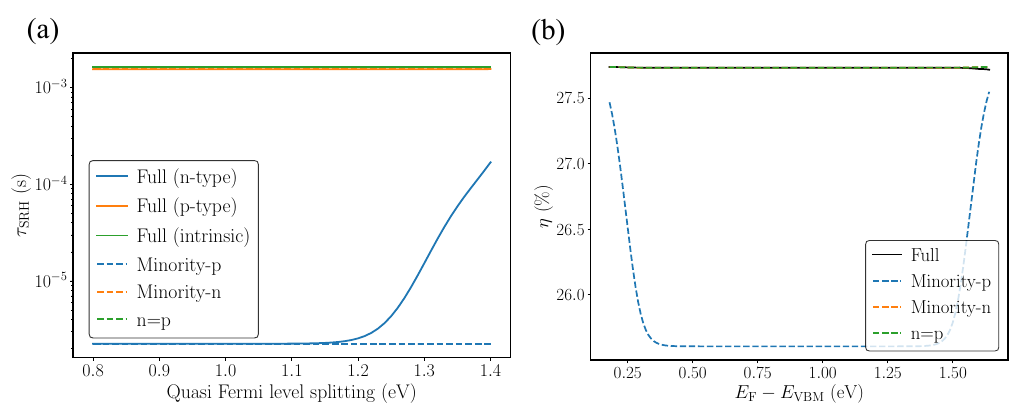}
    \caption{\footnotesize Same as Fig. \ref{fig:CISecomp}, for Ga$_{\mathrm{Cu}}$ defect in CuGaSe$_{2}$.}
    \label{fig:CGSecomp}
\end{figure*}

\begin{figure*}[!htbp]
    \centering
    \includegraphics[width=1.00\linewidth]{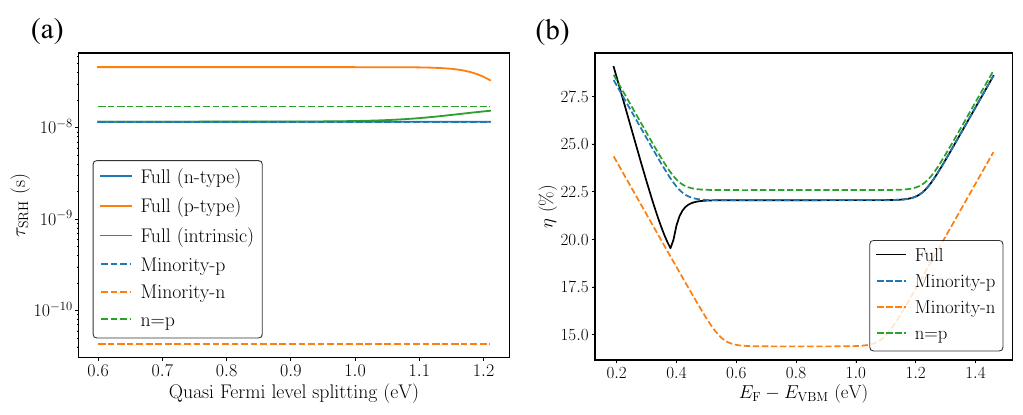}
    \caption{\footnotesize Same as Fig. \ref{fig:CISecomp}, for Sn$_{\mathrm{Zn}}$ defect in Cu$_{2}$ZnSnS$_{4}$.}
    \label{fig:CZTScomp}
\end{figure*}

\begin{figure*}[h!]
    \centering
    \includegraphics[width=1.00\linewidth]{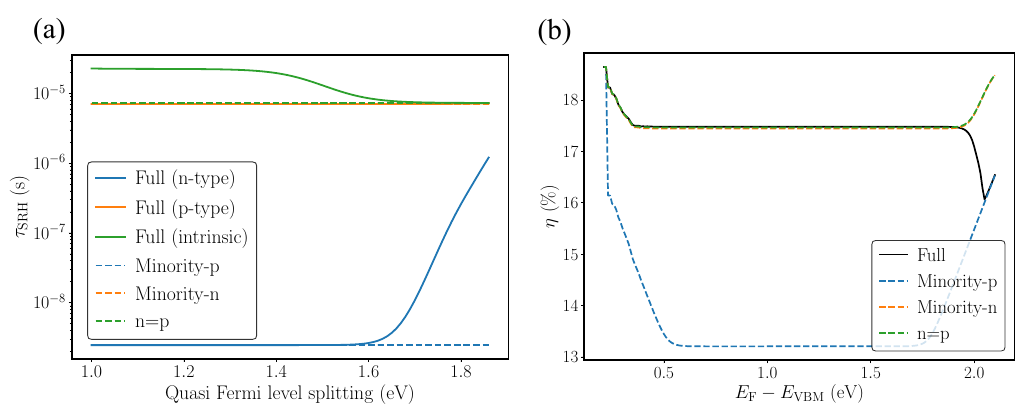}
    \caption{\footnotesize Same as Fig. \ref{fig:CISecomp}, for Ge$_{\mathrm{Zn}}$ defect in Cu$_{2}$GeSnS$_{4}$.}
    \label{fig:CZGScomp}
\end{figure*}

\begin{figure*}[h!]
    \centering
    \includegraphics[width=1.00\linewidth]{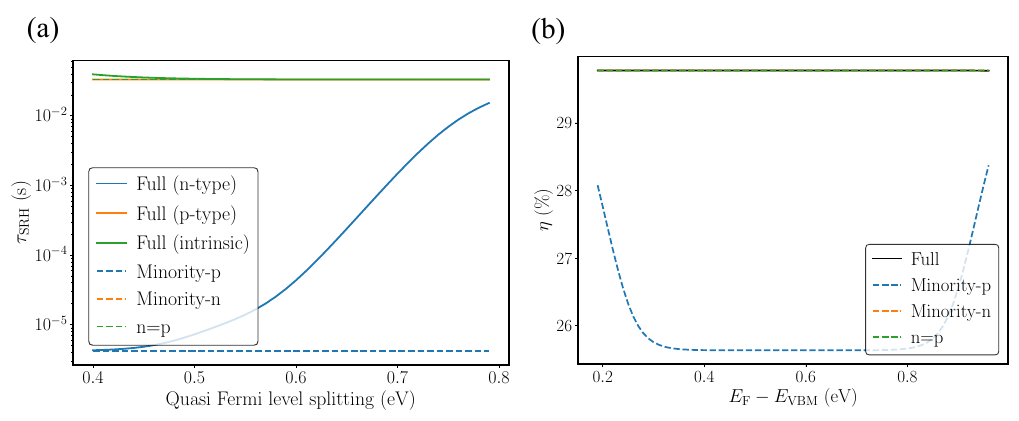}
    \caption{\footnotesize Same as Fig. \ref{fig:CISecomp}, for Sn$_{\mathrm{Zn}}$ defect in Cu$_{2}$ZnSnSe$_{4}$.}
    \label{fig:CZTSe_comp}
\end{figure*}

\clearpage

\subsection{Defect concentration distribution between charge states at different initial majority carrier types during photovoltaic operation}\label{cdefd}

If we use the full formalism as in our methodology (following eq.\ref{eqmulti6SI}-eq.\ref{eqtwoctl}) we can derive the defect charge state occupation ratios for a defect with two CTLs ($2/1$ and $1/0$) as,
\begin{equation}
    \frac{N_2}{N_1}=\frac{C_p^{1}p+C_n^{2}n^2}{C_n^{2}n+C_p^{1}p^2}
    \label{eq_N2N1}
\end{equation}
and 
\begin{equation}
    \frac{N_1}{N_0}=\frac{C_p^{0}p+C_n^{1}n^1}{C_n^{1}n+C_p^{0}p^1}
    \label{eq_N1N0}
\end{equation}.
Here $n^2$, $p^2$ are the free electron and hole concentrations if the Fermi level lies in the trap energy at CTL(2/1). Similarly $n^1$, $p^1$ are the electron and hole concentrations if the Fermi level lies in the trap energy at CTL(1/0). $C_n^q$ ($C_p^{q-1}$) is the electron (hole) capture coefficient of the defect in initial charge state $q$ ($q-1$).

In this section, we have plotted defect charge state distribution with respect to equilibrium Fermi level at three different operating conditions (low, moderate and high quasi Fermi level splitting). These plots help explain the defect charge state occupation and as such the rate determining transition at different initial and operating conditions. This behaviour is a clear differentiator between our formalism and typically used formalism which assumes same defect distribution irrespective of initial or operating conditions.
\begin{figure*}[!htbp]
    \centering
    \includegraphics[width=1.00\linewidth]{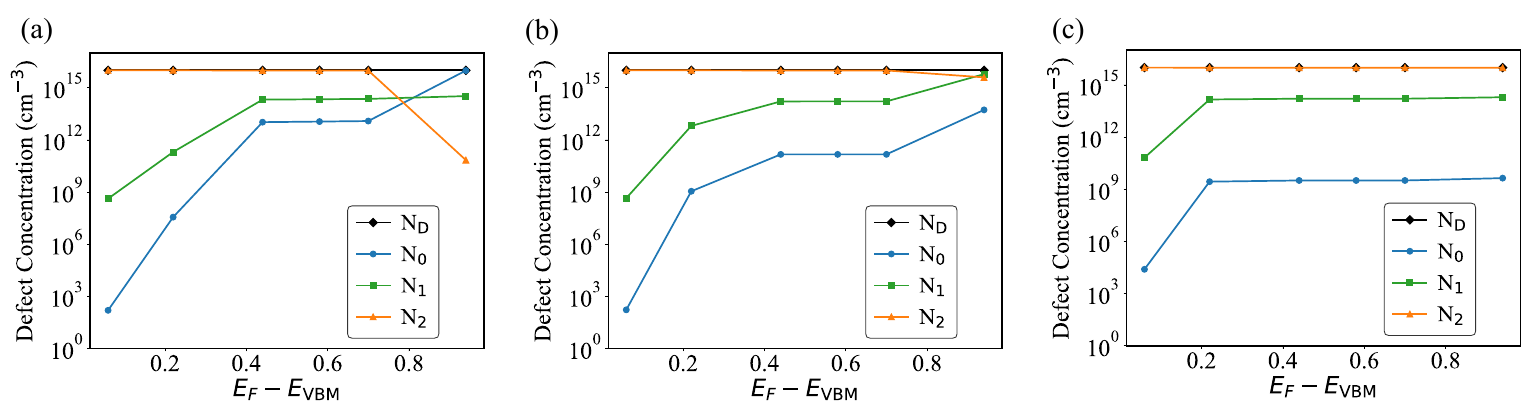}
    \caption{\footnotesize Variation of defect (D) charge state distribution between allowed charge states (here 2, 1, 0) with respect to initial condition (equilibrium Fermi level) during operation at (a) low quasi Fermi level splitting (0.5$V_\mathrm{oc}^\mathrm{rad}$), (b) moderate quasi Fermi level splitting (0.75$V_\mathrm{oc}^\mathrm{rad}$), (c) large quasi Fermi level splitting ($V_\mathrm{oc}^\mathrm{rad}$)) for In$_{\mathrm{Cu}}$ defect in CuInSe$_{2}$. Here total defect concentration is taken as N$_{\mathrm{D}}$=10$^{16}$ cm$^{-3}$. $V_\mathrm{oc}^\mathrm{rad}$ is the open circuit voltage with only radiative recombination considered.}
    \label{fig:ND1}
\end{figure*}

\begin{figure*}[!htbp]
    \centering
    \includegraphics[width=1.00\linewidth]{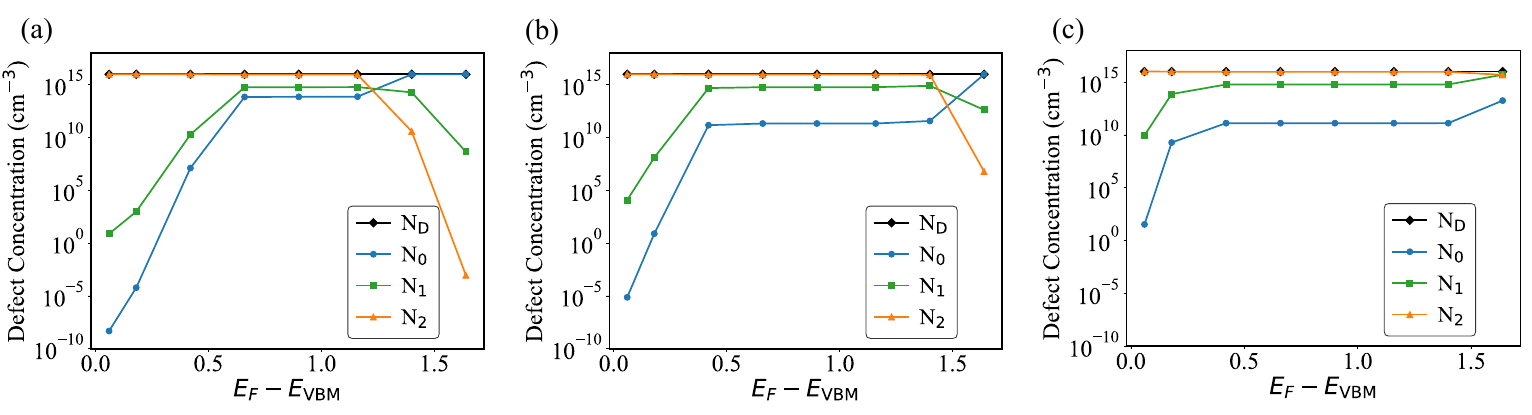}
    \caption{\footnotesize Same as Fig. \ref{fig:ND1}, for Ga$_{\mathrm{Cu}}$ defect in CuGaSe$_{2}$.}
    \label{fig:ND2}
\end{figure*}

\begin{figure*}[!htbp]
    \centering
    \includegraphics[width=1.00\linewidth]{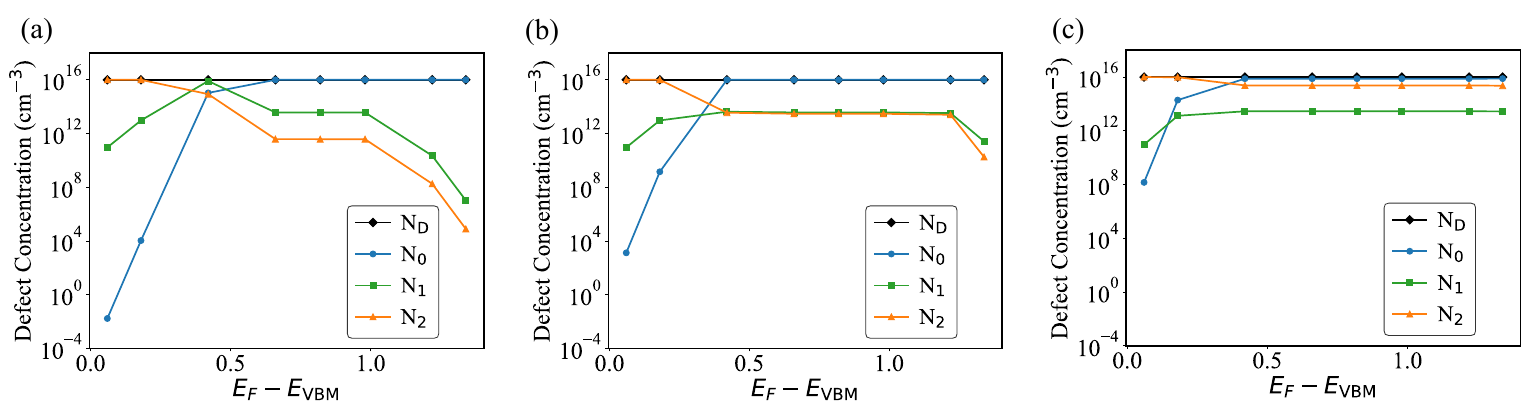}
    \caption{\footnotesize Same as Fig. \ref{fig:ND1}, for Sn$_{\mathrm{Zn}}$ defect in Cu$_{2}$ZnSnS$_{4}$.}
    \label{fig:ND3}
\end{figure*}

\begin{figure*}[!htbp]
    \centering
    \includegraphics[width=1.00\linewidth]{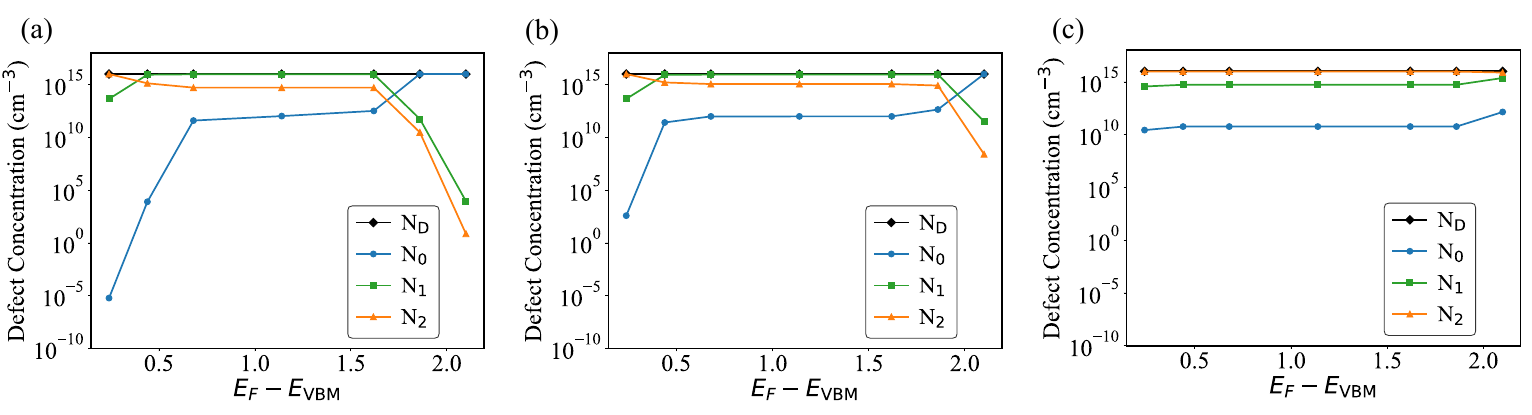}
    \caption{\footnotesize Same as Fig. \ref{fig:ND1}, for Ge$_{\mathrm{Zn}}$ defect in Cu$_{2}$GeSnS$_{4}$.}
    \label{fig:ND4}
\end{figure*}

\begin{figure*}[!htbp]
    \centering
    \includegraphics[width=1.00\linewidth]{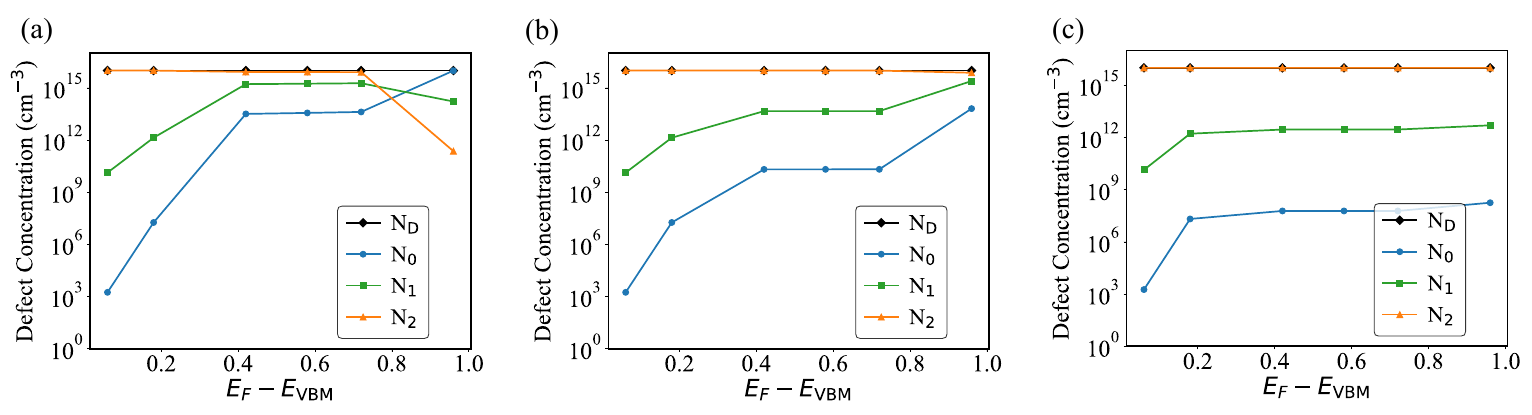}
    \caption{\footnotesize Same as Fig. \ref{fig:ND1}, for Sn$_{\mathrm{Zn}}$ defect in Cu$_{2}$ZnSnSe$_{4}$.}
    \label{fig:ND5}
\end{figure*}

\begin{figure*}[!htbp]
    \centering
    \includegraphics[width=1.00\linewidth]{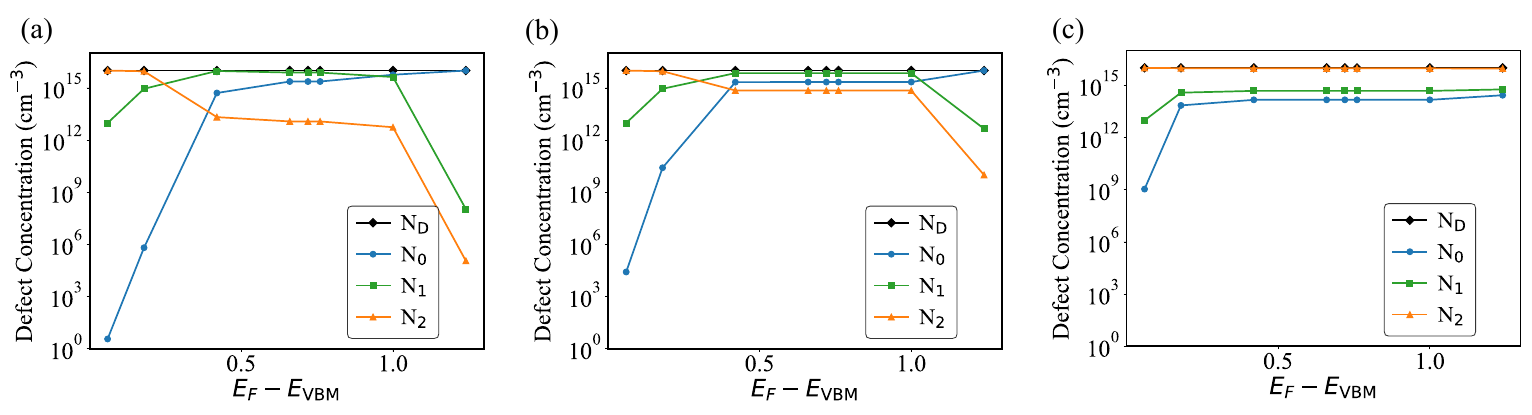}
    \caption{\footnotesize Same as Fig. \ref{fig:ND1}, for Ge$_{\mathrm{Zn}}$ defect in Cu$_{2}$GeSnSe$_{4}$.}
    \label{fig:ND6}
\end{figure*}

\begin{figure*}[!htbp]
    \centering
    \includegraphics[width=1.00\linewidth]{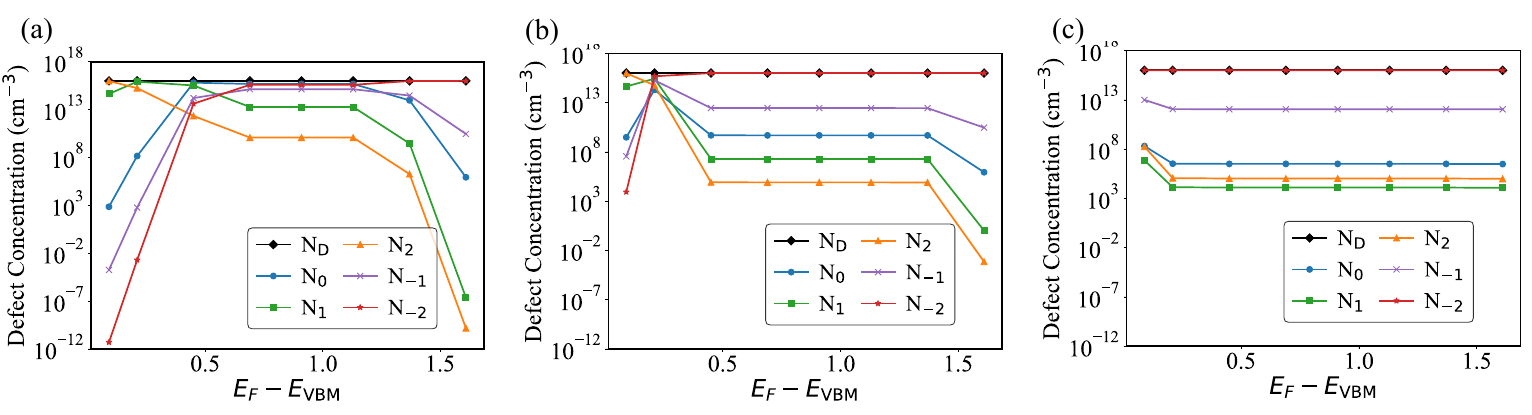}
    \caption{\footnotesize Variation of defect (D) charge state distribution between allowed charge states (2, 1, 0, -1, -2 here) with respect to initial condition (equilibrium Fermi level) during operation at (a) low quasi Fermi level splitting (0.5$V_\mathrm{oc}^\mathrm{rad}$), (b) moderate quasi Fermi level splitting (0.75$V_\mathrm{oc}^\mathrm{rad}$), (c) large quasi Fermi level splitting ($V_\mathrm{oc}^\mathrm{rad}$)) for Sn$_{\mathrm{Se}}$ defect in t-Se. Here total defect concentration is taken as N$_{\mathrm{D}}$=10$^{16}$ cm$^{-3}$. $V_\mathrm{oc}^\mathrm{rad}$ is the open circuit voltage with only radiative recombination considered.}
    \label{fig:ND7}
\end{figure*}

\clearpage

\makeatletter
\c@enumiv=0
\makeatother

%

%% file: main.bbl
\providecommand{\noopsort}[1]{}\providecommand{\singleletter}[1]{#1}
\begin{thebibliography}{46}%
\makeatletter
\providecommand \@ifxundefined [1]{%
 \@ifx{#1\undefined}
}%
\providecommand \@ifnum [1]{%
 \ifnum #1\expandafter \@firstoftwo
 \else \expandafter \@secondoftwo
 \fi
}%
\providecommand \@ifx [1]{%
 \ifx #1\expandafter \@firstoftwo
 \else \expandafter \@secondoftwo
 \fi
}%
\providecommand \natexlab [1]{#1}%
\providecommand \enquote  [1]{``#1''}%
\providecommand \bibnamefont  [1]{#1}%
\providecommand \bibfnamefont [1]{#1}%
\providecommand \citenamefont [1]{#1}%
\providecommand \href@noop [0]{\@secondoftwo}%
\providecommand \href [0]{\begingroup \@sanitize@url \@href}%
\providecommand \@href[1]{\@@startlink{#1}\@@href}%
\providecommand \@@href[1]{\endgroup#1\@@endlink}%
\providecommand \@sanitize@url [0]{\catcode `\\12\catcode `\$12\catcode `\&12\catcode `\#12\catcode `\^12\catcode `\_12\catcode `\%12\relax}%
\providecommand \@@startlink[1]{}%
\providecommand \@@endlink[0]{}%
\providecommand \url  [0]{\begingroup\@sanitize@url \@url }%
\providecommand \@url [1]{\endgroup\@href {#1}{\urlprefix }}%
\providecommand \urlprefix  [0]{URL }%
\providecommand \Eprint [0]{\href }%
\providecommand \doibase [0]{https://doi.org/}%
\providecommand \selectlanguage [0]{\@gobble}%
\providecommand \bibinfo  [0]{\@secondoftwo}%
\providecommand \bibfield  [0]{\@secondoftwo}%
\providecommand \translation [1]{[#1]}%
\providecommand \BibitemOpen [0]{}%
\providecommand \bibitemStop [0]{}%
\providecommand \bibitemNoStop [0]{.\EOS\space}%
\providecommand \EOS [0]{\spacefactor3000\relax}%
\providecommand \BibitemShut  [1]{\csname bibitem#1\endcsname}%
\let\auto@bib@innerbib\@empty
\bibitem [{\citenamefont {Freysoldt}\ \emph {et~al.}(2014)\citenamefont {Freysoldt}, \citenamefont {Grabowski}, \citenamefont {Hickel}, \citenamefont {Neugebauer}, \citenamefont {Kresse}, \citenamefont {Janotti},\ and\ \citenamefont {Walle}}]{Freysoldt2014}%
  \BibitemOpen
  \bibfield  {author} {\bibinfo {author} {\bibfnamefont {C.}~\bibnamefont {Freysoldt}}, \bibinfo {author} {\bibfnamefont {B.}~\bibnamefont {Grabowski}}, \bibinfo {author} {\bibfnamefont {T.}~\bibnamefont {Hickel}}, \bibinfo {author} {\bibfnamefont {J.}~\bibnamefont {Neugebauer}}, \bibinfo {author} {\bibfnamefont {G.}~\bibnamefont {Kresse}}, \bibinfo {author} {\bibfnamefont {A.}~\bibnamefont {Janotti}},\ and\ \bibinfo {author} {\bibfnamefont {C.~G. V.~D.}\ \bibnamefont {Walle}},\ }\bibfield  {title} {\bibinfo {title} {First-principles calculations for point defects in solids},\ }\href {https://doi.org/10.1103/RevModPhys.86.253} {\bibfield  {journal} {\bibinfo  {journal} {Reviews of Modern Physics}\ }\textbf {\bibinfo {volume} {86}},\ \bibinfo {pages} {253} (\bibinfo {year} {2014})}\BibitemShut {NoStop}%
\bibitem [{\citenamefont {Alkauskas}\ \emph {et~al.}(2016)\citenamefont {Alkauskas}, \citenamefont {McCluskey},\ and\ \citenamefont {Walle}}]{Alkauskas2016tut}%
  \BibitemOpen
  \bibfield  {author} {\bibinfo {author} {\bibfnamefont {A.}~\bibnamefont {Alkauskas}}, \bibinfo {author} {\bibfnamefont {M.~D.}\ \bibnamefont {McCluskey}},\ and\ \bibinfo {author} {\bibfnamefont {C.~G. V.~D.}\ \bibnamefont {Walle}},\ }\bibfield  {title} {\bibinfo {title} {Tutorial: Defects in semiconductors - combining experiment and theory},\ }\bibfield  {journal} {\bibinfo  {journal} {Journal of Applied Physics}\ }\textbf {\bibinfo {volume} {119}},\ \href {https://doi.org/10.1063/1.4948245} {10.1063/1.4948245} (\bibinfo {year} {2016})\BibitemShut {NoStop}%
\bibitem [{\citenamefont {Park}\ \emph {et~al.}(2018)\citenamefont {Park}, \citenamefont {Kim}, \citenamefont {Xie},\ and\ \citenamefont {Walsh}}]{Park2018}%
  \BibitemOpen
  \bibfield  {author} {\bibinfo {author} {\bibfnamefont {J.~S.}\ \bibnamefont {Park}}, \bibinfo {author} {\bibfnamefont {S.}~\bibnamefont {Kim}}, \bibinfo {author} {\bibfnamefont {Z.}~\bibnamefont {Xie}},\ and\ \bibinfo {author} {\bibfnamefont {A.}~\bibnamefont {Walsh}},\ }\bibfield  {title} {\bibinfo {title} {Point defect engineering in thin-film solar cells},\ }\href {https://doi.org/10.1038/s41578-018-0026-7} {\bibfield  {journal} {\bibinfo  {journal} {Nature Reviews Materials}\ }\textbf {\bibinfo {volume} {3}},\ \bibinfo {pages} {194} (\bibinfo {year} {2018})}\BibitemShut {NoStop}%
\bibitem [{\citenamefont {Dreyer}\ \emph {et~al.}(2018)\citenamefont {Dreyer}, \citenamefont {Alkauskas}, \citenamefont {Lyons}, \citenamefont {Janotti},\ and\ \citenamefont {Walle}}]{Dreyer2018}%
  \BibitemOpen
  \bibfield  {author} {\bibinfo {author} {\bibfnamefont {C.~E.}\ \bibnamefont {Dreyer}}, \bibinfo {author} {\bibfnamefont {A.}~\bibnamefont {Alkauskas}}, \bibinfo {author} {\bibfnamefont {J.~L.}\ \bibnamefont {Lyons}}, \bibinfo {author} {\bibfnamefont {A.}~\bibnamefont {Janotti}},\ and\ \bibinfo {author} {\bibfnamefont {C.~G. V.~D.}\ \bibnamefont {Walle}},\ }\bibfield  {title} {\bibinfo {title} {First-principles calculations of point defects for quantum technologies},\ }\href {https://doi.org/10.1146/annurev-matsci-070317} {\bibfield  {journal} {\bibinfo  {journal} {Annual Review of Materials Research}\ }\textbf {\bibinfo {volume} {48}},\ \bibinfo {pages} {1} (\bibinfo {year} {2018})}\BibitemShut {NoStop}%
\bibitem [{\citenamefont {Shockley}\ and\ \citenamefont {Read~Jr}(1952)}]{shockley1952statistics}%
  \BibitemOpen
  \bibfield  {author} {\bibinfo {author} {\bibfnamefont {W.}~\bibnamefont {Shockley}}\ and\ \bibinfo {author} {\bibfnamefont {W.}~\bibnamefont {Read~Jr}},\ }\bibfield  {title} {\bibinfo {title} {Statistics of the recombinations of holes and electrons},\ }\href@noop {} {\bibfield  {journal} {\bibinfo  {journal} {Physical review}\ }\textbf {\bibinfo {volume} {87}},\ \bibinfo {pages} {835} (\bibinfo {year} {1952})}\BibitemShut {NoStop}%
\bibitem [{\citenamefont {Dreyer}\ \emph {et~al.}(2020)\citenamefont {Dreyer}, \citenamefont {Alkauskas}, \citenamefont {Lyons},\ and\ \citenamefont {Walle}}]{Dreyer2020}%
  \BibitemOpen
  \bibfield  {author} {\bibinfo {author} {\bibfnamefont {C.~E.}\ \bibnamefont {Dreyer}}, \bibinfo {author} {\bibfnamefont {A.}~\bibnamefont {Alkauskas}}, \bibinfo {author} {\bibfnamefont {J.~L.}\ \bibnamefont {Lyons}},\ and\ \bibinfo {author} {\bibfnamefont {C.~G. V.~D.}\ \bibnamefont {Walle}},\ }\bibfield  {title} {\bibinfo {title} {Radiative capture rates at deep defects from electronic structure calculations},\ }\bibfield  {journal} {\bibinfo  {journal} {Physical Review B}\ }\textbf {\bibinfo {volume} {102}},\ \href {https://doi.org/10.1103/PhysRevB.102.085305} {10.1103/PhysRevB.102.085305} (\bibinfo {year} {2020})\BibitemShut {NoStop}%
\bibitem [{\citenamefont {Alkauskas}\ \emph {et~al.}(2014)\citenamefont {Alkauskas}, \citenamefont {Yan},\ and\ \citenamefont {Van~de Walle}}]{alkauskas2014first}%
  \BibitemOpen
  \bibfield  {author} {\bibinfo {author} {\bibfnamefont {A.}~\bibnamefont {Alkauskas}}, \bibinfo {author} {\bibfnamefont {Q.}~\bibnamefont {Yan}},\ and\ \bibinfo {author} {\bibfnamefont {C.~G.}\ \bibnamefont {Van~de Walle}},\ }\bibfield  {title} {\bibinfo {title} {First-principles theory of nonradiative carrier capture via multiphonon emission},\ }\href@noop {} {\bibfield  {journal} {\bibinfo  {journal} {Physical Review B}\ }\textbf {\bibinfo {volume} {90}},\ \bibinfo {pages} {075202} (\bibinfo {year} {2014})}\BibitemShut {NoStop}%
\bibitem [{\citenamefont {Hall}(1952)}]{hall1952electron}%
  \BibitemOpen
  \bibfield  {author} {\bibinfo {author} {\bibfnamefont {R.~N.}\ \bibnamefont {Hall}},\ }\bibfield  {title} {\bibinfo {title} {Electron-hole recombination in germanium},\ }\href@noop {} {\bibfield  {journal} {\bibinfo  {journal} {Physical review}\ }\textbf {\bibinfo {volume} {87}},\ \bibinfo {pages} {387} (\bibinfo {year} {1952})}\BibitemShut {NoStop}%
\bibitem [{\citenamefont {Henry}\ and\ \citenamefont {Lang}(1977)}]{henry1977nonradiative}%
  \BibitemOpen
  \bibfield  {author} {\bibinfo {author} {\bibfnamefont {C.}~\bibnamefont {Henry}}\ and\ \bibinfo {author} {\bibfnamefont {D.~V.}\ \bibnamefont {Lang}},\ }\bibfield  {title} {\bibinfo {title} {Nonradiative capture and recombination by multiphonon emission in gaas and gap},\ }\href@noop {} {\bibfield  {journal} {\bibinfo  {journal} {Physical Review B}\ }\textbf {\bibinfo {volume} {15}},\ \bibinfo {pages} {989} (\bibinfo {year} {1977})}\BibitemShut {NoStop}%
\bibitem [{\citenamefont {Rosier}\ and\ \citenamefont {Sah}(1971)}]{rosier1971thermal}%
  \BibitemOpen
  \bibfield  {author} {\bibinfo {author} {\bibfnamefont {L.}~\bibnamefont {Rosier}}\ and\ \bibinfo {author} {\bibfnamefont {C.}~\bibnamefont {Sah}},\ }\bibfield  {title} {\bibinfo {title} {Thermal emission and capture of electrons at sulfur centers in silicon},\ }\href@noop {} {\bibfield  {journal} {\bibinfo  {journal} {Solid-State Electronics}\ }\textbf {\bibinfo {volume} {14}},\ \bibinfo {pages} {41} (\bibinfo {year} {1971})}\BibitemShut {NoStop}%
\bibitem [{\citenamefont {Shi}\ and\ \citenamefont {Wang}(2012)}]{Shi2012}%
  \BibitemOpen
  \bibfield  {author} {\bibinfo {author} {\bibfnamefont {L.}~\bibnamefont {Shi}}\ and\ \bibinfo {author} {\bibfnamefont {L.~W.}\ \bibnamefont {Wang}},\ }\bibfield  {title} {\bibinfo {title} {Ab initio calculations of deep-level carrier nonradiative recombination rates in bulk semiconductors},\ }\bibfield  {journal} {\bibinfo  {journal} {Physical Review Letters}\ }\textbf {\bibinfo {volume} {109}},\ \href {https://doi.org/10.1103/PhysRevLett.109.245501} {10.1103/PhysRevLett.109.245501} (\bibinfo {year} {2012})\BibitemShut {NoStop}%
\bibitem [{\citenamefont {Wang}(2019)}]{Wang2019}%
  \BibitemOpen
  \bibfield  {author} {\bibinfo {author} {\bibfnamefont {L.}~\bibnamefont {Wang}},\ }\bibfield  {title} {\bibinfo {title} {Some recent advances in ab initio calculations of nonradiative decay rates of point defects in semiconductors},\ }\bibfield  {journal} {\bibinfo  {journal} {Journal of Semiconductors}\ }\textbf {\bibinfo {volume} {40}},\ \href {https://doi.org/10.1088/1674-4926/40/9/091101} {10.1088/1674-4926/40/9/091101} (\bibinfo {year} {2019})\BibitemShut {NoStop}%
\bibitem [{\citenamefont {Kim}\ \emph {et~al.}(2020{\natexlab{a}})\citenamefont {Kim}, \citenamefont {Márquez}, \citenamefont {Unold},\ and\ \citenamefont {Walsh}}]{Kim2020}%
  \BibitemOpen
  \bibfield  {author} {\bibinfo {author} {\bibfnamefont {S.}~\bibnamefont {Kim}}, \bibinfo {author} {\bibfnamefont {J.~A.}\ \bibnamefont {Márquez}}, \bibinfo {author} {\bibfnamefont {T.}~\bibnamefont {Unold}},\ and\ \bibinfo {author} {\bibfnamefont {A.}~\bibnamefont {Walsh}},\ }\bibfield  {title} {\bibinfo {title} {Upper limit to the photovoltaic efficiency of imperfect crystals from first principles},\ }\href {https://doi.org/10.1039/d0ee00291g} {\bibfield  {journal} {\bibinfo  {journal} {Energy and Environmental Science}\ }\textbf {\bibinfo {volume} {13}},\ \bibinfo {pages} {1481} (\bibinfo {year} {2020}{\natexlab{a}})}\BibitemShut {NoStop}%
\bibitem [{\citenamefont {Zhang}\ \emph {et~al.}(2021)\citenamefont {Zhang}, \citenamefont {Turiansky},\ and\ \citenamefont {de~Walle}}]{Zhang2021inorganicpv}%
  \BibitemOpen
  \bibfield  {author} {\bibinfo {author} {\bibfnamefont {X.}~\bibnamefont {Zhang}}, \bibinfo {author} {\bibfnamefont {M.~E.}\ \bibnamefont {Turiansky}},\ and\ \bibinfo {author} {\bibfnamefont {C.~G.~V.}\ \bibnamefont {de~Walle}},\ }\bibfield  {title} {\bibinfo {title} {All-inorganic halide perovskites as candidates for efficient solar cells},\ }\bibfield  {journal} {\bibinfo  {journal} {Cell Reports Physical Science}\ }\textbf {\bibinfo {volume} {2}},\ \href {https://doi.org/10.1016/j.xcrp.2021.100604} {10.1016/j.xcrp.2021.100604} (\bibinfo {year} {2021})\BibitemShut {NoStop}%
\bibitem [{\citenamefont {Dou}\ \emph {et~al.}(2023)\citenamefont {Dou}, \citenamefont {Falletta}, \citenamefont {Neugebauer}, \citenamefont {Freysoldt}, \citenamefont {Zhang},\ and\ \citenamefont {Wei}}]{Dou2023}%
  \BibitemOpen
  \bibfield  {author} {\bibinfo {author} {\bibfnamefont {B.}~\bibnamefont {Dou}}, \bibinfo {author} {\bibfnamefont {S.}~\bibnamefont {Falletta}}, \bibinfo {author} {\bibfnamefont {J.}~\bibnamefont {Neugebauer}}, \bibinfo {author} {\bibfnamefont {C.}~\bibnamefont {Freysoldt}}, \bibinfo {author} {\bibfnamefont {X.}~\bibnamefont {Zhang}},\ and\ \bibinfo {author} {\bibfnamefont {S.~H.}\ \bibnamefont {Wei}},\ }\bibfield  {title} {\bibinfo {title} {Chemical trend of nonradiative recombination in cu (in,ga)se2 alloys},\ }\bibfield  {journal} {\bibinfo  {journal} {Physical Review Applied}\ }\textbf {\bibinfo {volume} {19}},\ \href {https://doi.org/10.1103/PhysRevApplied.19.054054} {10.1103/PhysRevApplied.19.054054} (\bibinfo {year} {2023})\BibitemShut {NoStop}%
\bibitem [{\citenamefont {Wang}\ \emph {et~al.}(2024)\citenamefont {Wang}, \citenamefont {Kavanagh}, \citenamefont {Scanlon},\ and\ \citenamefont {Walsh}}]{Wang2024}%
  \BibitemOpen
  \bibfield  {author} {\bibinfo {author} {\bibfnamefont {X.}~\bibnamefont {Wang}}, \bibinfo {author} {\bibfnamefont {S.~R.}\ \bibnamefont {Kavanagh}}, \bibinfo {author} {\bibfnamefont {D.~O.}\ \bibnamefont {Scanlon}},\ and\ \bibinfo {author} {\bibfnamefont {A.}~\bibnamefont {Walsh}},\ }\bibfield  {title} {\bibinfo {title} {Upper efficiency limit of sb2se3 solar cells},\ }\href {https://doi.org/10.1016/j.joule.2024.05.004} {\bibfield  {journal} {\bibinfo  {journal} {Joule}\ }\textbf {\bibinfo {volume} {8}},\ \bibinfo {pages} {2105} (\bibinfo {year} {2024})}\BibitemShut {NoStop}%
\bibitem [{\citenamefont {Yee}\ \emph {et~al.}(2015)\citenamefont {Yee}, \citenamefont {Magyari-Köpe}, \citenamefont {Nishi}, \citenamefont {Bent},\ and\ \citenamefont {Clemens}}]{Yee2015}%
  \BibitemOpen
  \bibfield  {author} {\bibinfo {author} {\bibfnamefont {Y.~S.}\ \bibnamefont {Yee}}, \bibinfo {author} {\bibfnamefont {B.}~\bibnamefont {Magyari-Köpe}}, \bibinfo {author} {\bibfnamefont {Y.}~\bibnamefont {Nishi}}, \bibinfo {author} {\bibfnamefont {S.~F.}\ \bibnamefont {Bent}},\ and\ \bibinfo {author} {\bibfnamefont {B.~M.}\ \bibnamefont {Clemens}},\ }\bibfield  {title} {\bibinfo {title} {Deep recombination centers in c u2znsns e4 revealed by screened-exchange hybrid density functional theory},\ }\bibfield  {journal} {\bibinfo  {journal} {Physical Review B - Condensed Matter and Materials Physics}\ }\textbf {\bibinfo {volume} {92}},\ \href {https://doi.org/10.1103/PhysRevB.92.195201} {10.1103/PhysRevB.92.195201} (\bibinfo {year} {2015})\BibitemShut {NoStop}%
\bibitem [{\citenamefont {Kim}\ \emph {et~al.}(2018)\citenamefont {Kim}, \citenamefont {Park},\ and\ \citenamefont {Walsh}}]{Kim2018}%
  \BibitemOpen
  \bibfield  {author} {\bibinfo {author} {\bibfnamefont {S.}~\bibnamefont {Kim}}, \bibinfo {author} {\bibfnamefont {J.~S.}\ \bibnamefont {Park}},\ and\ \bibinfo {author} {\bibfnamefont {A.}~\bibnamefont {Walsh}},\ }\bibfield  {title} {\bibinfo {title} {Identification of killer defects in kesterite thin-film solar cells},\ }\href {https://doi.org/10.1021/acsenergylett.7b01313} {\bibfield  {journal} {\bibinfo  {journal} {ACS Energy Letters}\ }\textbf {\bibinfo {volume} {3}},\ \bibinfo {pages} {496} (\bibinfo {year} {2018})}\BibitemShut {NoStop}%
\bibitem [{\citenamefont {Li}\ \emph {et~al.}(2019)\citenamefont {Li}, \citenamefont {Yuan}, \citenamefont {Chen}, \citenamefont {Gong},\ and\ \citenamefont {Wei}}]{Li2019}%
  \BibitemOpen
  \bibfield  {author} {\bibinfo {author} {\bibfnamefont {J.}~\bibnamefont {Li}}, \bibinfo {author} {\bibfnamefont {Z.~K.}\ \bibnamefont {Yuan}}, \bibinfo {author} {\bibfnamefont {S.}~\bibnamefont {Chen}}, \bibinfo {author} {\bibfnamefont {X.~G.}\ \bibnamefont {Gong}},\ and\ \bibinfo {author} {\bibfnamefont {S.~H.}\ \bibnamefont {Wei}},\ }\bibfield  {title} {\bibinfo {title} {Effective and noneffective recombination center defects in cu 2 znsns 4 : Significant difference in carrier capture cross sections},\ }\href {https://doi.org/10.1021/acs.chemmater.8b03933} {\bibfield  {journal} {\bibinfo  {journal} {Chemistry of Materials}\ }\textbf {\bibinfo {volume} {31}},\ \bibinfo {pages} {826} (\bibinfo {year} {2019})}\BibitemShut {NoStop}%
\bibitem [{\citenamefont {Todorov}\ \emph {et~al.}(2017)\citenamefont {Todorov}, \citenamefont {Singh}, \citenamefont {Bishop}, \citenamefont {Gunawan}, \citenamefont {Lee}, \citenamefont {Gershon}, \citenamefont {Brew}, \citenamefont {Antunez},\ and\ \citenamefont {Haight}}]{todorov2017ultrathin}%
  \BibitemOpen
  \bibfield  {author} {\bibinfo {author} {\bibfnamefont {T.~K.}\ \bibnamefont {Todorov}}, \bibinfo {author} {\bibfnamefont {S.}~\bibnamefont {Singh}}, \bibinfo {author} {\bibfnamefont {D.~M.}\ \bibnamefont {Bishop}}, \bibinfo {author} {\bibfnamefont {O.}~\bibnamefont {Gunawan}}, \bibinfo {author} {\bibfnamefont {Y.~S.}\ \bibnamefont {Lee}}, \bibinfo {author} {\bibfnamefont {T.~S.}\ \bibnamefont {Gershon}}, \bibinfo {author} {\bibfnamefont {K.~W.}\ \bibnamefont {Brew}}, \bibinfo {author} {\bibfnamefont {P.~D.}\ \bibnamefont {Antunez}},\ and\ \bibinfo {author} {\bibfnamefont {R.}~\bibnamefont {Haight}},\ }\bibfield  {title} {\bibinfo {title} {Ultrathin high band gap solar cells with improved efficiencies from the world’s oldest photovoltaic material},\ }\href@noop {} {\bibfield  {journal} {\bibinfo  {journal} {Nature communications}\ }\textbf {\bibinfo {volume} {8}},\ \bibinfo {pages} {1} (\bibinfo {year} {2017})}\BibitemShut {NoStop}%
\bibitem [{\citenamefont {Nielsen}\ \emph {et~al.}(2024)\citenamefont {Nielsen}, \citenamefont {Schleuning}, \citenamefont {Karalis}, \citenamefont {Hemmingsen}, \citenamefont {Hansen}, \citenamefont {Chorkendorff}, \citenamefont {Unold},\ and\ \citenamefont {Vesborg}}]{Nielsen2024}%
  \BibitemOpen
  \bibfield  {author} {\bibinfo {author} {\bibfnamefont {R.~S.}\ \bibnamefont {Nielsen}}, \bibinfo {author} {\bibfnamefont {M.}~\bibnamefont {Schleuning}}, \bibinfo {author} {\bibfnamefont {O.}~\bibnamefont {Karalis}}, \bibinfo {author} {\bibfnamefont {T.~H.}\ \bibnamefont {Hemmingsen}}, \bibinfo {author} {\bibfnamefont {O.}~\bibnamefont {Hansen}}, \bibinfo {author} {\bibfnamefont {I.}~\bibnamefont {Chorkendorff}}, \bibinfo {author} {\bibfnamefont {T.}~\bibnamefont {Unold}},\ and\ \bibinfo {author} {\bibfnamefont {P.~C.}\ \bibnamefont {Vesborg}},\ }\bibfield  {title} {\bibinfo {title} {Increasing the collection efficiency in selenium thin-film solar cells using a closed-space annealing strategy},\ }\href {https://doi.org/10.1021/acsaem.4c00636} {\bibfield  {journal} {\bibinfo  {journal} {ACS Applied Energy Materials}\ }\textbf {\bibinfo {volume} {7}},\ \bibinfo {pages} {5209} (\bibinfo {year} {2024})}\BibitemShut {NoStop}%
\bibitem [{\citenamefont {Kim}\ \emph {et~al.}(2020{\natexlab{b}})\citenamefont {Kim}, \citenamefont {M{\'a}rquez}, \citenamefont {Unold},\ and\ \citenamefont {Walsh}}]{kim2020upper}%
  \BibitemOpen
  \bibfield  {author} {\bibinfo {author} {\bibfnamefont {S.}~\bibnamefont {Kim}}, \bibinfo {author} {\bibfnamefont {J.~A.}\ \bibnamefont {M{\'a}rquez}}, \bibinfo {author} {\bibfnamefont {T.}~\bibnamefont {Unold}},\ and\ \bibinfo {author} {\bibfnamefont {A.}~\bibnamefont {Walsh}},\ }\bibfield  {title} {\bibinfo {title} {Upper limit to the photovoltaic efficiency of imperfect crystals from first principles},\ }\href@noop {} {\bibfield  {journal} {\bibinfo  {journal} {Energy \& Environmental Science}\ }\textbf {\bibinfo {volume} {13}},\ \bibinfo {pages} {1481} (\bibinfo {year} {2020}{\natexlab{b}})}\BibitemShut {NoStop}%
\bibitem [{\citenamefont {Kim}\ \emph {et~al.}(2019{\natexlab{a}})\citenamefont {Kim}, \citenamefont {Hood},\ and\ \citenamefont {Walsh}}]{Kim2019}%
  \BibitemOpen
  \bibfield  {author} {\bibinfo {author} {\bibfnamefont {S.}~\bibnamefont {Kim}}, \bibinfo {author} {\bibfnamefont {S.~N.}\ \bibnamefont {Hood}},\ and\ \bibinfo {author} {\bibfnamefont {A.}~\bibnamefont {Walsh}},\ }\bibfield  {title} {\bibinfo {title} {Anharmonic lattice relaxation during nonradiative carrier capture},\ }\bibfield  {journal} {\bibinfo  {journal} {Physical Review B}\ }\textbf {\bibinfo {volume} {100}},\ \href {https://doi.org/10.1103/PhysRevB.100.041202} {10.1103/PhysRevB.100.041202} (\bibinfo {year} {2019}{\natexlab{a}})\BibitemShut {NoStop}%
\bibitem [{\citenamefont {Shockley}\ and\ \citenamefont {Queisser}(1961)}]{shockley1961detailed}%
  \BibitemOpen
  \bibfield  {author} {\bibinfo {author} {\bibfnamefont {W.}~\bibnamefont {Shockley}}\ and\ \bibinfo {author} {\bibfnamefont {H.~J.}\ \bibnamefont {Queisser}},\ }\bibfield  {title} {\bibinfo {title} {Detailed balance limit of efficiency of \textit{p-n} junction solar cells},\ }\href@noop {} {\bibfield  {journal} {\bibinfo  {journal} {J. Appl. Phys}\ }\textbf {\bibinfo {volume} {32}},\ \bibinfo {pages} {510} (\bibinfo {year} {1961})}\BibitemShut {NoStop}%
\bibitem [{\citenamefont {Tiedje}\ \emph {et~al.}(1984)\citenamefont {Tiedje}, \citenamefont {Yablonovitch}, \citenamefont {Cody},\ and\ \citenamefont {Brooks}}]{tiedje1984limiting}%
  \BibitemOpen
  \bibfield  {author} {\bibinfo {author} {\bibfnamefont {T.}~\bibnamefont {Tiedje}}, \bibinfo {author} {\bibfnamefont {E.}~\bibnamefont {Yablonovitch}}, \bibinfo {author} {\bibfnamefont {G.~D.}\ \bibnamefont {Cody}},\ and\ \bibinfo {author} {\bibfnamefont {B.~G.}\ \bibnamefont {Brooks}},\ }\bibfield  {title} {\bibinfo {title} {Limiting efficiency of silicon solar cells},\ }\href@noop {} {\bibfield  {journal} {\bibinfo  {journal} {IEEE Transactions on electron devices}\ }\textbf {\bibinfo {volume} {31}},\ \bibinfo {pages} {711} (\bibinfo {year} {1984})}\BibitemShut {NoStop}%
\bibitem [{\citenamefont {Crovetto}\ \emph {et~al.}(2020)\citenamefont {Crovetto}, \citenamefont {Kim}, \citenamefont {Fischer}, \citenamefont {Stenger}, \citenamefont {Walsh}, \citenamefont {Chorkendorff},\ and\ \citenamefont {Vesborg}}]{Crovetto2020}%
  \BibitemOpen
  \bibfield  {author} {\bibinfo {author} {\bibfnamefont {A.}~\bibnamefont {Crovetto}}, \bibinfo {author} {\bibfnamefont {S.}~\bibnamefont {Kim}}, \bibinfo {author} {\bibfnamefont {M.}~\bibnamefont {Fischer}}, \bibinfo {author} {\bibfnamefont {N.}~\bibnamefont {Stenger}}, \bibinfo {author} {\bibfnamefont {A.}~\bibnamefont {Walsh}}, \bibinfo {author} {\bibfnamefont {I.}~\bibnamefont {Chorkendorff}},\ and\ \bibinfo {author} {\bibfnamefont {P.~C.}\ \bibnamefont {Vesborg}},\ }\bibfield  {title} {\bibinfo {title} {Assessing the defect tolerance of kesterite-inspired solar absorbers},\ }\href {https://doi.org/10.1039/d0ee02177f} {\bibfield  {journal} {\bibinfo  {journal} {Energy and Environmental Science}\ }\textbf {\bibinfo {volume} {13}},\ \bibinfo {pages} {3489} (\bibinfo {year} {2020})}\BibitemShut {NoStop}%
\bibitem [{\citenamefont {Scaffidi}\ \emph {et~al.}(2023)\citenamefont {Scaffidi}, \citenamefont {Birant}, \citenamefont {Brammertz}, \citenamefont {de~Wild}, \citenamefont {Flandre},\ and\ \citenamefont {Vermang}}]{scaffidi2023ge}%
  \BibitemOpen
  \bibfield  {author} {\bibinfo {author} {\bibfnamefont {R.}~\bibnamefont {Scaffidi}}, \bibinfo {author} {\bibfnamefont {G.}~\bibnamefont {Birant}}, \bibinfo {author} {\bibfnamefont {G.}~\bibnamefont {Brammertz}}, \bibinfo {author} {\bibfnamefont {J.}~\bibnamefont {de~Wild}}, \bibinfo {author} {\bibfnamefont {D.}~\bibnamefont {Flandre}},\ and\ \bibinfo {author} {\bibfnamefont {B.}~\bibnamefont {Vermang}},\ }\bibfield  {title} {\bibinfo {title} {Ge-alloyed kesterite thin-film solar cells: previous investigations and current status--a comprehensive review},\ }\href@noop {} {\bibfield  {journal} {\bibinfo  {journal} {Journal of Materials Chemistry A}\ }\textbf {\bibinfo {volume} {11}},\ \bibinfo {pages} {13174} (\bibinfo {year} {2023})}\BibitemShut {NoStop}%
\bibitem [{\citenamefont {Liu}\ \emph {et~al.}(2023)\citenamefont {Liu}, \citenamefont {Duan}, \citenamefont {Ma}, \citenamefont {Tian}, \citenamefont {Li}, \citenamefont {Liang}, \citenamefont {Qin},\ and\ \citenamefont {Liu}}]{Liu2023}%
  \BibitemOpen
  \bibfield  {author} {\bibinfo {author} {\bibfnamefont {X.}~\bibnamefont {Liu}}, \bibinfo {author} {\bibfnamefont {K.}~\bibnamefont {Duan}}, \bibinfo {author} {\bibfnamefont {Z.}~\bibnamefont {Ma}}, \bibinfo {author} {\bibfnamefont {X.}~\bibnamefont {Tian}}, \bibinfo {author} {\bibfnamefont {F.}~\bibnamefont {Li}}, \bibinfo {author} {\bibfnamefont {S.}~\bibnamefont {Liang}}, \bibinfo {author} {\bibfnamefont {C.}~\bibnamefont {Qin}},\ and\ \bibinfo {author} {\bibfnamefont {X.}~\bibnamefont {Liu}},\ }\bibfield  {title} {\bibinfo {title} {Enhanced carrier management and photogenerated charge dynamics for selenium (se) film photovoltaics},\ }\bibfield  {journal} {\bibinfo  {journal} {Applied Physics Letters}\ }\textbf {\bibinfo {volume} {123}},\ \href {https://doi.org/10.1063/5.0155161} {10.1063/5.0155161} (\bibinfo {year} {2023})\BibitemShut {NoStop}%
\bibitem [{\citenamefont {P{\"a}ssler}(1976)}]{passler1976relationships}%
  \BibitemOpen
  \bibfield  {author} {\bibinfo {author} {\bibfnamefont {R.}~\bibnamefont {P{\"a}ssler}},\ }\bibfield  {title} {\bibinfo {title} {Relationships between the nonradiative multiphonon carrier-capture properties of deep charged and neutral centres in semiconductors},\ }\href@noop {} {\bibfield  {journal} {\bibinfo  {journal} {physica status solidi (b)}\ }\textbf {\bibinfo {volume} {78}},\ \bibinfo {pages} {625} (\bibinfo {year} {1976})}\BibitemShut {NoStop}%
\bibitem [{\citenamefont {Nielsen}\ \emph {et~al.}(2022)\citenamefont {Nielsen}, \citenamefont {Youngman}, \citenamefont {Moustafa}, \citenamefont {Levcenco}, \citenamefont {Hempel}, \citenamefont {Crovetto}, \citenamefont {Olsen}, \citenamefont {Hansen}, \citenamefont {Chorkendorff}, \citenamefont {Unold},\ and\ \citenamefont {Vesborg}}]{Nielsen2022}%
  \BibitemOpen
  \bibfield  {author} {\bibinfo {author} {\bibfnamefont {R.}~\bibnamefont {Nielsen}}, \bibinfo {author} {\bibfnamefont {T.~H.}\ \bibnamefont {Youngman}}, \bibinfo {author} {\bibfnamefont {H.}~\bibnamefont {Moustafa}}, \bibinfo {author} {\bibfnamefont {S.}~\bibnamefont {Levcenco}}, \bibinfo {author} {\bibfnamefont {H.}~\bibnamefont {Hempel}}, \bibinfo {author} {\bibfnamefont {A.}~\bibnamefont {Crovetto}}, \bibinfo {author} {\bibfnamefont {T.}~\bibnamefont {Olsen}}, \bibinfo {author} {\bibfnamefont {O.}~\bibnamefont {Hansen}}, \bibinfo {author} {\bibfnamefont {I.}~\bibnamefont {Chorkendorff}}, \bibinfo {author} {\bibfnamefont {T.}~\bibnamefont {Unold}},\ and\ \bibinfo {author} {\bibfnamefont {P.~C.}\ \bibnamefont {Vesborg}},\ }\bibfield  {title} {\bibinfo {title} {Origin of photovoltaic losses in selenium solar cells with open-circuit voltages approaching 1 v},\ }\href {https://doi.org/10.1039/d2ta07729a} {\bibfield  {journal} {\bibinfo  {journal} {Journal of Materials Chemistry A}\ }\textbf {\bibinfo {volume}
  {10}},\ \bibinfo {pages} {24199} (\bibinfo {year} {2022})}\BibitemShut {NoStop}%
\bibitem [{\citenamefont {Miller}\ \emph {et~al.}(2012)\citenamefont {Miller}, \citenamefont {Warren}, \citenamefont {Gunawan}, \citenamefont {Gokmen}, \citenamefont {Mitzi},\ and\ \citenamefont {Cohen}}]{miller2012electronically}%
  \BibitemOpen
  \bibfield  {author} {\bibinfo {author} {\bibfnamefont {D.~W.}\ \bibnamefont {Miller}}, \bibinfo {author} {\bibfnamefont {C.~W.}\ \bibnamefont {Warren}}, \bibinfo {author} {\bibfnamefont {O.}~\bibnamefont {Gunawan}}, \bibinfo {author} {\bibfnamefont {T.}~\bibnamefont {Gokmen}}, \bibinfo {author} {\bibfnamefont {D.~B.}\ \bibnamefont {Mitzi}},\ and\ \bibinfo {author} {\bibfnamefont {J.~D.}\ \bibnamefont {Cohen}},\ }\bibfield  {title} {\bibinfo {title} {Electronically active defects in the cu2znsn (se, s) 4 alloys as revealed by transient photocapacitance spectroscopy},\ }\href@noop {} {\bibfield  {journal} {\bibinfo  {journal} {Applied Physics Letters}\ }\textbf {\bibinfo {volume} {101}} (\bibinfo {year} {2012})}\BibitemShut {NoStop}%
\bibitem [{\citenamefont {Islam}\ \emph {et~al.}(2015)\citenamefont {Islam}, \citenamefont {Halim}, \citenamefont {Sakurai}, \citenamefont {Sakai}, \citenamefont {Kato}, \citenamefont {Sugimoto}, \citenamefont {Tampo}, \citenamefont {Shibata}, \citenamefont {Niki},\ and\ \citenamefont {Akimoto}}]{islam2015determination}%
  \BibitemOpen
  \bibfield  {author} {\bibinfo {author} {\bibfnamefont {M.~M.}\ \bibnamefont {Islam}}, \bibinfo {author} {\bibfnamefont {M.~A.}\ \bibnamefont {Halim}}, \bibinfo {author} {\bibfnamefont {T.}~\bibnamefont {Sakurai}}, \bibinfo {author} {\bibfnamefont {N.}~\bibnamefont {Sakai}}, \bibinfo {author} {\bibfnamefont {T.}~\bibnamefont {Kato}}, \bibinfo {author} {\bibfnamefont {H.}~\bibnamefont {Sugimoto}}, \bibinfo {author} {\bibfnamefont {H.}~\bibnamefont {Tampo}}, \bibinfo {author} {\bibfnamefont {H.}~\bibnamefont {Shibata}}, \bibinfo {author} {\bibfnamefont {S.}~\bibnamefont {Niki}},\ and\ \bibinfo {author} {\bibfnamefont {K.}~\bibnamefont {Akimoto}},\ }\bibfield  {title} {\bibinfo {title} {Determination of deep-level defects in cu2znsn (s, se) 4 thin-films using photocapacitance method},\ }\href@noop {} {\bibfield  {journal} {\bibinfo  {journal} {Applied Physics Letters}\ }\textbf {\bibinfo {volume} {106}} (\bibinfo {year} {2015})}\BibitemShut {NoStop}%
\bibitem [{\citenamefont {Ratz}\ \emph {et~al.}(2022)\citenamefont {Ratz}, \citenamefont {Nguyen}, \citenamefont {Brammertz}, \citenamefont {Vermang},\ and\ \citenamefont {Raty}}]{ratz2022relevance}%
  \BibitemOpen
  \bibfield  {author} {\bibinfo {author} {\bibfnamefont {T.}~\bibnamefont {Ratz}}, \bibinfo {author} {\bibfnamefont {N.~D.}\ \bibnamefont {Nguyen}}, \bibinfo {author} {\bibfnamefont {G.}~\bibnamefont {Brammertz}}, \bibinfo {author} {\bibfnamefont {B.}~\bibnamefont {Vermang}},\ and\ \bibinfo {author} {\bibfnamefont {J.-Y.}\ \bibnamefont {Raty}},\ }\bibfield  {title} {\bibinfo {title} {Relevance of ge incorporation to control the physical behaviour of point defects in kesterite},\ }\href@noop {} {\bibfield  {journal} {\bibinfo  {journal} {Journal of Materials Chemistry A}\ }\textbf {\bibinfo {volume} {10}},\ \bibinfo {pages} {4355} (\bibinfo {year} {2022})}\BibitemShut {NoStop}%
\bibitem [{\citenamefont {Xu}\ \emph {et~al.}(2021)\citenamefont {Xu}, \citenamefont {Yang}, \citenamefont {Chen},\ and\ \citenamefont {Gong}}]{Xu2021}%
  \BibitemOpen
  \bibfield  {author} {\bibinfo {author} {\bibfnamefont {Y.}~\bibnamefont {Xu}}, \bibinfo {author} {\bibfnamefont {J.~H.}\ \bibnamefont {Yang}}, \bibinfo {author} {\bibfnamefont {S.}~\bibnamefont {Chen}},\ and\ \bibinfo {author} {\bibfnamefont {X.~G.}\ \bibnamefont {Gong}},\ }\bibfield  {title} {\bibinfo {title} {Defect-assisted nonradiative recombination in cu2znsnse4: A comparative study with cu2znsn s4},\ }\bibfield  {journal} {\bibinfo  {journal} {Physical Review Materials}\ }\textbf {\bibinfo {volume} {5}},\ \href {https://doi.org/10.1103/PhysRevMaterials.5.025403} {10.1103/PhysRevMaterials.5.025403} (\bibinfo {year} {2021})\BibitemShut {NoStop}%
\bibitem [{\citenamefont {Han}\ \emph {et~al.}(2013)\citenamefont {Han}, \citenamefont {Sun}, \citenamefont {Bang}, \citenamefont {Zhang}, \citenamefont {Sun}, \citenamefont {Li},\ and\ \citenamefont {Zhang}}]{Han2013}%
  \BibitemOpen
  \bibfield  {author} {\bibinfo {author} {\bibfnamefont {D.}~\bibnamefont {Han}}, \bibinfo {author} {\bibfnamefont {Y.~Y.}\ \bibnamefont {Sun}}, \bibinfo {author} {\bibfnamefont {J.}~\bibnamefont {Bang}}, \bibinfo {author} {\bibfnamefont {Y.~Y.}\ \bibnamefont {Zhang}}, \bibinfo {author} {\bibfnamefont {H.~B.}\ \bibnamefont {Sun}}, \bibinfo {author} {\bibfnamefont {X.~B.}\ \bibnamefont {Li}},\ and\ \bibinfo {author} {\bibfnamefont {S.~B.}\ \bibnamefont {Zhang}},\ }\bibfield  {title} {\bibinfo {title} {Deep electron traps and origin of p-type conductivity in the earth-abundant solar-cell material cu2znsns4},\ }\bibfield  {journal} {\bibinfo  {journal} {Physical Review B - Condensed Matter and Materials Physics}\ }\textbf {\bibinfo {volume} {87}},\ \href {https://doi.org/10.1103/PhysRevB.87.155206} {10.1103/PhysRevB.87.155206} (\bibinfo {year} {2013})\BibitemShut {NoStop}%
\bibitem [{\citenamefont {Kim}\ \emph {et~al.}(2019{\natexlab{b}})\citenamefont {Kim}, \citenamefont {Park}, \citenamefont {Hood},\ and\ \citenamefont {Walsh}}]{kim2019lone}%
  \BibitemOpen
  \bibfield  {author} {\bibinfo {author} {\bibfnamefont {S.}~\bibnamefont {Kim}}, \bibinfo {author} {\bibfnamefont {J.-S.}\ \bibnamefont {Park}}, \bibinfo {author} {\bibfnamefont {S.~N.}\ \bibnamefont {Hood}},\ and\ \bibinfo {author} {\bibfnamefont {A.}~\bibnamefont {Walsh}},\ }\bibfield  {title} {\bibinfo {title} {Lone-pair effect on carrier capture in cu 2 znsns 4 solar cells},\ }\href@noop {} {\bibfield  {journal} {\bibinfo  {journal} {Journal of Materials Chemistry A}\ }\textbf {\bibinfo {volume} {7}},\ \bibinfo {pages} {2686} (\bibinfo {year} {2019}{\natexlab{b}})}\BibitemShut {NoStop}%
\bibitem [{\citenamefont {Mortensen}\ \emph {et~al.}(2024)\citenamefont {Mortensen}, \citenamefont {Larsen}, \citenamefont {Kuisma}, \citenamefont {Ivanov}, \citenamefont {Taghizadeh}, \citenamefont {Peterson}, \citenamefont {Haldar}, \citenamefont {Dohn}, \citenamefont {Sch{\"a}fer}, \citenamefont {J{\'o}nsson} \emph {et~al.}}]{mortensen2024gpaw}%
  \BibitemOpen
  \bibfield  {author} {\bibinfo {author} {\bibfnamefont {J.~J.}\ \bibnamefont {Mortensen}}, \bibinfo {author} {\bibfnamefont {A.~H.}\ \bibnamefont {Larsen}}, \bibinfo {author} {\bibfnamefont {M.}~\bibnamefont {Kuisma}}, \bibinfo {author} {\bibfnamefont {A.~V.}\ \bibnamefont {Ivanov}}, \bibinfo {author} {\bibfnamefont {A.}~\bibnamefont {Taghizadeh}}, \bibinfo {author} {\bibfnamefont {A.}~\bibnamefont {Peterson}}, \bibinfo {author} {\bibfnamefont {A.}~\bibnamefont {Haldar}}, \bibinfo {author} {\bibfnamefont {A.~O.}\ \bibnamefont {Dohn}}, \bibinfo {author} {\bibfnamefont {C.}~\bibnamefont {Sch{\"a}fer}}, \bibinfo {author} {\bibfnamefont {E.~{\"O}.}\ \bibnamefont {J{\'o}nsson}}, \emph {et~al.},\ }\bibfield  {title} {\bibinfo {title} {Gpaw: An open python package for electronic structure calculations},\ }\href@noop {} {\bibfield  {journal} {\bibinfo  {journal} {The Journal of Chemical Physics}\ }\textbf {\bibinfo {volume} {160}} (\bibinfo {year} {2024})}\BibitemShut {NoStop}%
\bibitem [{\citenamefont {Larsen}\ \emph {et~al.}(2017)\citenamefont {Larsen}, \citenamefont {Mortensen}, \citenamefont {Blomqvist}, \citenamefont {Castelli}, \citenamefont {Christensen}, \citenamefont {Du{\l}ak}, \citenamefont {Friis}, \citenamefont {Groves}, \citenamefont {Hammer}, \citenamefont {Hargus} \emph {et~al.}}]{larsen2017atomic}%
  \BibitemOpen
  \bibfield  {author} {\bibinfo {author} {\bibfnamefont {A.~H.}\ \bibnamefont {Larsen}}, \bibinfo {author} {\bibfnamefont {J.~J.}\ \bibnamefont {Mortensen}}, \bibinfo {author} {\bibfnamefont {J.}~\bibnamefont {Blomqvist}}, \bibinfo {author} {\bibfnamefont {I.~E.}\ \bibnamefont {Castelli}}, \bibinfo {author} {\bibfnamefont {R.}~\bibnamefont {Christensen}}, \bibinfo {author} {\bibfnamefont {M.}~\bibnamefont {Du{\l}ak}}, \bibinfo {author} {\bibfnamefont {J.}~\bibnamefont {Friis}}, \bibinfo {author} {\bibfnamefont {M.~N.}\ \bibnamefont {Groves}}, \bibinfo {author} {\bibfnamefont {B.}~\bibnamefont {Hammer}}, \bibinfo {author} {\bibfnamefont {C.}~\bibnamefont {Hargus}}, \emph {et~al.},\ }\bibfield  {title} {\bibinfo {title} {The atomic simulation environment—a python library for working with atoms},\ }\href@noop {} {\bibfield  {journal} {\bibinfo  {journal} {Journal of Physics: Condensed Matter}\ }\textbf {\bibinfo {volume} {29}},\ \bibinfo {pages} {273002} (\bibinfo {year} {2017})}\BibitemShut {NoStop}%
\bibitem [{\citenamefont {Kangsabanik}\ \emph {et~al.}(2022)\citenamefont {Kangsabanik}, \citenamefont {Svendsen}, \citenamefont {Taghizadeh}, \citenamefont {Crovetto},\ and\ \citenamefont {Thygesen}}]{kangsabanik2022indirect}%
  \BibitemOpen
  \bibfield  {author} {\bibinfo {author} {\bibfnamefont {J.}~\bibnamefont {Kangsabanik}}, \bibinfo {author} {\bibfnamefont {M.~K.}\ \bibnamefont {Svendsen}}, \bibinfo {author} {\bibfnamefont {A.}~\bibnamefont {Taghizadeh}}, \bibinfo {author} {\bibfnamefont {A.}~\bibnamefont {Crovetto}},\ and\ \bibinfo {author} {\bibfnamefont {K.~S.}\ \bibnamefont {Thygesen}},\ }\bibfield  {title} {\bibinfo {title} {Indirect band gap semiconductors for thin-film photovoltaics: High-throughput calculation of phonon-assisted absorption},\ }\href@noop {} {\bibfield  {journal} {\bibinfo  {journal} {Journal of the American Chemical Society}\ }\textbf {\bibinfo {volume} {144}},\ \bibinfo {pages} {19872} (\bibinfo {year} {2022})}\BibitemShut {NoStop}%
\bibitem [{\citenamefont {Heyd}\ \emph {et~al.}(2003)\citenamefont {Heyd}, \citenamefont {Scuseria},\ and\ \citenamefont {Ernzerhof}}]{heyd2003hybrid}%
  \BibitemOpen
  \bibfield  {author} {\bibinfo {author} {\bibfnamefont {J.}~\bibnamefont {Heyd}}, \bibinfo {author} {\bibfnamefont {G.~E.}\ \bibnamefont {Scuseria}},\ and\ \bibinfo {author} {\bibfnamefont {M.}~\bibnamefont {Ernzerhof}},\ }\bibfield  {title} {\bibinfo {title} {Hybrid functionals based on a screened coulomb potential},\ }\href@noop {} {\bibfield  {journal} {\bibinfo  {journal} {The Journal of chemical physics}\ }\textbf {\bibinfo {volume} {118}},\ \bibinfo {pages} {8207} (\bibinfo {year} {2003})}\BibitemShut {NoStop}%
\bibitem [{\citenamefont {Freysoldt}\ \emph {et~al.}(2009)\citenamefont {Freysoldt}, \citenamefont {Neugebauer},\ and\ \citenamefont {Van~de Walle}}]{freysoldt2009fully}%
  \BibitemOpen
  \bibfield  {author} {\bibinfo {author} {\bibfnamefont {C.}~\bibnamefont {Freysoldt}}, \bibinfo {author} {\bibfnamefont {J.}~\bibnamefont {Neugebauer}},\ and\ \bibinfo {author} {\bibfnamefont {C.~G.}\ \bibnamefont {Van~de Walle}},\ }\bibfield  {title} {\bibinfo {title} {Fully ab initio finite-size corrections for charged-defect supercell calculations},\ }\href@noop {} {\bibfield  {journal} {\bibinfo  {journal} {Physical review letters}\ }\textbf {\bibinfo {volume} {102}},\ \bibinfo {pages} {016402} (\bibinfo {year} {2009})}\BibitemShut {NoStop}%
\bibitem [{\citenamefont {Gjerding}\ \emph {et~al.}(2021)\citenamefont {Gjerding}, \citenamefont {Skovhus}, \citenamefont {Rasmussen}, \citenamefont {Bertoldo}, \citenamefont {Larsen}, \citenamefont {Mortensen},\ and\ \citenamefont {Thygesen}}]{gjerding2021atomic}%
  \BibitemOpen
  \bibfield  {author} {\bibinfo {author} {\bibfnamefont {M.}~\bibnamefont {Gjerding}}, \bibinfo {author} {\bibfnamefont {T.}~\bibnamefont {Skovhus}}, \bibinfo {author} {\bibfnamefont {A.}~\bibnamefont {Rasmussen}}, \bibinfo {author} {\bibfnamefont {F.}~\bibnamefont {Bertoldo}}, \bibinfo {author} {\bibfnamefont {A.~H.}\ \bibnamefont {Larsen}}, \bibinfo {author} {\bibfnamefont {J.~J.}\ \bibnamefont {Mortensen}},\ and\ \bibinfo {author} {\bibfnamefont {K.~S.}\ \bibnamefont {Thygesen}},\ }\bibfield  {title} {\bibinfo {title} {Atomic simulation recipes: A python framework and library for automated workflows},\ }\href@noop {} {\bibfield  {journal} {\bibinfo  {journal} {Computational Materials Science}\ }\textbf {\bibinfo {volume} {199}},\ \bibinfo {pages} {110731} (\bibinfo {year} {2021})}\BibitemShut {NoStop}%
\bibitem [{\citenamefont {Mortensen}\ \emph {et~al.}(2020)\citenamefont {Mortensen}, \citenamefont {Gjerding},\ and\ \citenamefont {Thygesen}}]{mortensen2020myqueue}%
  \BibitemOpen
  \bibfield  {author} {\bibinfo {author} {\bibfnamefont {J.~J.}\ \bibnamefont {Mortensen}}, \bibinfo {author} {\bibfnamefont {M.}~\bibnamefont {Gjerding}},\ and\ \bibinfo {author} {\bibfnamefont {K.~S.}\ \bibnamefont {Thygesen}},\ }\bibfield  {title} {\bibinfo {title} {Myqueue: Task and workflow scheduling system},\ }\href@noop {} {\bibfield  {journal} {\bibinfo  {journal} {Journal of Open Source Software}\ }\textbf {\bibinfo {volume} {5}},\ \bibinfo {pages} {1844} (\bibinfo {year} {2020})}\BibitemShut {NoStop}%
\bibitem [{\citenamefont {Zhang}\ \emph {et~al.}(2017)\citenamefont {Zhang}, \citenamefont {Shi}, \citenamefont {Yang}, \citenamefont {Zhao}, \citenamefont {Xu},\ and\ \citenamefont {Wang}}]{zhang2017first}%
  \BibitemOpen
  \bibfield  {author} {\bibinfo {author} {\bibfnamefont {H.-S.}\ \bibnamefont {Zhang}}, \bibinfo {author} {\bibfnamefont {L.}~\bibnamefont {Shi}}, \bibinfo {author} {\bibfnamefont {X.-B.}\ \bibnamefont {Yang}}, \bibinfo {author} {\bibfnamefont {Y.-J.}\ \bibnamefont {Zhao}}, \bibinfo {author} {\bibfnamefont {K.}~\bibnamefont {Xu}},\ and\ \bibinfo {author} {\bibfnamefont {L.-W.}\ \bibnamefont {Wang}},\ }\bibfield  {title} {\bibinfo {title} {First-principles calculations of quantum efficiency for point defects in semiconductors: The example of yellow luminance by gan: Cn+ on and gan: Cn},\ }\href@noop {} {\bibfield  {journal} {\bibinfo  {journal} {Advanced Optical Materials}\ }\textbf {\bibinfo {volume} {5}},\ \bibinfo {pages} {1700404} (\bibinfo {year} {2017})}\BibitemShut {NoStop}%
\bibitem [{\citenamefont {Turiansky}\ \emph {et~al.}(2021)\citenamefont {Turiansky}, \citenamefont {Alkauskas}, \citenamefont {Engel}, \citenamefont {Kresse}, \citenamefont {Wickramaratne}, \citenamefont {Shen}, \citenamefont {Dreyer},\ and\ \citenamefont {Van~de Walle}}]{turiansky2021nonrad}%
  \BibitemOpen
  \bibfield  {author} {\bibinfo {author} {\bibfnamefont {M.~E.}\ \bibnamefont {Turiansky}}, \bibinfo {author} {\bibfnamefont {A.}~\bibnamefont {Alkauskas}}, \bibinfo {author} {\bibfnamefont {M.}~\bibnamefont {Engel}}, \bibinfo {author} {\bibfnamefont {G.}~\bibnamefont {Kresse}}, \bibinfo {author} {\bibfnamefont {D.}~\bibnamefont {Wickramaratne}}, \bibinfo {author} {\bibfnamefont {J.-X.}\ \bibnamefont {Shen}}, \bibinfo {author} {\bibfnamefont {C.~E.}\ \bibnamefont {Dreyer}},\ and\ \bibinfo {author} {\bibfnamefont {C.~G.}\ \bibnamefont {Van~de Walle}},\ }\bibfield  {title} {\bibinfo {title} {Nonrad: Computing nonradiative capture coefficients from first principles},\ }\href@noop {} {\bibfield  {journal} {\bibinfo  {journal} {Computer Physics Communications}\ }\textbf {\bibinfo {volume} {267}},\ \bibinfo {pages} {108056} (\bibinfo {year} {2021})}\BibitemShut {NoStop}%
\bibitem [{\citenamefont {Kim}\ \emph {et~al.}(2020{\natexlab{c}})\citenamefont {Kim}, \citenamefont {Hood}, \citenamefont {van Gerwen}, \citenamefont {Whalley},\ and\ \citenamefont {Walsh}}]{kim2020carriercapture}%
  \BibitemOpen
  \bibfield  {author} {\bibinfo {author} {\bibfnamefont {S.}~\bibnamefont {Kim}}, \bibinfo {author} {\bibfnamefont {S.~N.}\ \bibnamefont {Hood}}, \bibinfo {author} {\bibfnamefont {P.}~\bibnamefont {van Gerwen}}, \bibinfo {author} {\bibfnamefont {L.~D.}\ \bibnamefont {Whalley}},\ and\ \bibinfo {author} {\bibfnamefont {A.}~\bibnamefont {Walsh}},\ }\bibfield  {title} {\bibinfo {title} {Carriercapture. jl: Anharmonic carrier capture},\ }\href@noop {} {\bibfield  {journal} {\bibinfo  {journal} {Journal of Open Source Software}\ }\textbf {\bibinfo {volume} {5}},\ \bibinfo {pages} {2102} (\bibinfo {year} {2020}{\natexlab{c}})}\BibitemShut {NoStop}%
\end{thebibliography}%


\begin{thebibliography}{6}

\bibitem{S1}
A. Alkauskas, Q. Yan, and C. G. Van de Walle, First-principles theory of nonradiative carrier capture via multiphonon emission, \textit{Physical Review B} \textbf{90}, 075202 (2014).

\bibitem{S2}
R. Pässler, Relationships between the nonradiative multiphonon carrier-capture properties of deep charged and neutral centres in semiconductors, \textit{physica status solidi (b)} \textbf{78}, 625 (1976).

\bibitem{S3}
S. Kim, J. A. Márquez, T. Unold, and A. Walsh, Upper limit to the photovoltaic efficiency of imperfect crystals from first principles, \textit{Energy \& Environmental Science} \textbf{13}, 1481 (2020).

\bibitem{S4}
W. Shockley and H. J. Queisser, Detailed balance limit of efficiency of \textit{p-n} junction solar cells, \textit{J. Appl. Phys} \textbf{32}, 510 (1961).

\bibitem{S5}
J. Kangsabanik, M. K. Svendsen, A. Taghizadeh, A. Crovetto, and K. S. Thygesen, Indirect band gap semiconductors for thin-film photovoltaics: High-throughput calculation of phonon-assisted absorption, \textit{Journal of the American Chemical Society} \textbf{144}, 19872 (2022).

\bibitem{S6}
T. Tiedje, E. Yablonovitch, G. D. Cody, and B. G. Brooks, Limiting efficiency of silicon solar cells, \textit{IEEE Transactions on Electron Devices} \textbf{31}, 711 (1984).

\end{thebibliography}
